\documentclass[manuscript,screen]{acmart} 
\usepackage{amsmath,amsfonts}
\usepackage{algorithm,algorithmic,setspace}
\usepackage{graphicx}
\usepackage{xcolor}
\usepackage{multirow}
\usepackage{subcaption}
\usepackage{lscape}
\usepackage{tabularx}
\usepackage{geometry}
\usepackage{longtable}
\usepackage{tablefootnote}

\newcolumntype{C}[1]{>{\centering\arraybackslash}m{#1}}
\AtBeginDocument{%
  \providecommand\BibTeX{{%
    \normalfont B\kern-0.5em{\scshape i\kern-0.25em b}\kern-0.8em\TeX}}}

\setcopyright{acmcopyright}
\copyrightyear{2022}
\acmYear{2022}
\acmDOI{00.0000/1122445.1122456}

\begin{document}

\title{A Survey on Password Guessing}



\author{Lam Tran}
\email{halam189@gmail.com}
\orcid{0000-0002-8080-7271}
\affiliation{%
	\institution{Kongju National University}
	\city{Kongju}
	\country{South Korea}
}

\author{Thuc Nguyen}
\email{ndthuc@fit.hcmus.edu.vn}
\orcid{0000-0003-0524-8841}
\affiliation{%
	\institution{Ho Chi Minh University of Science}
	\city{Ho Chi Minh}
	\country{Vietnam}
}

\author{Changho Seo}
\email{chseo@kongju.ac.kr}
\orcid{0000-0002-0779-3539}
\affiliation{%
	\institution{Kongju National University}
	\city{Kongju}
	\country{South Korea}
}

\author{Hyunil Kim}
\authornote{Corresponding Authors: Hyunil Kim, Deokjai Choi}
\email{hyunil89@kongju.ac.kr}
\orcid{0000-0002-4018-4540}
\affiliation{%
	\institution{Kongju National University}
	\city{Kongju}
	\country{South Korea}
}

\author{Deokjai Choi}
\email{dchoi@jnu.ac.kr}
\orcid{0000-0001-9502-9882}
\authornotemark[1]
\affiliation{%
	\institution{Chonnam National University}
	\city{Gwangju}
	\country{South Korea}
}

%
%
%
%
%
%

\renewcommand{\shortauthors}{Lam Tran, et al.}

\newcommand{\ie}{\textit{i.e.}}
\newcommand{\eg}{\textit{e.g.}}
\newcommand{\tb}[1]{\textbf{#1}}
\newcommand{\ti}[1]{\textit{#1}}
\newcommand{\cm}[1]{\textcolor{blue}{#1}}
\newcommand{\ea}{\textit{et al.}}
\newcommand{\quotes}[1]{``#1''}
\newcommand\tab[1][0.5cm]{\hspace*{#1}}


\begin{abstract}
	Text password has served as the most popular method for user authentication so far, and is not likely to be totally replaced in foreseeable future. 
	Password authentication offers several desirable properties (e.g., low-cost, highly available, easy-to-implement, reusable). However, it suffers from a critical security issue mainly caused by the inability to memorize complicated strings of humans. 
	Users tend to choose easy-to-remember passwords which are not uniformly distributed in the key space.
	Thus, user-selected passwords are susceptible to guessing attacks. 
	In order to encourage and support users to use strong passwords, it is necessary to simulate automated password guessing methods to determine the passwords' strength and identify weak passwords. 
	A large number of password guessing models have been proposed in the literature. 
	However, little attention was paid to the task of providing a systematic survey which is necessary to review the state-of-the-art approaches, identify gaps, and avoid duplicate studies. 
	Motivated by that, we conduct a comprehensive survey on all password guessing studies presented in the literature from 1979 to 2022. 
	We propose a generic methodology map to present an overview of existing methods. 
	Then, we explain each representative approach in detail. 
	The experimental procedures and available datasets used to evaluate password guessing models are summarized, and the reported performances of representative studies are compared. 
	Finally, the current limitations and the open problems as future research directions are discussed. 
	We believe that this survey is helpful to both experts and newcomers who are interested in password security.
\end{abstract}



\maketitle

\section{Introduction}\label{sec_intro}

Text password has served as the most popular user authentication method so far.
The attempted user is verified by checking whether he knows the password (\ie, secret text string) previously registered by the legitimate user \cite{lamport1981password}.
Password offers a low-cost, highly available, easy-to-implement, and reusable authentication mechanism. 
However, due to the inability to memorize high-entropy strings, users usually select easy-to-memorize passwords (\eg, simple, familiar strings) \cite{ur2015added,bonneau2012linguistic,wang2019birthday,wei2018password,yang2021studies,shay2010encountering}, and use similar passwords for different services \cite{pearman2017let,haque2013study,gaw2006password}.
So, human-chosen passwords are not uniformly distributed in the exponentially large key space, but relatively gather in few clusters.
Thus, the real-world passwords are effectively modeled by attacker, making the authentication systems susceptible to guessing attacks.
As a result, password is usually the weakest point in the whole security chain \cite{taneski2019systematic}.
Despite the notorious weaknesses, it is not likely that passwords will be replaced in a foreseeable future \cite{clair2006password,herley2011research,bonneau2015passwords}.
Thus, significant effort has been invested in improving password strength against guessing attacks.
In the early time, password policy was widely adopted to prevent users from using weak passwords. 
A password policy consists of a set of heuristic rules specifying some properties that a password must satisfy (\eg, having at least $8$ characters, containing both lower-case and upper-case letters) \cite{summers2004password,wang2015emperor}.
However, prior studies showed that password policy could not improve the password strength in practice as the users tend to bypass policies with some tricks such as using some common patterns (\eg, first character as an upper-case letter) \cite{ur2015added}, including personal information (\eg, birth dates, name, email...) \cite{veras2012visualizing,wang2019birthday,bonneau2012birthday,bryant2006user,krishnamurthy2009leakage}, using same/similar passwords for multiple services \cite{stobert2014password,wash2016understanding,haque2013study,gaw2006password,habib2018user}, using leet speak (\eg, \quotes{$\mathsf{password}$} to \quotes{$\mathsf{p4s5w0rd}$}) \cite{golla2016security}.
An over-strictly policy, on the other hand, leads to impossible-to-memorize passwords, and users will store them somewhere which causes other security issues \cite{zhao2013all,gaw2006password,stobert2018password,inglesant2010true}.
Moreover, password policies are also unable to educate the users on how weak their passwords are and how to construct strong passwords.

In order to encourage and provide soundness advice that supports users to use strong passwords, it is necessary to simulate automated password guessing attacks, to measure how strong a password is and which weakness it suffers \cite{ur2015measuring,melicher2016fast}.
In other words, simulating password guessing (as close as possible to the real attacker's behavior) is critical to ensure the security of passwords against real guessing attacks.
Specifically, such password guessing methods can be used to construct a proactive password strength meter that detects weak passwords in the password registering step \cite{kelley2012guess,golla2018accuracy}. 
The administrators can also rely on that password guessing method to periodically evaluate the password strength of existing accounts.
Since the first official report of Morris \ea \ \cite{morris1979password}, many password guessing studies have been proposed in the literature.
In the early stage, the research effort mainly focused on discovering the potential information sources that can assist in providing high-hitting-chance guesses (\eg, dictionary \cite{bonneau2012linguistic,chou2013password}, keyboard \cite{schweitzer2011visualizing,li2014large}, leaked password database \cite{zhang2010security}, language property \cite{li2014large}). %
The later works leveraged modern machine learning and deep learning techniques to develop more sophisticated and robust password guessing methods (\eg, Probabilistic Context-free Grammar \cite{weir2009password,chou2013password}, Markov chain \cite{narayanan2005fast,durmuth2015omen}, Long Short-term Memory Network \cite{xu2017password,fang2018password}, Generative Adversarial Network \cite{hitaj2019passgan,guo2021pggan}).
Despite the merits of existing studies, the automated password guessing models are still far from the real-world attacker.

With the increase of data breaches occurring continuously (see Table \ref{tab_recent_leaked} for examples), educating user on password composition to improve the resistance against guessing attack is more critical.
For the advancement of a research topic, literature reviews are necessary to provide important concepts and summarize the existing advances of the topic.
This allows newly participated researchers to quickly obtain background knowledge, identify gaps, and avoid duplicative studies.
Moreover, understanding how the password guessing methods work also helps the administrator develops effective defensive mechanisms. 
However, little effort has been invested in the task of providing a systematic review of the password guessing studies. 
Motivated by that, in this study, we conduct a systematic survey on existing password guessing studies that provides not only an overview of the research field, but also the detail of each subsequent approach. 
Specifically, this survey aims to answer the following questions.
\ti{(Q1)} What are the ethical uses of password guessing in practice, and how can a password guessing model be used to improve the security of password authentication? 
\ti{(Q2)} What information sources can be inferred to construct a password guessing model? 
In other words, what are the vulnerabilities in the password composing behavior of users that enable effective password guessing? 
\ti{(Q3)} Which algorithms have been developed to produce candidate passwords from the information sources?
\ti{(Q4)} How are the effectiveness (\ie, performance) and reliability of existing password guessing models? How are they evaluated?
\ti{(Q5)} What are the open challenges that need more future research attention?
The answers of these questions form the main contributions of this survey, which are outlined as follows:
\begin{itemize}
	\item We conduct the first comprehensive and systematic review of all password guessing studies proposed in the literature from 1979 to 2022.
	
	\item We briefly summarized the evolution of password guessing and present its ethical applications.
	
	\item 
	We conduct a methodology map of password guessing models to provide an overview of the research field.
	
	\item We systematically present each existing strategy and technique for password guessing.
	
	\item We also present the available datasets and procedures for evaluating a password guessing model.
	Then, the performance of representative researches are summarized.
	\item 
	Finally, the current limitations in password guessing approaches and the open problems as future research directions are then discussed.
	
\end{itemize}
 
The remaining parts of the paper are organized as follows.
In Section \ref{sec_history_rel_work}, we briefly summarize the evolution of password guessing techniques, discuss the ethical applications, and review the related works of password security research.
Section \ref{sec_guessing_review} presents a systematic review of existing password guessing studies. 
In this section, we propose a methodology map to provide an overview of password guessing methods.
Then, each approach depicted on the map is described in detail.
In Section \ref{sssec_future_work}, we discuss the remaining challenges and open problems that need more future research effort.
Finally, a conclusion is drawn in Section \ref{sec_conclusion}.

\begin{table}
	\centering
	\caption{Some of recent data breaches that include passwords (source: \href{https://haveibeenpwned.com/}{https://haveibeenpwned.com/})}
	\label{tab_recent_leaked}
	\def\arraystretch{1.2}
	\begin{tabular}{|lcrcC{7cm}|} \hline
		\multirow{2}{*}{\ti{\tb{Service}}}& \ti{\tb{Leaked}} & \ti{\tb{\centering No. of }} &\ti{\tb{Data}}& \multirow{2}{*}{\ti{\tb{Other Information}}} \\
		& \ti{\tb{Time }} & \ti{\tb{\centering Records}} &\ti{\tb{Type}}&   \\
		\hline
		BrandNew Tube\tablefootnote{\href{https://brandnewtube.com/}{https://brandnewtube.com/}} &
		Aug, 2022& 
		$350$K&
		Plaintext &
		Email, gender, IP address, private message, username\\
		\hline
		START\tablefootnote{\href{https://start.ru/}{https://start.ru/}} &
		Aug, 2022&
		$44$M&
		Hashed 	&
		Email, country\\
		\hline
		%
		%
		%
		%
		SitePoint\tablefootnote{\href{https://www.sitepoint.com/}{https://www.sitepoint.com/}} &
		June, 2020&
		$1$M&
		Hashed&
		Email, IP address, name, username\\
		\hline
		Tuned Global\tablefootnote{\href{https://www.tunedglobal.com/}{https://www.tunedglobal.com/}} &
		Jan, 2021&
		$985$K&
		Plaintext&
		Email, names, physical addresses, phone number\\
		\hline
		%
		%
		%
		%
		Unico Campania\tablefootnote{\href{https://www.unicocampania.it/}{https://www.unicocampania.it/}} &
		Aug, 2020&
		$166$K&
		Plaintext&
		Email address\\
		\hline 
		
%
		Vakinha\tablefootnote{\href{https://www.vakinha.com.br/}{https://www.vakinha.com.br/}} &
		Jun, 2020&
		$4.77$M&
		Hashed&
		Email, IP address, name, birth date, Phone number\\
		\hline
		
		Vedantu\tablefootnote{\href{https://www.vedantu.com/}{https://www.vedantu.com/}} &
		Jul, 2019&
		$687$K&
		Hashed&
		Email, gender, IP address, name, phone number, spoken language...\\
		\hline
		
	\end{tabular}
\end{table} 

\section{History and Related Work}\label{sec_history_rel_work}
In this section, we briefly present the evolution history of password guessing research.
Then, we introduce some ethical applications of password guessing in practice.
Finally, we discuss existing surveys on related research of password security.

\subsection{The Evolution of Password Guessing}

Password guessing attack has been used in practice for a long time \cite{branka2022ken}.
The first official report of password guessing was conducted by Morris and Thompson in 1979 \cite{morris1979password} which described how to guess hashed passwords using common words in dictionaries and some simple mangling rules.
After that, dictionary-based password guessing was widely used as the core component for proactive password checking (\eg, \cite{bishop1990proactive,spafford1992opus,bergadano1998high}) to prevent users from choosing weak passwords.
Oechslin \ea \ \cite{oechslin2003making} presented the \ti{rainbow table attack} which pre-computed hash values of a large number of candidate passwords, then used such hash values to match with the attacked passwords' hash values.
By pre-computing the hash table, this approach is much faster compared to computing the hash value  online, especially in the old days, when the processing speed was slow and the passwords were protected by storing their hashed values.
However, the later systems mostly combined a salted value with a password before hashing \cite{scarfone2009guide}.
That made the rainbow table attack become inapplicable.
So, the password guessing strategy was changed to producing high-matching-chance guesses by studying how user select their passwords.

Observing that the distribution of letters in passwords is inherited from the distribution of letters in the user's native language, Narayanan \ea \ \cite{narayanan2005fast} proposed the first probabilistic approach which used Markov chain \cite{fine1998hierarchical} to model the password distribution from dictionary words.
This is also the first machine learning-based password guessing model, which opened a brand new approach for this research.
Weir \ea \ \cite{weir2009password} proposed a new password guessing approach by using probabilistic context-free grammar (PCFG) to model the password distribution from a real password database.
This is also the first study that used real passwords as the referenced information to construct a password guessing model.
After this study, real passwords leaked from some services and published on the Internet were widely used for this task. 
Besides, some other information sources have been confirmed to be useful for password guessing (\eg, keyboard pattern \cite{chou2012password,kavrestad2020analyzing}, language property \cite{rao2013effect,veras2014semantic,zhang2020csnn}).
Thus, the later password guessing model usually utilized multiple information sources (\eg, \cite{houshmand2015next,fang2018password,cheng2021improved}).
Recently, the success of deep learning techniques in other areas attracted the attention of the password security community.
Melicher \ea \ \cite{melicher2016fast} proposed the first deep learning-based password guessing model which used Long Short-term Memory (LSTM) network to estimate the password distribution from a leaked password database.
This approach quickly attracted more effort (\eg, \cite{xu2017password, fang2018password,zhang2018password,li2019password}).
Deep Generative Networks were also presented and showed promising performance (\eg, Generative Adversarial Networks \cite{hitaj2019passgan, pasquini2021improving, guo2021pggan}).
Some studies proposed hybrid approaches that combined deep learning and conventional techniques (\eg, \cite{zhang2020preliminary, zhang2018password, liu2018genpass}) to leverage the advantages of both directions.

\subsection{Application of Password Guessing}

Password guessing has served as the heart of several password security applications (\eg, password strength meter, password-enhanced advisor, lost password recovery).
A \ti{password strength meter} (PSM) provides an estimated score (in graphical or text form) that reasonably reflects the strength of a user-selected password against a guessing attack \cite{golla2018accuracy}.
PSMs are widely deployed in practice to encourage users to use strong (\ie, hard-to-guess) passwords \cite{carnavalet2015large,golla2018accuracy,pereira2020evaluating}.
A PSM can operate in two modes: \ti{proactive} and \ti{reactive}.
A proactive PSM is activated at the time users register or change their passwords.
The new passwords must satisfy the minimum strength assessed by the PSM to be used. 
A reactive PSM is periodically launched to check for weak passwords over all the accounts, usually performed or scheduled by the administrator.
It has been shown that PSMs influence users' choices when composing passwords \cite{proctor2002improving,egelman2013does,ur2012does}.
However, early PSM mostly relied on common sense and experts’ opinions to design some password policies (\eg, \cite{helkala2011password}).
Such approaches do not actually improve the password's strength \cite{campbell2011impact,inglesant2010true,lee2022password}.
In order to create positive impacts, it is critical that the strength estimated by the PSMs closely reflects the effort needed by the attacker to break the password \cite{weir2010testing,kelley2012guess,golla2018accuracy}. 
An ineffective guessing technique would overestimate the password strength and allow weak passwords to be used.
This implies that a practical PSM should be built on top of a password guessing model that closely simulates real-attack strategies used in practice \cite{dell2015monte,golla2018accuracy}.

\ti{Password-enhanced advisors} (PEAs) are PSMs that additionally suggest stronger passwords (\eg, \cite{huh2015surpass,ur2017design,melicher2017better}), or give actionable feedback (\ie, guidance) \cite{kim2014yourpassword,ur2017design,zhang2013password,woo2018guidedpass} from the user-inputting passwords. 
It has been shown that the detailed feedback from PEA help users to create more secure passwords than the PSM with only a strength indicator, without significant impact on password memorability \cite{ur2017design,shay2015spoonful}.
%
%
In order to construct an effective PEA, it is necessary to identify the strength of an inputting password as well as understand its weakness (\ie, how a practical password guessing method can crack it) to suggest improvements.
%
%
Moreover, besides metering and identifying the weakness of input passwords, a PEA must ensure that suggested passwords or the passwords produced from the advice could not be effectively modeled by the attacker.
Again, a solid background of password guessing attacks is necessary to validate the security impacts of PEAs.

Besides using for user authentication, password is widely used for protecting sensitive data stored on personal devices using modern encryption techniques \cite{aumasson2017serious}.
When losing a password, it is unable to retrieve the encrypted data.
In addition, the user is unable to reset the password by contacting the administrator like online service authentication.
In this case, password recovery tools built from the password guessing technique are useful.

\subsection{Related Work}
 

Up to the present, passwords have served as the key components in most security infrastructures.
In addition, despite the merits of the security community, a new mechanism that can totally replace the password is not likely to be available in near future \cite{clair2006password,herley2011research,bonneau2015passwords}.
Thus, password security has received significant research attention.
The behaviors of user in composing and managing passwords have been analyzed thoroughly.
It has been shown that the passwords chosen by users are far from random, as users usually compose them from memorable words \cite{bonneau2012linguistic,kuo2006human,rao2013effect}, user information \cite{veras2012visualizing,wang2019birthday,bonneau2012birthday,bryant2006user,krishnamurthy2009leakage}, and keyboard pattern \cite{schweitzer2011visualizing,yang2021studies}.
In order to bypass the password policy, some common patterns are usually adopted to compose passwords (\eg, a special symbol or sequence of numbers are often appended after a dictionary word; if there is an upper-case letter, it is often the first letter of the word) \cite{tatli2015cracking}.
Managing multiple passwords for different services is even more challenging for users.
Thus, a large number of users added simple modifications to create new passwords or completely reused their passwords across different services \cite{stobert2014password,wash2016understanding,haque2013study,gaw2006password}.
In addition, the password also reflects the category of the service (\ie, name of the service or semantic information is usually included in the passwords) \cite{wei2018password}.

To prevent the user from using weak passwords, significant research effort has been invested in designing password guessing and password metering methods.
Along with the original research, some literature and experimental reviews on these topics were also published (\eg, PSM \cite{de2014very,yang2016comparing,hu2017password,golla2018accuracy}, password guessing \cite{han2014password,zhang2020deep,yu2022deep}).
De \ea \ \cite{de2014very} presented experimental analyses on $13$ password checkers used by $11$ well-known web services (\ie, Apple, Dropbox, Drupal, eBay, FedEx, Google, Microsoft, PayPal, Skype, Twitter, and Yahoo) to study their behaviors.
The study showed that these meters were highly inconsistent (\ie, they output different scores for the same password; weak passwords were scored strong, or even very strong).
Yang \ea \ \cite{yang2016comparing} 
provided a summarization of existing password strength meters proposed in the literature (\eg, \cite{castelluccia2012adaptive,ma2014study,ur2015measuring,weir2010testing,wheeler2016zxcvbn}), and conducted experimental analyses with real password datasets.
%
%
Hu \ea \ \cite{hu2017password} reviewed various metrics that are used to measure password strength.
They also presented a comparison on the strengths and weaknesses of each metric.
Golla \ea \ \cite{golla2018accuracy} proposed a set of requirements that a strength meter needs to fulfill to accurately measure password strength.
They also defined a number of tests to validate these requirements.
Then, the proposed requirements and tests were used to validate the accuracy of $19$ existing password strength meters.


%
Few surveys of password guessing studies have been presented in the literature. 
%
%
Han \ea \ \cite{han2014password} conducted a short survey on password guessing and the countermeasure methods proposed before 2014.
Zhang \ea \ \cite{zhang2020deep} briefly reviewed the password guessing and metering research which leveraged deep learning techniques to generate the guesses.
Yu \ea \ \cite{yu2022deep} also summarized the password guessing studies that adopted deep learning techniques.
Besides, some studies conducted large-scale experimental analyses on the performance of existing password guessing methods with large password datasets (\eg, \cite{ji2015zero,biesner2021advances,shi2021understanding}).
%
Despite their merits, answers of the questions raised in Section \ref{sec_intro} could not be found in such surveys.
These surveys mostly focused on summarizing each existing research separately without providing a systematic overview of password guessing studies.
They also lack several important aspects that needed to be considered when constructing a password guessing model (\eg, input data, experimental procedure, evaluation dataset, guessing performance).
Moreover, some important studies were not included (\eg, \cite{wang2016targeted,zhang2020preliminary}).

\section{Systematic Review of Password Guessing}\label{sec_guessing_review}

\subsection{Systematic Overview}

\begin{figure*}[t]
	\centering
	\includegraphics[scale=0.0405]{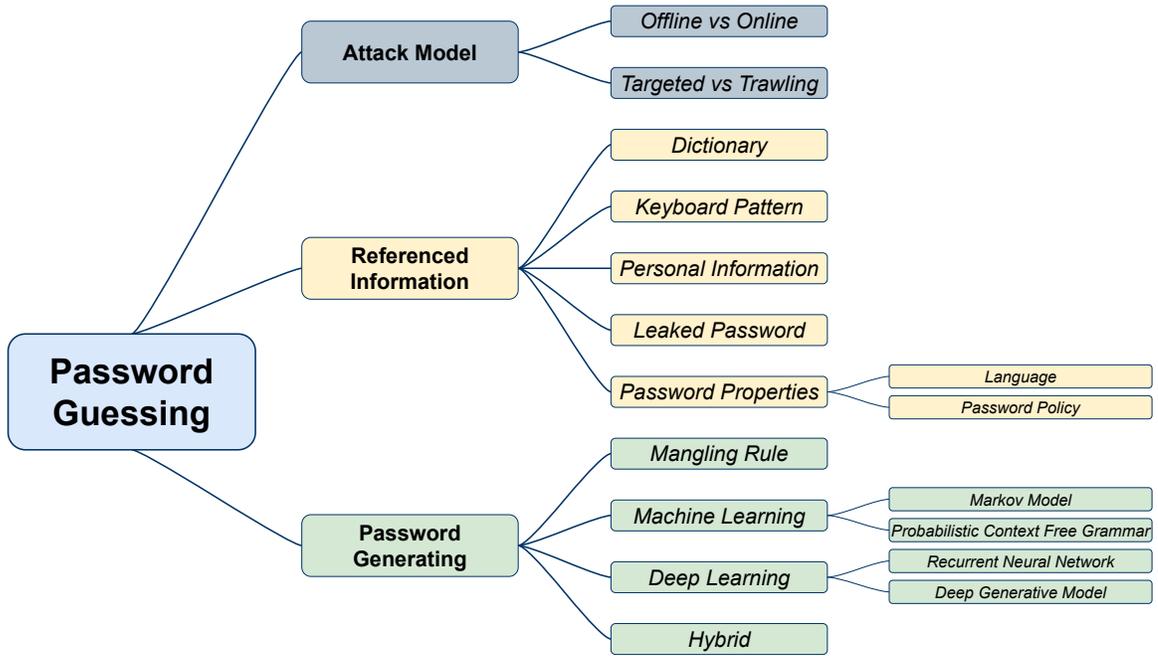}
	\caption{The systematic overview of existing password guessing models.}
	\label{fig_overall_pg}
\end{figure*}

A systematic overview of existing password guessing approaches is depicted in Figure \ref{fig_overall_pg}.
Overall, constructing a password guessing model involves performing three main tasks:
\ti{(a)} identifying the security model;
\ti{(b)} acquiring the useful referenced information;
\ti{(c)} and designing a password generating method.
First, to identify the security model, two questions need to be answered regarding the attack settings: \ti{(i)} the model will operate in online or offline mode; and \ti{(ii)} the model will guess the passwords of a targeted user or a group of users.
The answers of these questions decide the reference information that needs to be obtained and the suitable password generating algorithms.
Second, some information sources are gathered, which will be inferred to generate the candidate passwords for guessing.
Third, the algorithm for generating candidate passwords is designed, which can range from mangling rules to machine learning, or deep learning techniques.
%
%
Some models only output the guessing passwords, meanwhile, some additionally provide the probability of the candidate passwords to enable trying the high-likelihood passwords first.
Table \ref{tab_pwd_guess_list} presents an overview of the setting, referenced information, and password generation algorithm of each password guessing study proposed in the literature. 
In the following sections, we discuss existing password guessing approaches based on the above principles in detail.


%


\subsection{Security Model} 
When constructing a password guessing model, it is important to specify two settings of the attack: online or offline, targeted or trawling.

\subsubsection{Online or Offline}



Depending on the attack's purpose and condition, password guessing could be either performed in online or offline mode.
Each mode faces different challenges, and usually adopts dedicated strategies. 
The \textit{online} password guessing tries to authenticate as the claimed user (\ie, the user being attacked) to a system by guessing the password of that user.
To perform this mode, the attacker needs to repeatedly interact with the targeted system until success. 
In this scenario, the attacker is limited by the number of guesses which usually can not exceed some threshold configured by the target system.
When continuously producing a certain number of incorrect guesses, the targeted account, or the attacker’s IP address may be locked forever or during a specific time \cite{alsaleh2011revisiting}.
In addition, online guessing usually targets a remote host, thus, the attacking time is slowed down by the network delay.
So, the overall strategy used in this mode is to pick the most likely password first.
And most of the attacker's efforts are to determine such passwords.
The common approach is exploiting the target's information such as user ID, birthdate, and leaked passwords of the targeted user (\eg, \cite{zhang2010security,li2016study,wang2016targeted,pal2019beyond,zhang2020preliminary}).

On the other hand, \textit{offline} password guessing attempts to crack unknown passwords when the password hashes or databases are accessible.
Because of requiring the hashed value, offline password guessing is only applicable in some special situations (\eg, the target's password file is leaked), thus, is not considered a real security threat nowadays.
In practice, this attack mode is commonly used by the legitimate user to recover his lost password, or the administrator to check the password strength of the users (\eg, proactive password checking \cite{kelley2012guess, ur2015measuring, melicher2016fast, wang2016fuzzypsm}).
In this mode, the attacker has several important advantages.
First, the attack is not slowed down by the delay of network transmission.
%
%
Second, the limitation of guessing times is also relaxed to be bounded by the computation capability, so, millions of guesses can be attempted until the correct password is found.
As a result, the generic strategy of this mode is to build a password generating model which encodes as much as possible the password space from several information sources (\eg, 
dictionary \cite{narayanan2005fast, chou2013password,rao2013effect}, 
leaked password \cite{weir2009password,zhang2010security,chou2013password,das2014tangled,li2014large,durmuth2015omen,xu2017password,zhang2018password,fang2018password,liu2018genpass,hitaj2019passgan,li2019password,zhang2020preliminary,xia2019genpass,zhang2020csnn,pasquini2021improving,pasquini2021reducing,guo2021dynamic,guo2021pggan,zhou2022password}, 
keyboard pattern \cite{chou2013password,li2014large, houshmand2015next}, 
%
%
language \cite{li2014large}).
Then, the built model is used to generate the guessed passwords, usually in the order of decreasing probability. 
The model's effectiveness is decided by not only the hitting chance of candidate passwords, but also the number of unique candidates the model can generate.

\subsubsection{Targeted vs Trawling}

Typically, a password guessing session could be configured to target a specific user (\ie, targeted) or a group of users (\ie, trawling). 
The targeted guessing is usually performed when the attacked user has been specified (\eg, attacking an online account, or recovering the user's lost password).
The overall strategy of this setting is to try the passwords that are likely to be used by the targeted user. 
So, it is crucial to examine in-depth all the information about the user such as leaked passwords, user personal information, and password reuse behavior.
Such information has been shown to effectively narrow down the guessing space and produce the guess with a high-hitting chance \cite{zhang2010security,das2014tangled,wang2016targeted,li2016study,pal2019beyond,zhang2020preliminary,xie2020new}.

The trawling attack, on the contrary, aims to identify as much as possible the passwords of a group of users.
Nowadays, online services usually require the user identity along with the corresponding password to authenticate the claimer.
Thus, trawling password guessing has limited applications in practice.
It is just applicable when the password storage/hashed values have been compromised.
Most of the existing trawling password guessing is used for \eg, proactive password checking \cite{melicher2016fast}, or adversarial modeling-based password metering \cite{melicher2016fast,wheeler2016zxcvbn, pal2019beyond}.
For this mode, the use of personal information of a specific user (\eg, birthdate, email) is not effective.
Instead, this mode focuses on the passwords that are likely to be used by many users by examining the password distribution of a group of users or inferring some related information sources (\eg, leaked password dataset, dictionary, common words). 

In practice, when interacting with a remote server/service to authenticate as a claimed user (online mode), the attacker has to submit the identity of the claimed user first, then provide the guessed password.
This means, the online password guessing can only target a specific user for each guess.
Thus, the online mode is performed only for the targeted attack (\ie, the online mode cannot go with the trawling attack).
%
%
%
On the other hand, the trawling attack is only performed in offline mode, 
in which the model just tries to guess as many passwords as possible given a set of hashed passwords.
%
%
%
So, in most of the studies, online guessing and targeted attack refer to the same scenario when the guessed passwords are used to crack a specific account, and the number of attempts is restricted by the setting of the targeted service.
Also, the trawling and offline are used to indicate the scenario in which the guessed passwords can be used to compromise many passwords from their hashed values, and the attack is just limited by the computational capability of hardware.
In this paper, we use online and targeted interchangeably, and so on for trawling and offline.

\begin{table*}[htp]
	\centering
	\caption{An overall summary on the security model, information source, password generating algorithm, and the output of the existing password guessing studies.}
	\label{tab_pwd_guess_list}
	\def\arraystretch{1.2}
	\begin{tabular}{|c|cc|cccccc|c|} \hline
		\multirow{2}{*}{\ti{\tb{Study}}}& \multicolumn{2}{|c|}{\ti{\tb{Mode}}} &  \multicolumn{6}{|c|}{\ti{\tb{Referenced Information}}} & \multirow{2}{*}{\ti{\tb{Generation Algorithm}}} \\ \cline{2-9}
		&\multicolumn{1}{|c|}{\rotatebox{90}{\ti{Online}}}&\multicolumn{1}{|c|}{\rotatebox{90}{\ti{Offline}}}&\multicolumn{1}{|c|}{\rotatebox{90}{\ti{Dictionary}}}&\multicolumn{1}{|c|}{\rotatebox{90}{\ti{Leaked Password}}}&\multicolumn{1}{|c|}{\rotatebox{90}{\ti{Keyboard Pattern}}}&\multicolumn{1}{|c|}{\rotatebox{90}{\ti{User Information}}}&\multicolumn{1}{|c|}{\rotatebox{90}{\ti{Language}}}&\multicolumn{1}{|c|}{\rotatebox{90}{\ti{Password Policy}}}&\\\hline 
		
		
		
		%
		\multicolumn{1}{|l|}{Narayanan \ea, 2005 \cite{narayanan2005fast}} &&$\checkmark$  &$\checkmark$&&&&$\checkmark$&&Markov \\\hline
		%
		%
		\multicolumn{1}{|l|}{Weir \ea, 2009 \cite{weir2009password}} &&$\checkmark$&$\checkmark$&$\checkmark$&&&&&PCFG \\\hline		
		\multicolumn{1}{|l|}{Zhang \ea, 2010 \cite{zhang2010security}}&$\checkmark$&$\checkmark$&&$\checkmark$&&&&&Mangling Rule \\\hline
		%
		\multicolumn{1}{|l|}{Chou \ea, 2012 \cite{chou2012password}}&&$\checkmark$&&&$\checkmark$&&&&Dictionary \\\hline
		\multicolumn{1}{|l|}{Chou \ea, 2013 \cite{chou2013password}} &&$\checkmark$&$\checkmark$&$\checkmark$&$\checkmark$&&&&PCFG \\\hline
		%
		\multicolumn{1}{|l|}{Rao \ea, 2013 \cite{rao2013effect}}&&$\checkmark$&$\checkmark$&&&&$\checkmark$&&Mangling Rule \\\hline
		%
		\multicolumn{1}{|l|}{Das \ea, 2014 \cite{das2014tangled}}		&$\checkmark$&&&$\checkmark$&$\checkmark$&&$\checkmark$&&Mangling Rule \\\hline
		%
		%
		\multicolumn{1}{|l|}{Veras \ea, 2014 \cite{veras2014semantic}}		&&$\checkmark$&$\checkmark$&&&&$\checkmark$&&PCFG \\\hline
		%
		%
		\multicolumn{1}{|l|}{Li \ea, 2014 \cite{li2014large}}&&$\checkmark$&$\checkmark$&$\checkmark$&$\checkmark$&&$\checkmark$&&PCFG \\\hline
		%
		\multicolumn{1}{|l|}{Durmuth\ea, 2015 \cite{durmuth2015omen}}&&$\checkmark$&&$\checkmark$&&&&&Markov \\\hline
		%
		\multicolumn{1}{|l|}{Houshmand \ea, 2015 \cite{houshmand2015next}}&&$\checkmark$&$\checkmark$&$\checkmark$&$\checkmark$&&$\checkmark$&&PCFG \\\hline
		%
		\multicolumn{1}{|l|}{Tatli \ea, 2015 \cite{tatli2015cracking}}&&$\checkmark$&$\checkmark$&$\checkmark$&$\checkmark$&&&&Mangling Rule \\\hline
		
		\multicolumn{1}{|l|}{Wang \ea, 2016 \cite{wang2016targeted}}&$\checkmark$&&$\checkmark$&$\checkmark$&$\checkmark$&$\checkmark$&$\checkmark$&$\checkmark$&PCFG \\\hline
		%
		%
		\multicolumn{1}{|l|}{Li \ea, 2016 \cite{li2016study}}&$\checkmark$&$\checkmark$&$\checkmark$&$\checkmark$&&$\checkmark$&&&PCFG \\\hline
		%
		\multicolumn{1}{|l|}{Melicher \ea, 2016 \cite{melicher2016fast}}&&$\checkmark$&&$\checkmark$&&&&&LSTM \\\hline
		%
		%
		%
		%
		\multicolumn{1}{|l|}{Xu \ea, 2016 \cite{xu2017password}}&&$\checkmark$&&$\checkmark$&&&&&LSTM \\\hline
		%
		%
		\multicolumn{1}{|l|}{Zhang \ea, 2018 \cite{zhang2018password}}&&$\checkmark$&&$\checkmark$&&&&&PCFG $\&$ BiLSTM \\\hline
		%
		\multicolumn{1}{|l|}{Fang \ea,2018 \cite{fang2018password}}&&$\checkmark$&$\checkmark$&$\checkmark$&$\checkmark$&&$\checkmark$&&LSTM \\\hline
		%
		\multicolumn{1}{|l|}{Liu \ea, 2018 \cite{liu2018genpass}}&&$\checkmark$&&$\checkmark$&&&&&PCFG $\&$ LSTM \\\hline
		%
		\multicolumn{1}{|l|}{Hita \ea, 2019 \cite{hitaj2019passgan}}&&$\checkmark$&&$\checkmark$&&&&&GAN \\\hline
		%
		\multicolumn{1}{|l|}{Pal \ea, 2019 \cite{pal2019beyond}}&$\checkmark$&&&$\checkmark$&&&&&Encoder - Decoder (RNN) \\\hline
		%
		\multicolumn{1}{|l|}{Li \ea, 2019 \cite{li2019password}}&&$\checkmark$&&$\checkmark$&&&$\checkmark$&&BiLSTM \\\hline
		%
		\multicolumn{1}{|l|}{Xia \ea, 2019 \cite{xia2019genpass}}&&$\checkmark$&&$\checkmark$&&&&&PCFG $\&$LSTM \\\hline
		%
		\multicolumn{1}{|l|}{Luo \ea, 2019 \cite{luo2019recurrent}}&&$\checkmark$&&$\checkmark$&&$\checkmark$&&&LSTM \\\hline
		%
		\multicolumn{1}{|l|}{Zhang \ea, 2020 \cite{zhang2020preliminary}}&$\checkmark$&$\checkmark$&&$\checkmark$&&$\checkmark$&&&PCFG $\&$ LSTM \\\hline
		%
		%
		%
		\multicolumn{1}{|l|}{Zhang \ea, 2020 \cite{zhang2020csnn}}&&$\checkmark$&&$\checkmark$&&&$\checkmark$&&LSTM \\\hline
		%
		%
		\multicolumn{1}{|l|}{Pasquini \ea, 2021 \cite{pasquini2021improving}}&&$\checkmark$&&$\checkmark$&&&&&GAN-WAEs, WAE \\\hline
		%
		\multicolumn{1}{|l|}{Pasquini \ea, 2021 \cite{pasquini2021reducing}}&&$\checkmark$&$\checkmark$&$\checkmark$&&&&&Mangling Rule $\&$ CNN \\\hline
		%
		\multicolumn{1}{|l|}{Guo \ea, 2021 \cite{guo2021pggan}}&&$\checkmark$&&$\checkmark$&&&&&GAN \\\hline
		%
		\multicolumn{1}{|l|}{Cheng \ea, 2021 \cite{cheng2021improved}}&&$\checkmark$&$\checkmark$&$\checkmark$&$\checkmark$&&$\checkmark$&&PCFG \\\hline
		%
		%
		\multicolumn{1}{|l|}{Zhou \ea, 2022 \cite{zhou2022password}}&&$\checkmark$&&$\checkmark$&&&&&GAN \\\hline
		
		\hline
	\end{tabular}
\end{table*}

\subsection{Referenced Information}

Textual passwords have been shown not to be random due to the inability of the human to memorize complicated strings, but they usually contain some identifiable information (\eg, personal information \cite{bryant2006user,wang2019birthday,veras2012visualizing,bonneau2012birthday}, service information \cite{wei2018password}, keyboard pattern \cite{schweitzer2011visualizing}, linguistic properties \cite{bonneau2012linguistic,rao2013effect}).
In addition, passwords (used for different services) of a user are usually identical or share some commonality \cite{florencio2007large,haque2013study,wash2016understanding,pearman2017let,gaw2006password}.
Thus, instead of guessing passwords randomly, most existing models exploited some information sources to infer the password distribution and produce high-likelihood guesses.
This section summarizes the common information sources that have been exploited to construct password guessing models.

\subsubsection{Dictionary}
Dictionary has been used for textual password guessing from a very early time, with the first official report presented by Morris and Thompson in \cite{morris1979password}.
In that study, hashed values were directly computed from words in some dictionaries, and matched with the hashed passwords.
Despite the simplicity, this approach showed alarming results when it cracked one-third of $3,289$ passwords in a very short time processing and limited computing resource.
They additionally suggested some simple methods to extend the word list by using backward-spelled words, first/last names, street names, city names, valid license plate numbers, telephone numbers, etc.
The success of dictionary-based password cracking comes from the fact that meaningful words are much easy to be memorized than random character sequences.
Thus, entire or part of passwords is usually constructed from ordinary or well-known words \cite{bonneau2012linguistic}.
Besides directly using the words from dictionaries as the guessed passwords, some studies applied mangling rules to turn each ordinary word into a group of related words for guessing  \cite{chou2013password,wang2016targeted,pasquini2021reducing,wheeler2016zxcvbn}.
This strategy is used in practical password recovery tools such as Hashcat \cite{jens2009hashcat}, John the Ripper \cite{peslyak2014john}.
Moreover, some password guessing models were built by capturing the letter distribution of a dictionary with native language processing techniques 
(\eg, Markov \cite{narayanan2005fast,wang2016targeted}, 
Parts of Speech Tagging \cite{rao2013effect}, 
Probabilistic Context Free Grammar (PCFG) \cite{veras2014semantic,houshmand2015next,wang2016targeted,li2016study}).

\subsubsection{Real Password}\label{sssec_real_password}

\begin{table}[htp]
	\centering
	\caption{The popular leaked password databases that have been used to construct password guessing models. \ti{(Note that, as passwords are sensitive, we do not include the download links of these datasets. Most of them are available on the Internet.)}}
	\label{tab_leaked_password}
	\def\arraystretch{1.15}
	\begin{tabular}{|lcrcccc|} \hline
		\multirow{2}{*}{\ti{\tb{Dataset}}}& \ti{\tb{Leaked}} & \ti{\tb{\centering No. of }} &\ti{\tb{Data}}& \multirow{2}{*}{\ti{\tb{Language}}} & \ti{\tb{User}} & \ti{\tb{Used in}} \\
		& \ti{\tb{Time }} & \ti{\tb{\centering Passwords}} &\ti{\tb{Type}}& & \ti{\tb{Info}} & \ti{\tb{Study}} \\
		 \hline
		000webhost 	
		&Oct, 2015
		& $15,251,073$
		&Plaintext 
		&Chinese
		&Email
		& \cite{wang2016targeted,cheng2021improved}\\
		\hline
		
		\multirow{2}{*}{12306}	
		&\multirow{2}{*}{2014}
		& \multirow{2}{*}{$\sim 130,000$}
		&\multirow{2}{*}{Plaintext} 
		&\multirow{2}{*}{Chinese}
		& Name;
		& \multirow{2}{*}{\cite{wang2016targeted,li2016study,zhang2020preliminary,zhang2020csnn}}\\
		
		& 
		&
		& 
		& 
		&real ID& \\
		\hline
		
		126 &Dec, 2011& $6,392,568$&Plaintext &Chinese&Email& \cite{wang2016targeted}	\\
		\hline 
		
		178.com 	 &Dec, 2011& $\sim 9,072,824$&Plaintext&Chinese&Username&\cite{li2014large}\\
		\hline
		
		7k7k 	&2011& $19,138,270$&Plaintext&Chinese&Email&\cite{li2014large}\\
		\hline
		
		Animoto &Aug, 2020 &$\sim 8\cdot 10^6$ &Hashed&English&User info& \cite{pasquini2021reducing}\\
		\hline
		
		Clixsense	&$-$& $1,628,894$&Plaintext&English&$-$& \cite{li2019password,cheng2021improved}\\ 
		\hline
		
		\multirow{2}{*}{CSDN} 	 & \multirow{2}{*}{Dec, 2011}& \multirow{2}{*}{$\sim 6,428,629$}&\multirow{2}{*}{Plaintext}& \multirow{2}{*}{Chinese}&Email,&\multirow{2}{*}{\cite{li2014large,wang2016targeted,zhang2020csnn,zhang2018password,fang2018password,cheng2021improved}}\\ 
		&& & && Username&\\	
		\hline
		
		Dodonew &Dec, 2011& $\sim 16,258,891$&Plaintext &Chinese&Email& \cite{wang2016targeted,zhang2018password,zhang2020csnn,cheng2021improved}	\\
		\hline
		
		Duduniu 	 & 2011	& $16,282,969$&Plaintext&Chinese&Username&\cite{li2014large}		\\ 
		\hline
		\multirow{2}{*}{Duowan} &\multirow{2}{*}{2018} &\multirow{2}{*}{$\sim 4,982,730$} &Plaintext&\multirow{2}{*}{Chinese}&User ID, Email, & \multirow{2}{*}{\cite{cheng2021improved}}\\
		&&&\& Hashed&&User name&\\
		\hline
		
		Facebook 	&$-$&$-$& Plaintext&$-$&Email& \cite{durmuth2015omen,xu2017password,fang2018password}\\
		\hline
		
		\multirow{2}{*}{Finnish} &\multirow{2}{*}{$-$}&\multirow{2}{*}{$38,432$} &Plaintext&\multirow{2}{*}{Finland}&\multirow{2}{*}{None}&\multirow{2}{*}{\cite{weir2009password}}\\
		&&&\&Hashed&&&\\
		\hline
		
		Hotmail 	&2009&8,930 &Plaintext&English&None& \cite{pasquini2021improving}\\ 
		\hline
		
		JingDong 	&$-$& $390,000$&Plaintext &Chinese&$-$& \cite{zhang2018password}\\
		\hline 
		
		LinkedIn &2012& $\sim 60,065,486$&Hashed &English&Email& \cite{hitaj2019passgan,xia2019genpass,pasquini2021improving,pasquini2021reducing,zhou2022password}	\\ 
		\hline 
		
		MyHeritage 	&Oct, 2017&$92,283,889$ &Hashed&$-$&Email&\cite{pasquini2021reducing}\\
		\hline
		
		\multirow{2}{*}{MySpace} & \multirow{2}{*}{2006} &$37,000$& \multirow{2}{*}{Plaintext}  &\multirow{2}{*}{English}&\multirow{2}{*}{None}& \cite{weir2009password,chou2013password,durmuth2015omen,xu2017password}\\ 
			&&$\sim 67,042$ &&&& \cite{pasquini2021improving,li2019password,xia2019genpass}\\
		\hline 
		Netease &Oct, 2015&$1,220,088,121$&Plaintext&$-$&$-$& \cite{zhang2020csnn}	\\
		\hline
		Phpbb   &Jan, 2009& $\sim 184,380$&Plaintext&English&None& \cite{chou2013password,li2019password,xia2019genpass,zhang2018password,pasquini2021improving,pasquini2021reducing}\\ %
		\hline
		
		\multirow{3}{*}{RockYou} &\multirow{3}{*}{2009}&\multirow{3}{*}{$\sim 32,603,388$}& \multirow{3}{*}{Plaintext}&\multirow{3}{*}{English}&\multirow{3}{*}{None}& \cite{chou2013password,wang2016targeted,li2019password,xia2019genpass}\\
		&& & &&&\cite{li2014large,tatli2015cracking,durmuth2015omen,guo2021dynamic,pasquini2021improving,cheng2021improved}\\
		&& &&&& \cite{xu2017password,zhang2018password,fang2018password,hitaj2019passgan,pasquini2021reducing,zhou2022password}\\
		\hline 
				
		Rootkit &Feb, 2011& $69,418$&Plaintext &English&User info& \cite{wang2016targeted}	\\
		\hline	
					
		SilentWhisper & &$\sim 7,480$ &Plaintext&$-$&None& \cite{weir2009password}\\
		\hline
		
		Tianya 	&Dec, 2011	& $30,179,474$&Plaintext&Chinese&Username&\cite{li2014large}\\
		\hline

		Xato 	&Feb, 2015& $9,997,772$&Plaintext &English&$-$& \cite{wang2016targeted}\\
		\hline 
		
		Xiaomi 	&May, 2014& $8,281,385$&Hashed&Chinese&$-$& \cite{wang2016targeted}\\
		\hline 
		
		Yahoo 		&July, 2012	& $442,837$&Plaintext&English&Username&\cite{li2014large,zhang2018password,wang2016targeted,li2019password}\\ 
		\hline
		
		Youku 	&2016&$100,759,591$ &Plaintext&Chinese&Email& \cite{pasquini2021improving,pasquini2021reducing}\\
		\hline
		
		Zomato	&May, 2017 &$17,000,000$ &Hashed&$-$&Email& \cite{pasquini2021improving,pasquini2021reducing}\\
		\hline
		
		Zooks	&Jan, 2020&$\sim2.9\cdot 10^7$ &Plaintext&$-$&User info& \cite{pasquini2021reducing}\\ 
		\hline
		
		\multicolumn{7}{c}{`\ti{$-$': the corresponding information was not specified.} }\\
	\end{tabular}
\end{table}


%
Real-world password (\textit{aka}, leaked password, disclosed password) datasets, which were hacked by attackers or were privately collected inside an organization (for research), have been shown as the most effective information source to build password guessing models.
The effectiveness of this approach comes from two facts: \ti{(i)} users usually use the same or similar passwords for different services  \cite{gaw2006password,florencio2007large,haque2013study,stobert2014password,wash2016understanding,pearman2017let}; 
\ti{(ii)} some common passwords are widely used by many users (\eg, \quotes{$\mathsf{123456}$}, \quotes{$\mathsf{111111}$}, \quotes{$\mathsf{password}$}) \cite{wang2016targeted,han2015regional}.
A leaked password dataset reveals various kinds of sensitive information which is useful to construct effective password guessing models such as user behavior on password forming, password policy, and user information.
Such observations were confirmed firstly by Weir \ea \ in \cite{weir2009password}.
In that study, the authors leveraged leaked passwords to derive the underlying structures (\ie, grammar) of these passwords with PCFG, then, used such structures to generate the guessed passwords.
Their experiments showed an improved performance of $28\%$ to $129\%$ over the pure dictionary-based password guessing with John the Ripper.
After that, using leaked passwords available on the Internet for constructing password guessing models received more research attention. 
\ti{MySpace} was one of the first leaked password datasets that has been used for password guessing research.
The dataset was collected by a fishing attack in 2006, comprised of about $37,000$ to $67,042$ plaintext passwords of MySpace's users, later, used in some password guessing studies (\eg, \cite{weir2009password,chou2013password,durmuth2015omen,xu2017password,pasquini2021improving,li2019password,xia2019genpass}).
However, the dataset does not contain other information (\eg, username, email...) of the leaked accounts.
\ti{RockYou} is a dataset consisting of more than $32$ million passwords of a company named Rockyou, leaked in 2009, and later, used in many researches (\eg, \cite{chou2013password,wang2016targeted,li2019password,xia2019genpass,li2014large,durmuth2015omen,guo2021dynamic,pasquini2021improving,xu2017password,zhang2018password,fang2018password,hitaj2019passgan,pasquini2021reducing,zhou2022password}).
Like MySpace, all the passwords of Rockyou were originally stored in plaintext, but there was no additional information (\eg, username, email...) of the users.
\ti{LinkedIn} is another leaked password dataset of English-speaking users which consists of more than $60$ million salted passwords along with the user's emails.
The plaintext passwords of this dataset were obtained by using John the Ripper and HashCat to recover from the salted values.
Thus, this dataset only contains passwords that are crackable by rule-based tools. 
\ti{CSDN} is a leaked password dataset comprised of more than $6$ million passwords of Chinese-speaking users (programmers) leaked in 2011.
Along with the leaked passwords, this dataset also contains other types of information such as username, and email address.
Instead of using leaked passwords of other users, Zhang \ea \ \cite{zhang2010security} measured the security of new passwords of a group of users when their old passwords were given.
They leveraged a password database of their university community that consisted of $51,141$ unsalted MD5 passwords from $10,374$ users.
Their experiment showed alarming security risks when the old passwords were leaked.
Specifically, with the offline guessing setting, about $41\%$ of new passwords were cracked based on the old passwords in under $3$ seconds.
For the online setting, about $17\%$ of accounts can be broken within $5$ guesses.
Furthermore, the effectiveness of using leaked passwords was also confirmed in targeted guessing attacks (\eg, \cite{pal2019beyond,wang2016targeted}).
Besides the above datasets, many other real password databases of different sites have been leaked, and then, leveraged for password guessing studies (Table \ref{tab_leaked_password}). 

Although rich information could be inferred from a leaked password dataset, using such datasets to construct a password cracking model has several challenges.
First, the adopted password policies (\eg, password length, character set...) strongly impact the password distribution created by the user \cite{proctor2002improving,ur2012does,shay2015spoonful}.
Thus, in order to construct an effective password guessing model, the referenced password dataset should share as much as possible some commonalities with the targeted passwords in terms of password policy and other factors (\eg, language).
%
However, such a requirement is not easy to fulfill because password policies (and other factors) are diverse between different organizations. 
Moreover, most of the leaked databases do not contain the adopted password policies that users had to follow when creating them. 
Second, these leaked databases were illegally obtained and distributed, thus, there is no assurance regarded to the reliability of these data.
Fractions of passwords could be fake records, or from dummy accounts which can create biases for password studying.

\subsubsection{User Information}
It has been widely studied that user usually constructs their passwords from some familiar information sources, and personal information (\eg, birthdate, name, email...) are the most common \cite{wang2019birthday,veras2012visualizing,bonneau2012birthday,bryant2006user}.
Thus, the personal information of the targeted user has been exploited as a reliable source for guessing the passwords of that user, especially effective in targeted guessing attacks \cite{li2016study,wang2016targeted}.
Li \ea \ \cite{li2016study} considered six types of common personal information including name, birthdate, email address, phone number, account name, and ID number to construct a password guessing model with PCFG (so, called personal-PCFG). 
Their experiment on 000webhost dataset demonstrated that: in offline mode, the personal-PCFG model cracked passwords much faster than PCFG; and in online mode, it achieved a higher success rate.
Wang \ea \ \cite{wang2016targeted} considered different types of personal information (\eg, old password, name, birthdate, phone, gender, age...), and incorporated them with other information sources (\eg, dictionary, leaked password) to build an online targeted password guessing model, named TarGuess.
Zhang \ea \ \cite{zhang2020preliminary} constructed a hybrid model that combined TarGuess \cite{wang2016targeted} (which fully exploited personal information) with Long Short-term Memory (LSTM) network to enable the generation of a large number of passwords.
Their experiment showed that, in the offline setting, the proposed architecture can successfully crack more passwords than using each individual model.
%
%
Overall, personal information has been shown highly effective and is the mandatory source for constructing a targeted password guessing model.
%
%


\subsubsection{Keyboard Pattern}
Typing with a special pattern on the keyboard can generate character sequences that look random because they do not follow common character orders such as alphabet, numeric, or language dictionary words.
The user usually adopted a keyboard pattern (\eg, \quotes{$\mathsf{zxcvbn}$}, \quotes{$\mathsf{1qa2ws3ed4rf}$}) when creating a password, as the obtained string looks like a strong password, and is easy for the user to reproduce (\eg, \cite{schweitzer2011visualizing,li2014large,yang2021studies}).
From that observation, several studies have exploited the common keyboard patterns for password guessing (\eg, \cite{kavrestad2020analyzing,chou2012password,chou2013password,houshmand2015next,wheeler2016zxcvbn,fang2018password}).
In \cite{chou2012password}, the authors described a framework to model the passwords formed by the adjacent keys (\eg, \quotes{$\mathsf{gbnjuy}$}) and parallel keys (\eg, \quotes{$\mathsf{tregfd}$}).
That framework also supports modeling the passwords generated when using printable keys with functional keys (\eg, \quotes{$\mathsf{@ \# \$ QWE}$}).
The keyboard patterns then were widely combined with other information sources such as dictionaries and leaked passwords to increase the guessable password space \cite{chou2013password,houshmand2015next,wheeler2016zxcvbn,cheng2021improved}. 

\subsubsection{Auxiliary Information}
Besides the above information sources, some other information could be used to narrow down the guessing space and improve the hitting chance of the candidate passwords (\eg, native language, password policy). 
Many studies have confirmed that passwords inherit many properties of the language spoken by the user.
For instance, to guess the password of a Chinese user, it is more effective to use a Chinese dictionary (\eg, Pinyins) than using an English dictionary, and vice versa \cite{li2014large,fang2018password,zhang2020csnn, zhang2020preliminary,cheng2021improved}.
Moreover, passwords (especially the long ones) were commonly composed with the semantic patterns of spoken language inside. 
This fact enables the adoption of Part-of-speech (POS) tagging \cite{voutilainen2003part} to exploit the semantic patterns of spoken language inside the passwords to improve the guessing effectiveness \cite{rao2013effect,veras2014semantic}.
Besides language, password creation policy is also the important information that effectively assists in the generation of high-hitting-chance candidate passwords \cite{wang2016targeted}.
Setting certain policies is a common method employed by many services/applications in order to prevent the user from using a weak password.
It is straightforward that, those passwords which do not satisfy the policies of a target service, should not be used for guessing because the users can not use such passwords on that service.

\subsection{Password Generation Algorithm}
Upon gathering the referenced information, an algorithm is used to generate candidate passwords.
In the early time, mangling rules based on common sense and experts' knowledge are used.
The later studies leveraged machine learning and deep learning to automatically infer the password distribution and generate guesses.
This section covers different approaches of password generation proposed in the literature.
\subsubsection{Mangling Rule}

Mangling rule is known as the earliest and most popular password guessing approach.
A mangling rule defines a transformation on one or multiple input strings (\ie, passwords, dictionary words) to produce new passwords.
For example, applying the rule \quotes{\ti{capitalize the first letter}} to \quotes{$\mathsf{lucky}$} will output \quotes{$\mathsf{Lucky}$}.
This strategy exploits the fact that users tend to \ti{(i)} follow common patterns to compose passwords, and \ti{(ii)} adopt a predictable pool of mangling functions to create new passwords from the old ones (when changing passwords).
%
%
Mangling rule attack is simple and easy to implement, yet, it poses a substantial threat to password security. 
Moreover, it can be processed in parallel, and easy for upgrading (\eg, adding new words or rules is easier compared to re-training a machine learning or deep learning model).
However, the effectiveness of this strategy completely depends on the input word list and the mangling rules.
While the input word list could be easy to obtain from dictionaries, common words (\eg, street name, person name), or leaked passwords, manually defining the mangling rules is a time-costing task and requires experts' knowledge.
For example, the mangling rules used in popular password guessing tools like Hashcat, John the Ripper have been conducted through continuous updating for a long time.
To provide an effective guess, the mangling rules should capture the users' behavior in composing or reusing the passwords.
Thus, instead of manually designing such rules, some studies suggest inferring the rules from the leaked passwords. 
Zhang \ea \ \cite{zhang2010security} proposed the first strategy to learn them automatically from the leaked passwords.
Specifically, their method requires the knowledge of several old passwords of the targeted user.
From these passwords, the mangling behavior of the targeted user is inferred and used to guess the future password.
Das \ea \ \cite{das2014tangled} surveyed the password reuse behavior of a group of volunteers.
The mangling rules were derived from the reuse behavior and used to predict the future password of the observed user.
Tatli \ea \ \cite{tatli2015cracking} manually examined the leaked password datasets to identify the common patterns (\ie, composing rules) that users usually followed when composing passwords (\eg, a special symbol (exclamation mark, dot...) or sequence of numbers are often appended after a dictionary word; if there is an upper-case letter, it is often the first letter of a dictionary word).
They divided the identified patterns into $10$ categories as appending, prefixing, inserting, repeating, sequencing, replacing, reversing, capitalizing, special-format, and mixed Patterns.
Then, they developed \ti{pbp-generator} (pattern-based password generator), a tool to generate candidate passwords from the identified patterns and given dictionaries.

While most of the prior researches focused on generating effective rules by investigating the password-forming behavior of users, Pasquini \ea \ \cite{pasquini2021reducing} proposed Adaptive Mangling Rule (AdaMs) which paid attention to the later step as optimizing the combination between mangling rules and words.
AdaMs relied on intuition as \quotes{users usually select a dictionary word first, then the mangling rule conditioned on the word}.
Each mangling rule is only applicable on a small subset of words (\ie, combining the rule on these words will produce guessing hits with high probability), strictly defined by the users' habits on password composing.
In other words, each pair of mangling rule and word has a certain compatible level decided by the user's behavior.
Instead of applying all mangling rules to all dictionary words, the authors used residual neural network \cite{he2016deep} to model the compatibility.
Only the pairs having sufficient compatibility are used for guessing.
To train the deep network, a set of mangling rules $\mathcal{R}$, a set of words $\mathcal{W}$, and a set of leaked passwords $\mathcal{X}$ are required. 
For each word $w_i \in \mathcal{W}$, the deep network outputs a compatibility vector
\begin{equation}
\mathbf{y}_i = \begin{bmatrix}
y_{i1}&\ldots& y_{ij}& \ldots& y_{i|\mathcal{R}|}
\end{bmatrix},
\end{equation}
where $y_{ij}=\mathsf{P}(r_j(w_i) \in \mathcal{X})$ is the compatibility of $r_j$ and $w_i$ (\ie, the probability of applying rule $r_j$ on the word $w_i$ will result to a password in the set $\mathcal{X}$).
Upon training, the model is used to estimate the compatibility of a set of rules given a dictionary word $w_i$.
Given a set of compatibility vector $\mathbf{y}_i$, the rule $r_j$ is applied to $w_i$ if $y_{ij} > \delta$, where $\delta \in [0, 1)$ is a threshold defined by the user. 

\subsubsection{Markov Model}

\begin{figure*}[t]
	\centering
	\includegraphics[scale=0.07]{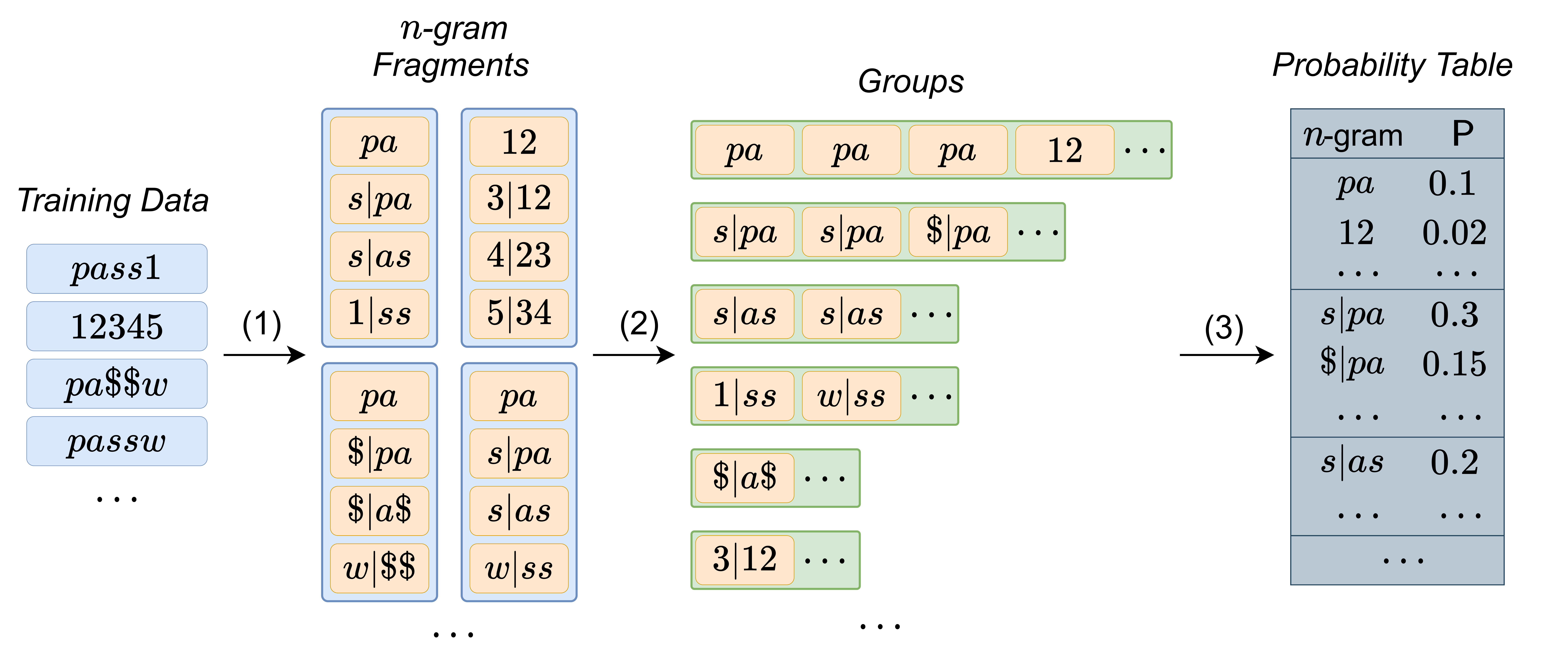}
	\caption{The overall training process of a 3-gram Markov-based password guessing model.}
	\label{fig_markov_train}
\end{figure*}

Before the emergence of deep learning techniques, Markov model was known as the most dominant approach for natural language processing.
From the observation that users usually form the password from their native language for easy memorizing \cite{bonneau2012linguistic,rao2013effect}, some studies proposed using Markov model for password guessing (\eg, \cite{narayanan2005fast,durmuth2015omen,guo2021dynamic}).
The core concept of Markov-based password guessing is to exploit the interconnection between consecutive characters in the passwords to predict the next character based on previous characters (\ie, conditional characters). 
This is inspired by the fact that adjacent letters in human-generated passwords are not independently chosen, but follow certain regularities inherited from human languages (\eg, in English, \quotes{$\mathsf{th}$} is much more likely than \quotes{$\mathsf{tq}$}, and the letter \quotes{$\mathsf{e}$} is likely to be after \quotes{$\mathsf{th}$}).
Specifically, for a password of length $m$ (denoted as $c_1 c_2 \ldots c_m$), the Markov model estimates its occurrence probability as
\begin{equation}\label{eq_markov_proba}
\mathsf{P}(c_1 c_2 \ldots c_m) = \mathsf{P}(c_1 c_2 \ldots c_{n})
\prod_{i=n+1}^m \mathsf{P}(c_i|c_{i-n+1}\ldots c_{i}), 
\end{equation}
where $n$ defines the number of historical characters used to predict the next character, which is also referred to as the \ti{order} of the model.
The combination of $n$ history characters and the predicted character forms $(n+1)$-gram fragment, and the described model is also known as $(n+1)$-gram Markov model.
Before using for password guessing, the initial probabilities $\mathsf{P}(c_1 c_2 \ldots c_{n})$ and the transition probabilities $\mathsf{P}(c_i|c_{i-n+1}\ldots c_{i})$ are inferred from the real password dataset through a training process. 
Specifically, given a training dataset, each password is split into multiples $(n+1)$-gram fragments.
All the $(n+1)$-gram fragments (obtained from all passwords) are grouped by the conditional characters (\ie, the first $n$ characters of each fragment).
Then, the probability of each $(n+1)$-gram (considered within each group) is estimated.
Figure \ref{fig_markov_train} depicts an example of training a $2$-order Markov model for password guessing.

Narayanan \ea \ \cite{narayanan2005fast} conducted the first study that uses Markov model for password guessing.
They proposed \ti{Markovian filter} which captures the similarity between passwords and user's native language.
Such filters are used to efficiently enumerate all candidate passwords of a given length accepted by a deterministic automaton without searching through the entire password space.
Durmuth \ea \ \cite{durmuth2015omen} proposed OMEN (Ordered Markov ENumerator) - a Markov-based password generative model that enumerates passwords in the order of decreasing probability.
While the prior studies used Markov model for trawling password guessing, Wang \ea \ \cite{wang2016targeted} showed that Markov model is also effective for targeted password guessing when it is trained with the targeted user's personal information.
%
%
%
%
Recently, Guo \ea \ \cite{guo2021dynamic} proposed dynamic Markov model to address the problem of high repetition rate in prior Markov-based password guessing models.
Upon generating a candidate password for guessing, the dynamic Markov model reduces the probability of this password. 
Thus, the likelihood of generating this password in subsequent attempts is reduced.

\subsubsection{Probabilistic Context Free Grammar}\label{sssec_pcfg}

Probabilistic Context Free Grammar (PCFG) has been widely used for modeling natural language (\eg, \cite{knight2005overview}). 
Several works have shown that, PCFG is also effective for password guessing \cite{weir2009password,chou2013password,li2014large,veras2014semantic,houshmand2015next,li2016study,cheng2021improved}.
This approach is inspired from the observation that user usually follows typical patterns when composing passwords to comply with stringent password creation policies (\eg, must include letter, digit, and special symbol). 
%
%
For instance, \quotes{$\mathsf{password123!@\# }$} and \quotes{$\mathsf{secret214!()}$} share the same pattern as a letter string followed by some digits, and ended with special symbols.
This is similar to the fact that a typical English sentence usually consists of a subject, a verb, and an object.
PCFG approach tries to model such password patterns (structures) from the real password dataset. %
Such patterns are then used to generate candidates for password guessing.

\begin{figure*}[t]
	\centering
	\includegraphics[scale=0.065]{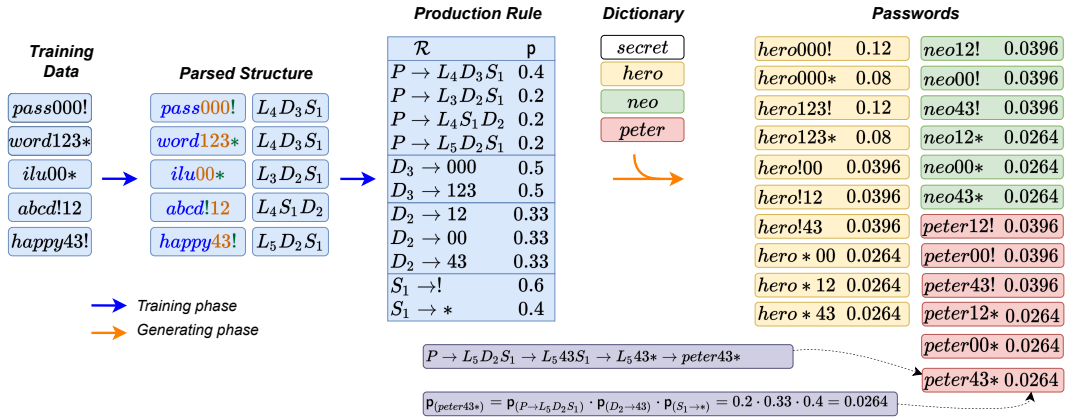}
	\caption{An illustration training and generating password following the PCFG model proposed by Wei \ea \ \cite{weir2009password}, whereas, the probabilities of generated passwords are computed according to the pre-terminal probability order.}
	\label{fig_pcfg}
\end{figure*}

The first PCFG-based password guessing model was proposed by Weir \ea \ in \cite{weir2009password}.
A PCFG is defined by a quintuple $\mathcal{G}=(P,\mathcal{N},\mathcal{T},\mathcal{R},\mathsf{p})$, 
where $\mathcal{N}$ is a set of non-terminals (\textit{aka}, variables), $P\in \mathcal{N}$ is the start symbol, $\mathcal{T}$ is a set of terminals, $\mathcal{R}$ is a set of production rules $\mathcal{N}\rightarrow (\mathcal{N}\cup \mathcal{T})^*$, and $p$ is a function that gives the probability of each rule in $\mathcal{R}$ such that, for any non-terminal $X \in \mathcal{N}$, the sum of probabilities associated with all rules having $X$ in the left side is $1$ (\ie, $\forall X \in \mathcal{N}$, $\sum\limits_{\alpha \in (\mathcal{N}\cup \mathcal{T})^*}p_{(X \rightarrow\alpha)}=1$). 
The set of variables (non-terminals) $\mathcal{N}$ specify the character types from which the passwords are constructed. 
The model in \cite{weir2009password} supports 3 types of basic characters as alphabet letters (no discrimination between lowercase and uppercase letters), digits, and special symbols.
The variable set is defined as $\mathcal{N}=\{P, L, D, S\}$, where $L$ represents the alphabet letter (\ie, $\{\mathsf{a,b,c\ldots}\}$), $D$ means the number digit (\ie, $\{\mathsf{0,1,2\ldots}\}$), and $S$ stands for the special symbol (\ie, $\{!@\#\ldots\}$).
Each variable is subscripted by an integer number to specify the character length.
A group of passwords having the same structure is represented by a \ti{non-terminal form} (\ie, a string that contains at least 1 non-terminal).
For instance, the non-terminal form $L_5D_2S_1$ represents the passwords formed from $5$ alphabet letters, $2$ digits, and $1$ special symbol. 
The start symbol $P$ is a special variable, representing all the passwords that can be generated by the PCFG.
The actual password is represented by a terminal form.
A production rule transforms a non-terminal form to either a terminal or another non-terminal form.
Given a non-terminal form, production rules are recursively applied until only the terminals are left.
Each production rule has an associated probability of occurrence such that the sum of probabilities associated with all rules having the same left side is $1$.
The probability of a terminal/non-terminal form is the product of probabilities of all the rules used in its derivation. 
For example, given a PCFG depicted in Figure \ref{fig_pcfg}, the non-terminal form $L_543*$ represents all passwords having $5$ alphabet letters followed by the digit string $43$ and the symbol $*$.
This non-terminal form can be derived from $P$ via a chain of production rules $P\rightarrow L_5D_2S_1$; $D_2 \rightarrow 43$; and $S_1 \rightarrow *$, (\ie, $P\rightarrow L_5D_2S_1 \rightarrow L_543S_1 \rightarrow L_543*$).
The probabilities associated to the rules are $\mathsf{P}_{(P\rightarrow L_5D_2S_1)}=0.2$, $\mathsf{P}_{(D_2 \rightarrow 43)}=0.33$, and $\mathsf{P}_{(S_1 \rightarrow *)}=0.4$.
Then, the probability of the non-terminal form $L_543*$ is computed as $\mathsf{p}_{(L_543*)}=\mathsf{p}_{(P\rightarrow L_5D_2S_1)} \cdot \mathsf{p}_{(D_2 \rightarrow 43)} \cdot\mathsf{p}_{(S_1 \rightarrow *)}=0.2 \cdot 0.33 \cdot 0.4 = 0.0264$.

A PCFG-based password guessing model operates in two phases as follows. \ti{Training}: Given a training dataset (\eg, leaked passwords dataset), this phase infers the base structures (\ie, non-terminal forms), the set of production rules, and the probabilities of each entity. 
Specifically, each password in the training set is parsed to identify the types of characters (\eg, letters, digits, special symbols).
Then, the base structures and production rules are recorded, and their corresponding probabilities are estimated.
In addition, the probabilities of digit strings and special strings are also computed.
%
%
However, in most existing models, the learning phase does not include determining probabilities of the alphabet non-terminal strings (\ie, $L$), but treats them like terminal strings.
%
%
\ti{Generating}: A dictionary is fed into the trained PCFG to generate the candidate passwords, usually in the decreasing order of probability.
The probabilities associated to the generated passwords could be computed in either \ti{pre-terminal probability order} or \ti{terminal probability order} \cite{weir2009password}.
With the pre-terminal order, the probability of a password is determined only from the probability of the non-terminal form that the password is derived from (\eg, $\mathsf{p}_{(peter43*)}=\mathsf{p}_{(L_543*)}$).
On the other hand, by the terminal order, the probability of a password is computed from both the non-terminal form and the probability of the alphabet string (\eg, $\mathsf{p}_{(peter43*)}=\mathsf{p}_{(L_543*)}\cdot \mathsf{p}_{(peter)}$).
This approach requires the probability of alphabet string which could be obtained in several ways as assigning equal probability to all words having the same length; assigning based on the frequency according to some prior sources like language statistics, or leaked password statistics. 
Figure \ref{fig_pcfg} depicts an example of PCFG described in \cite{weir2009password}, in which the pre-terminal order is used to compute the probabilities of the generated passwords.

Inspired by the work of Weir \ea \ \cite{weir2009password}, several improvements of PCFG-based password guessing model have been proposed (\eg, \cite{chou2013password,li2014large,veras2014semantic, houshmand2015next,li2016study,cheng2021improved}).
Specifically, Chou \ea \ \cite{chou2013password} use additional variables to enable the modeling of lowercase, uppercase letters, and keyboard patterns.
Li \ea \ \cite{li2014large} improve the traditional PCFG by inserting Pinyins into the dictionary.
Thus, the model is able to guess Chinese passwords.
Their work also allows modeling both uppercase and lowercase letters.
The study of Houshmand \ea \ \cite{houshmand2015next} proposes two improvements.
First, they use a new variable $K$ to model the keyboard pattern (\eg, \quotes{$\mathsf{asdfgh}$}, \quotes{$\mathsf{zaqwsx}$}). 
Such patterns could not be generated using the method described in \cite{weir2009password}.
Second, they support modeling passwords with multi-word phrases (\eg, \quotes{$\mathsf{iloveyou}$}, \quotes{$\mathsf{tingting}$}) and non-dictionary alphabet patterns (\eg, \quotes{$\mathsf{abcdef}$}, \quotes{$\mathsf{xxyyzz}$}) by additionally adopting several variables.
Veras \ea \ \cite{veras2014semantic} proposed a password guessing model by using PCFG built on top of Part of Speech Tagging \cite{voutilainen2003part}.
Specifically, they applied Part of Speech Tagging to segment the passwords into semantic and syntactic categories instead of only semantic as in \cite{weir2009password}.
The syntactic categories clearly specify the role of a word in the spoken language (\eg, \quotes{$\mathsf{love}$} can be either a verb or a noun depending on its role in a given clause).
This helps further narrow down the search space when performing a guessing attack.
%
%
In their PCFG, the terminal symbols $\mathcal{T}$ include the source corpora (\ie, dictionary word) and the learned gap segments (\ie, unknown word, sequence of special symbols).
And the set of semantics (\eg, number, special symbols) and syntactic categories of identified words (\eg, noun, verb, adjective) form the non-terminal set $\mathcal{N}$.
Li \ea \ \cite{li2016study} propose personal-PCFG which additionally employs several variables to model different kinds of personal information (\eg, $A$ for account identity, $B$ for birthdate, $C$ for phone number, $E$ for email address, $I$ for identifier number, and $N$ for personal name).
While most prior studies focused on enhancing the variable set to model different types of passwords, Cheng \ea \ \cite{cheng2021improved} focus on improving the semantic parsing step (\ie, splitting the training password into segments).
Specifically, semantic parsing requires effectively identifying the words inside passwords.
In the prior approaches, this task is done by matching the possible substrings of the passwords with a list of standard words (\eg, dictionary, keyboard pattern, popular words).
For the word that is not included in the list, it can not be identified by matching.
The parsing method in \cite{cheng2021improved} addresses this problem through an iterative process that updates the list of standard words with new words identified in the training passwords.
To identify new words, they use the concepts of \ti{cohesion} and \ti{freedom} of words which were proposed in \cite{he2001bootstrap}.
The cohesion evaluates the internal association of a given string. %
A complete word should have large cohesion.
On the other hand, freedom measures the independence of a substring from its context.
Large freedom means the string likely composes of several words.
Wang \ea \ \cite{wang2016targeted} proposed \ti{TarGess} framework which consists of four targeted password guessing models (named \ti{TarGuess I} to \ti{TarGuess VII}) based on PCFG to address different practical scenarios of targeted password guessing.
%
%
%
\ti{TarGuess-I} aims to online guess a targeted user's password by exploiting some personal information of that user (\eg, name, birthday, not gender).
The \ti{TarGuess-II} algorithm guesses a user's password given a leaked password of that user.
\ti{TarGuess-III} and \ti{IV} used both one leaked password and some personal information of a user to guess the unknown password of that user.
The authors also noted that the ideas of these models can be implemented using other algorithms (\eg, Markov). 

\subsubsection{Deep Learning}
Inspired by the success of deep learning in different fields (\eg, Agriculture \cite{kamilaris2018deep}, Medical \cite{litjens2017survey}, Robotics \cite{karoly2020deep}), many studies have adopted deep learning techniques to password guessing (\eg, Recurrent Neural Network \cite{xu2017password,fang2018password,luo2019recurrent,zhang2020csnn}, Generative Adversarial Network \cite{hitaj2019passgan,guo2021pggan,pasquini2021improving}, Convolutional Neural Network \cite{pasquini2021reducing}, Transformer \cite{li2019password}). 

\textbf{\textit{Recurrent Neural Network}}: Recurrent neural network (RNN) refers to a class of artificial neural networks that are specialized for sequential data (\eg, time-series signals, natural language data).
An RNN processes the sequential data step-by-step in a way that the output of a step is affected by both the input of that step and the outputs of previous steps.
It is well-known that standard RNN suffers from the vanishing gradient problem and is ineffective for modeling long sequential data \cite{hochreiter1998vanishing}.
%
%
Long Short-Term Memory (LSTM) is an RNN architecture specifically designed to address the vanishing gradient problem \cite{hochreiter1997long}.
Inspired by the success of LSTM in various sequential data types, some studies adopted LSTM for password modeling (\eg, \cite{melicher2016fast,xu2017password,fang2018password,li2019password,luo2019recurrent,zhang2020csnn}).

\begin{figure*}[t]
	\centering
	\includegraphics[scale=0.07]{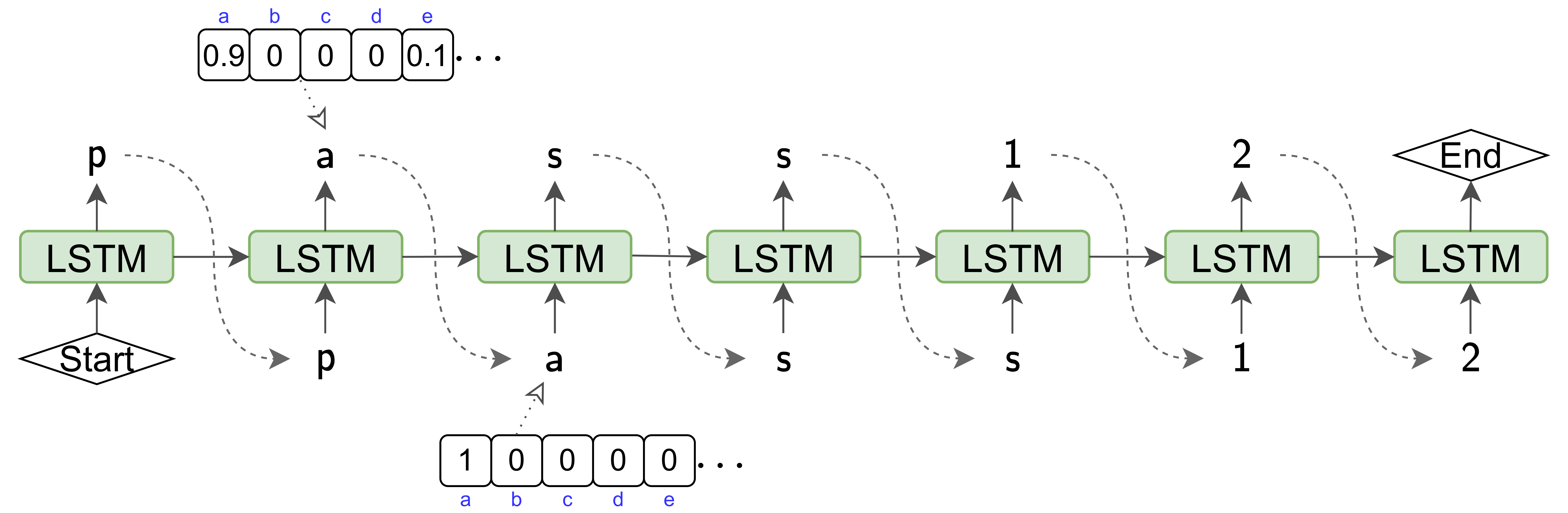}
	\caption{An example of using LSTM proposed in \cite{melicher2016fast} to generate the password \quotes{$\mathsf{pass12}$}. In each step, the network outputs one character (represented by a one-hot encoding vector) which is used as the input of next step to generate next character.}
	\label{fig_lstm_pg}
\end{figure*}

The first LSTM-based password model was proposed by Melicher \ea \ \cite{melicher2016fast}.
In that study, the password guessing task is performed by using an LSTM network (having a 3-layer LSTM and 2 fully connected layers) to repeatedly predict the next character of the password given preceding character(s).
For example, to generate the password \quotes{$\mathsf{pass12}$}, the LSTM first predicts the probability of `$\mathsf{p}$'.
Then, it uses `$\mathsf{p}$' to forecast the occurrence of `$\mathsf{a}$'.
Then, \quotes{$\mathsf{pa}$} is used to predict `$\mathsf{s}$'. 
The process is repeated until the \ti{password-ending} symbol is encountered.
To be processed by LSTM, each character of a password is transformed into a numerical vector using \ti{one-hot encoding}. 
For example, to represent a character from the space of $26$ alphabet letters and $10$ number digits, a vector of length $36$ is used, in which the value of component indexed by the character is $1$ and all other components are $0$.
In this case, the output (at each step) of LSTM is also a vector of length $36$.
Each component of the output vector contains the predicted probability of the corresponding character.
The probability of the predicted password is the production of all characters' probabilities.
Figure \ref{fig_lstm_pg} illustrates an example of generating the password \quotes{$\mathsf{pass12}$} using LSTM. 
Fang \ea \ \cite{fang2018password} proposed another password guessing model by combining semantic analysis and LSTM.
Specifically, before processing by LSTM network, the training passwords are split into segments by their semantic (\eg, dictionary word, keyboard pattern, number, special symbol).
Then, the segments are used as the basic data for each step of LSTM (instead of characters as in \cite{melicher2016fast}).
The idea of combining semantic analysis and LSTM was also used in \cite{zhang2020csnn} to target the Chinese passwords.
In that study, the Chinese syllabus is used to parse the training passwords to obtain the password structures and segments.
Then, the segments and structures were used for training the LSTM network.
Luo \ea \ \cite{luo2019recurrent} proposed an LSTM-based password guessing that works on frames of fixed length.
%
In each step, this LSTM receives a segment of $T$ characters as input, and uses it to guess the next character of the password, where $T$ is the model parameter specified by the user.
The output of each step is also a segment of $T$ characters, in which, the last character is the predicted one, and the first $T-1$ characters are the last $T-1$ characters of the input. 
For example, given a password as \quotes{$\mathsf{passw12}$} and $T=3$, the first step of LSTM will get \quotes{$\mathsf{pas}$} as input and output \quotes{$\mathsf{ass}$}.
In the second step, \quotes{$\mathsf{ass}$} is input to get \quotes{$\mathsf{ssw}$}.
The process is repeated to complete all characters of the password.
In addition, the model also integrates the group information of the targeted user (\eg, game, programmer, marriage, writing, social) to enhance the hitting probability.
Li \ea \ \cite{li2019password} proposed an improved RNN model for password guessing with Bidirectional LSTM (BiLSTM).
Bidirectional RNN processes the information flow in both the backward and forward directions \cite{schuster1997bidirectional}.
Thus, the model is capable of incorporating information from both the past and the future to produce more meaningful features at each step.
Besides, the authors leveraged and transformed the learned knowledge in BERT (Bidirectional Encoder Representations from Transformer) \cite{devlin2018bert} to their model using knowledge distillation \cite{hinton2015distilling}.

Pal \ea \ \cite{pal2019beyond} proposed \ti{pass2path}, an LSTM deep network following the Encoder-Decoder architecture \cite{sutskever2014sequence}, to model the user's behavior on reusing passwords.
This study introduced a different approach compared to prior methods which mostly focused on modeling password distribution.
Specifically, given a password, pass2path produces some \ti{transformation paths} that users will likely perform in practice in order to get new passwords. 
Each transformation path involves a sequence of substituting, inserting, or deleting a character of the password to obtain a new password. 
The pass2path's architecture comprises two main sub-networks: an Encoder and a Decoder, each one is a 2-layer LSTM with residual connections.
The operation of pass2path is similar to a typical seq2seq model \cite{sutskever2014sequence}. 
Given an input password (each password's character is represented by a one-hot encoding vector), the Encoder maps it to a latent vector denoted as $\mathbf{v}_0$.
Then, the Decoder gets $\mathbf{v}_0$ and a special \ti{beginning-of-sequence} symbol as the inputs, to output a probability distribution over the set of transformations.
The most probable transformation is selected and fed to the Decoder in the next iteration to get the next transformation set, until the \ti{end-of-sequence} symbol is returned.
%
%
To obtain a set of transformation paths, beam search \cite{wilt2010comparison} of size $q$ is used (at each step $q$ most probable transformations are used). 
The obtained transformation paths are then applied to the password to generate the candidate passwords.

%
%

\textbf{\textit{Deep Generative Models}}:
Deep generative models (DGMs) are neural networks trained to parametrize the distribution of unknown high-dimensional data, given a set of training data samples.
%
%
Upon successfully training, the DGMs can be used to generate new samples that comply with the learned data distribution. 
There are two common approaches for training a DGM: Generative Adversarial Networks (GAN) \cite{goodfellow2014generative} and Autoencoder (AE) \cite{kingma2013auto}.
%
%
%

\ti{Generative Adversarial Network (GAN)} is a generic architecture for training a generative model, proposed by Ian Goodfellow \ea \ \cite{goodfellow2014generative}.
Given a training dataset $\mathcal{X}=\{\mathbf{x}_1, \mathbf{x}_2,\ldots,\mathbf{x}_n\}$, GAN can automatically discover the regularities or patterns in a training dataset $\mathcal{X}$ such that the trained model can generate new data samples that have the same distribution as the training set.
GAN's architecture involves two sub-networks: the \ti{generator} and the \ti{discriminator}.
The generator takes as input a fixed-length random vector $\mathbf{z}$ and tries to create a faked data sample that is as close to the real in $\mathcal{X}$ as possible. 
The discriminator tries to discriminate whether a given data sample is real or generated by the generator. %
In the training process, the task of data distribution estimation to parametrize the generator is transformed into the binary classification problem which discriminates the real and fake data samples, performed by the discriminator.
Formally, the generator and discriminator are trained simultaneously following a minimax game with the loss function:
\begin{equation}\label{eq_gan_loss}
\min\limits_{\theta_G} \max\limits_{\theta_D} E_{\mathbf{x}}[\log(\mathsf{f}(\mathbf{x};\theta_D))] + E_{\mathbf{z}}[\log(1-\mathsf{f}(\mathsf{g}(\mathbf{z};\theta_G);\theta_D))],
\end{equation}
where $\theta_G$ and $\theta_D$ mean the parameters of the generator and discriminator, respectively; 
$\mathsf{f}(\mathbf{x};\theta_D)$ represents the probability estimated by the discriminator that a real data $\mathbf{x}$ is real; 
$\mathsf{g}(\mathbf{z};\theta_G)$ means generating a fake data sample by the generator when inputting a random vector $\mathbf{z}$; 
and $\mathsf{f}(\mathsf{g}(\mathbf{z};\theta_G);\theta_D))$ represents the probability estimated by the discriminator that the fake data is real.
Once trained, the generator can output the data samples that have the same distribution as the training set $\mathcal{X}$ when a random vector is fed as its input. 

\tb{\ti{Autoencoder}}: Autoencoder is another deep generative model that comprises two deep networks: an \ti{encoder} and a \ti{decoder}.
The encoder network maps the input data $\mathbf{x}$ into a vector $\mathbf{z}$ in latent space which is usually smaller than the data space.
The decoder gets the latent vector $\mathbf{z}$ and tries to obtain a reconstructed version $\mathbf{x}'$.
Basically, an AE network is optimized by minimizing the distance between a reconstructed vector $\mathbf{x}'$ and its original version $\mathbf{x}$,
\begin{equation}
\min \limits_{\theta_E,\theta_D} \sum_{x\in\mathcal{X}} \mathsf{d}(\mathbf{x},\mathbf{x}'),
\end{equation}
where $\theta_E$ and $\theta_D$ respectively mean the parameters of the encoder and decoder, $\mathcal{X}$ are the training dataset, and $\mathsf{d}$ is the desired distance measurement function.
%
%
%
Once trained, the decoder can be used as a deep generator network which gets a latent vector as input and outputs a data sample.
However, the basic AE architecture as described above is unable to produce new content. 
So, several regularization methods have been proposed, which force the latent space to follow a chosen prior distribution (\eg, Variational Autoencoder (VAE) \cite{kingma2013auto}, Adversarial Autoencoder (AAE) \cite{makhzani2015adversarial}, Wasserstein Autoencoder (WAE) \cite{tolstikhin2017wasserstein}). 

\begin{figure*}[t]
	\begin{subfigure}{1.00\textwidth}
		\centering
		\includegraphics[scale=0.09]{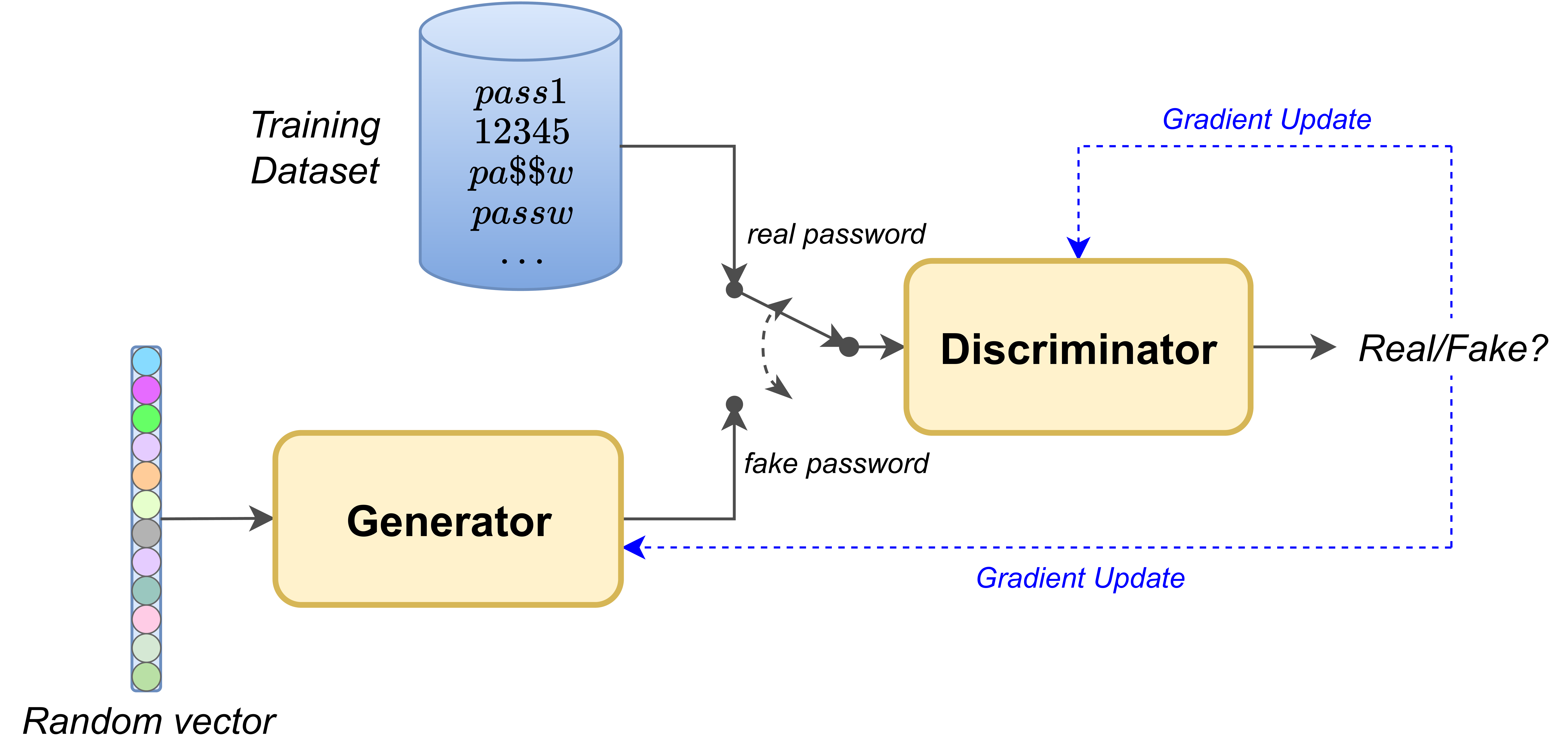}
		\caption{ }
		\label{fig_gan_pg}
	\end{subfigure}%
	\\
	\begin{subfigure}{1.0\textwidth}
		\centering
		\includegraphics[scale=0.09]{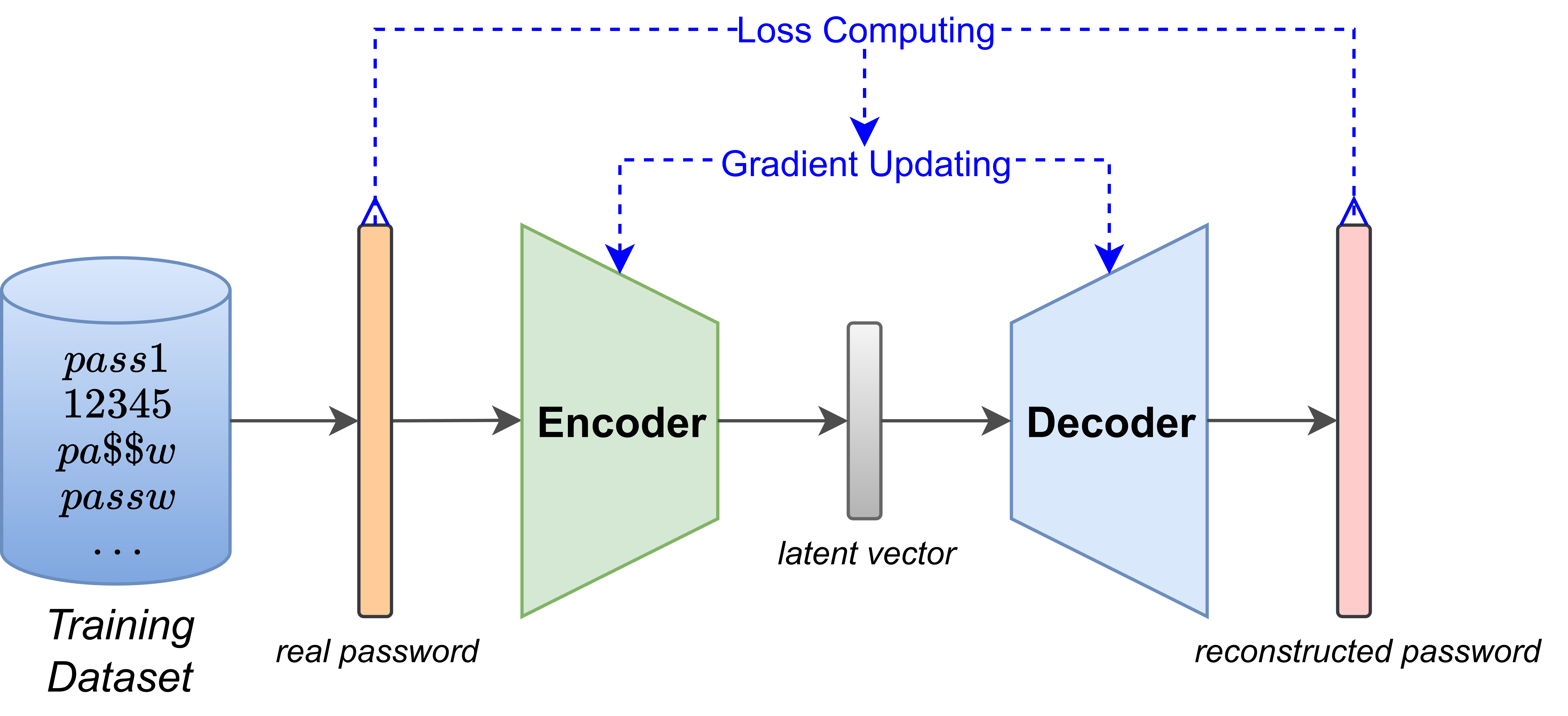}
		\caption{ }
		\label{fig_ae_pg}
	\end{subfigure}%
	\caption{The generative models for password guessing. (a) Generative Adversarial Network. After training, the \ti{Generator} is used to sample new passwords for guessing; (b) Autoencoder. Upon trained, the \ti{Decoder} can generate candidate passwords.}
	\label{fig_generative_pg}
\end{figure*}

%
Inspired by the success of DGMs on various tasks, several DGMs have been proposed for password guessing (\eg, GAN-based models \cite{hitaj2019passgan,guo2021pggan,pasquini2021improving,zhou2022password}, AE-based models \cite{pasquini2021improving}). 
%
%
%
In such studies, the leaked passwords are used to train a deep generative network following GAN or AE. 
%
%
%
Figures \ref{fig_generative_pg} illustrate the use of GAN and AE for training a deep password generator.
Upon training, the generator (GAN) or decoder (AE) is used to sample the candidate passwords.
%
%
%
Hitaj \ea \ \cite{hitaj2019passgan} proposed PASSGAN which is the first password guessing model using GAN to autonomously learn the password distribution from a leaked password dataset.
In that study, the authors followed the improved Wasserstein GANs (IWGAN) \cite{gulrajani2017improved}, and used the stack of 5 residual blocks \cite{he2016deep} as the backbones for the generator and discriminator. 
The model was trained with the leaked password dataset.
To be processed by the deep model, each character was represented by a one-hot encoding vector.
Then, each password was represented by a matrix formed from the one-hot encoding vectors of its characters.
However, only passwords of $10$ characters or less were considered in that study.
%
%
Guo \ea \ \cite{guo2021pggan} proposed PGGAN which aimed to address the high duplication rate (\ie, some passwords are repeatedly generated) in the existing GAN-based password guessing models. 
The PGGAN architecture composes of the generator, discriminator, and controller. 
The functions of the generator and discriminator are similar to those in the original GAN.
The controller receives an input string which is either a password from the generator or a random-generated string, and tries to tell the origin of this string.
By combining the feedback from the discriminator and controller, the generator is trained to output the passwords that are close to the distribution of training dataset and also close to the uniform distribution. 
The above goals contradict each other, thus, a suitable balance should be specified.
Zhou \ea, \cite{zhou2022password} proposed G-Pass which is another GAN-based password generation model. 
G-Pass used LSTM network for the generator, and multiple convolutional layers with multiple filter sizes for the discriminator.
In addition, they used Gumbel-Softmax \cite{jang2016categorical} to address the difficulty of gradient computing when working with categorical data (\eg, characters of passwords).
Pasquini \ea \ \cite{pasquini2021improving} proposed two deep generative models, one used improved Wasserstein GAN \cite{gulrajani2017improved}, and another relied on Wasserstein Autoencoders \cite{tolstikhin2017wasserstein}.
To overcome the problem of discrete representation of the characters, they used stochastic smoothing which adds small-magnitude noise to the one-hot-encoding vector of each character.
The authors also utilized the geometric relation of vectors in the latent space to propose the concept of password locality.
Specifically, in the latent space, vectors of different classes of passwords are organized into different zones.
This enables conditional password guessing, which allows the generation of passwords in a specific class by inputting vectors from a specific zone in the latent space.


\subsubsection{Hybrid Model}

%

There are several hybrid approaches that combined different techniques to build a password guessing model (\eg, PCFG and BiLSTM \cite{zhang2018password}; GENPass (PCFG and LSTM) \cite{liu2018genpass,xia2019genpass}, PaMLGuess (PCFG and LSTM) \cite{zhang2020preliminary}).
Zhang \ea \ \cite{zhang2018password} proposed a password guessing model by combining PCFG and BiLSTM.
In the training phase, PCFG was used to parse the passwords to a collection of basic structures (non-terminal forms) and strings (common words) sorted by probability. 
The strings were used to train a BiLSTM network.
%
In the password generating process, the strings generated by BiLSTM were combined with the password structures (obtained from the training phase) to sample candidate passwords.
The authors in \cite{liu2018genpass, xia2019genpass} proposed GENPass, a password generation method built from the ideas of PCFG and LSTM.
GENPass is designed to generate a generic candidate password list which is learned from different leaked datasets.
First, the combination of PCFG and LSTM (called PL) was used to learn, then generate passwords following a specific dataset.
Using PCFG, the pre-terminal form of each training password was obtained (\eg, $S_1L_4D_3$ from \quotes{$\mathsf{!pass135}$}) (see Section \ref{sssec_pcfg}).
From the pre-terminal forms, an LSTM network was trained to generate high-probable pre-terminal forms for password guessing.
Each PL model was trained separately using its own dataset.
Upon being trained, all the PL models were combined (\ie, their outputs were fused to provide a generic list of the generated non-terminal forms). 
In this step, a discriminator module is used to filter the passwords that do not have a consistent probability of appearance in all the PL models. 
Zhang \ea \ \cite{zhang2020preliminary} combined Targess I (PCFG) \cite{wang2016targeted} and LSTM to construct a password guessing model.
Given the training password dataset, the basic password structures and the strings (words) were extracted according to the variables defined in Targess I. 
The strings were used to train LSTM network.
For generating the candidate passwords, the LSTM network samples new strings which will be fed to the password structures.

%
%
%
%
%
%
%
%


\subsection{Guessing Performance}

This section discusses the performance of existing password guessing models. 
First, we summarize methods and procedures used to experimentally evaluate the guessing performance of existing studies.
Then, we present the criteria and metrics used in the evaluation.
Finally, we summarize the experimental guessing performances reported in the original papers.

\subsubsection{Evaluation Method}
%

To evaluate the performance of a password guessing model, a set of passwords $\mathcal{T}$ were used as the targets to be attacked.
As passwords are extremely sensitive, it is impractical to collect and establish public benchmark datasets by leveraging contributions from volunteers.
Users may contribute unrealistic passwords (\ie, those that they do not actually use) which raises a concern about the reliability of research. 
Moreover, it is also challenging to manage and share the collected datasets so that they are only used for research.
Most existing studies leveraged the real password datasets leaked by some attacks and made public on the Internet.
There are a large number of leaked password datasets, each one can have thousands to millions of real passwords which have served as an irreplaceable source for password modeling research (see Section \ref{sssec_real_password} for the list of these datasets).
However, only relying on the leaked password datasets faces several limitations (see \ref{sssec_future_work}).
Some studies leveraged the real password datasets obtained from their organizations (\eg, ONYEN from California University \cite{zhang2010security}, CMU password dataset \cite{mazurek2013measuring}). 
Such datasets usually contain additional user information such as email, user ID.
However, these datasets could not be shared for public research due to user privacy.

Given a target set $\mathcal{T}$, the guessing model (under evaluation) will generate candidate passwords to be tried against $\mathcal{T}$.
Depending on the security model (\ie, online or offline), the subsequent evaluation procedure will have some differences.
For an offline password guessing model, a sequence of candidate passwords $\mathcal{S} =\{s_1, s_2, \ldots, s_N\}$ are generated (usually following the order of descending probability, \ie, $\mathsf{p}_{(s_1)}> \mathsf{p}_{(s_2)}>...>\mathsf{p}_{(s_N)}$).
Then, each candidate password in $\mathcal{S}$ is attempted against all the passwords in $\mathcal{T}$. 
After each iteration, the matched passwords in $\mathcal{T}$ are removed from the set (and are considered cracked).
The number of cracked passwords (after trying all candidate passwords) is used to determine the model performance.
As the offline password attack does not limit the number of trials, a large number of candidate passwords (\eg, millions of passwords) can be generated, depending on the model's capability.

In the online setting, the procedure is slightly different.
For each target $t$ in $\mathcal{T}$, the model will generate a sequence of candidate passwords $\mathcal{S}^t =\{s^t_1, s^t_2, \ldots, s^t_n\}$, also following the descending probability order. 
However, $\mathcal{S}^t$ are usually generated by leveraging both the \ti{global} and \ti{target's information}. %
The global information is similar for all targets (\eg, password policy, native language) while the target's information is different for each user (\eg, name, birthdate, old password, password-forming behavior).
So, the password sequences used to attack different targets are not guaranteed to be similar.
%
%
Moreover, for an online attack, the number of candidates is usually small (\eg, 5).

\subsubsection{Evaluation Criteria}\label{sssec_eva_criteria}
A password guessing model can be evaluated based on several criteria such as guessing performance, running time, generation capacity, duplicate rate, and model size.

\tb{Guessing Performance}: (\ie, guessing accuracy) is the most important criterion to evaluate a password guessing model.
Several metrics have been defined for this purpose. 
The most common one is the $\beta$-\ti{success rate} which was formally described by Boztas \cite{boztas1999entropies}.
$\beta$-\ti{success rate} shows the expected ratio of passwords being cracked when guessing with $\beta$ candidates generated by the guessing model.
Let $p_i$ be defined as $p_i=\frac{|\hat{\mathcal{T}}_i|}{|\mathcal{T}|}$, where $\hat{\mathcal{T}}_i$ means a subset of $\mathcal{T}$ which are cracked by the guess $i$.
Then, the $\beta$-\ti{success rate} is determined by:
\begin{equation}\label{eq_beta_success_rate}
\lambda_{\beta}(\mathcal{T}) = \sum_i^{\beta}p_i.
\end{equation}
From the definition, $\lambda_{\beta}(\mathcal{T})$ was originally designed for the evaluation of an offline model as it is computed by accumulating the success rates when using a (fixed) sequence of candidate passwords to guess a set of targets $\mathcal{T}$.
In practice, $\lambda_{\beta}(\mathcal{T})$ can also be used to measure the performance of an online model. In this setting, it is determined by
\begin{equation}\label{eq_on_beta_success_rate}
\Lambda_{\beta}(\mathcal{T}) = \frac{|\hat{\mathcal{T}}_{(\beta)}|}{|\mathcal{T}|},
\end{equation}
where $\hat{\mathcal{T}}_\beta$ is the set of users cracked within $\beta$ guesses.
Overall, $\Lambda_{\beta}(\mathcal{T})$ can be used to measure the performance of both the online and offline models, and has been widely used in prior password guessing research.
The main difference between two settings is how to determine the set of users cracked within $\beta$ guesses (\ie, $\hat{\mathcal{T}}_\beta$).
In the online mode, a sequence of $\beta$ candidate passwords is optimally generated for each targeted user in $\mathcal{T}$.
So, there could be $|\mathcal{T}|$ different sequences being generated.
On the other hand, the offline setting uses only one sequence of passwords to attack all passwords in $\mathcal{T}$.
In addition, $\beta$ is usually small (\eg, $\beta=3$, $\beta=5$) for the online setting while it can reach thousands of million in an offline attack.
Note that, in most existing password guessing studies, $\beta$-success rate was represented in form of \ti{percent}, instead of the ratio.
%

Another metric used to evaluate the guessing performance is $\alpha$-\ti{work-factor}, first formalized by Pliam \cite{pliam2000incomparability}, which shows the fixed number of guesses (per account) required to crack $\alpha$ passwords in $\mathcal{T}$:
\begin{equation}\label{eq_alpha_work_factor}
	\mu_{\alpha}(\mathcal{T}) =  \min\{j\bigg\rvert\sum_{i=1}^j p_i\ge \alpha\},
\end{equation}
where $p_i$ is the fraction of passwords in $\mathcal{T}$ cracked by the guess $i$.
Similar to $\beta$-\ti{success rate}, the $\alpha$-\ti{work-factor} can be used either online or offline as explained above.

\tb{Running Time:} measures how long the model needs to generate a candidate password.
In the case of online password guessing, only a small number of guesses are allowed.
Moreover, the password generation time is much smaller compared to the network delay (online guessing usually targets a remote service).
Thus, the running time is not important and is mostly ignored for an online password guessing model.
However, this criterion is worth considering for an offline model due to several reasons.
First, an offline password cracking model usually performs millions of guesses.
Second, the cracking speed mainly relies on the password generation speed and the operation of hashing, however, no network delay.
Third, one of the main uses of an offline guessing model is proactive password checking which usually runs on the client side having low computing resources.
Thus, having a low running time is a considerable advantage of an offline password guessing model.

\tb{Other Criteria:} besides, there are several other criteria that need to be aware of when designing a password guessing model.

\ti{Model size} is an important criterion when using the password guessing model as a proactive password checker deployed on the client side \cite{wheeler2016zxcvbn}.
The small-size model can be quickly transferred to and installed in the client device.
Moreover, such models are also ideal for mobile devices which usually have low storage capacity.

\ti{Generation capacity} refers to the number of unique passwords a model can generate.
A password guessing model has an upper bound $B$ on the number of unique passwords that it can generate (\eg, $B=10^{10}$).
Upon reaching this limitation, some models can not generate any more new candidate passwords without changing their referenced information (\eg, Markov, PCFG, mangling rule).
On the other hand, some models will generate repeated passwords upon reaching their upper bound (\eg, GAN-based model)
As an online attack is limited by a small number of guesses, the generation capacity is mostly ignored when designing an online password guessing model.
However, offline password guessing is only limited by computing speed and generation capacity.
So, besides arranging the candidate passwords in the decreasing order of hitting chances, the offline password guessing model also notices the generation capacity.
This could be enhanced by increasing the amount of referenced information, or the model parameters.

\ti{Duplicate rate}: refers to the portion of passwords that the model generates more than once (\ie, the passwords that are repeatedly produced by the model).
In the online attack, a small number of candidate passwords with high hitting-chance are generated. 
Thus, there is no duplication on the generated passwords.
In the offline setting, the model can produce millions of passwords and duplication is easy to occur.
The duplicate passwords prolong the attacking time and reduce the guessing performance.
Thus, some studies are also aware of reducing the duplicate rate of the generated candidate passwords \cite{guo2021pggan}.

\subsubsection{Evaluation Results}
%

%
\begin{table}[ht]
	\centering
	\caption{Experiment procedure and reported performance of the existing online password guessing models.}
	\label{table_onpg_per}
	\def\arraystretch{1.2}
	\begin{tabular}{|cccccc|} \hline
		\multirow{2}{*}{\ti{\tb{Method}}}
		& \multirow{2}{*}{\ti{\tb{Target Set}}} 
		& \multicolumn{2}{c}{\ti{\tb{Referenced Data}}} 
		& \ti{\tb{No. Guesses}}
		& \ti{\tb{$\beta$-success Rate}} 
		\\ \cline{3-4}

		&
		& \ti{Global}
		& $^{(\S)}$\ti{Target's Infor}
		&\ti{\tb{$(\beta)$}}
		& $(\Lambda_\beta)$ \\ 
		\hline
		
		\multirow{2}{*}{Zhang \ea \ \cite{zhang2010security}} 
		& $7,752$ users	
		& \multirow{2}{*}{$-$} 		
		& \multirow{2}{*}{P}		
		& $5$ 
		& $0.13$ 
		\\ \cline{5-6}
		& of ONYEN 
		& 
		&  
		& $20$ 
		& $0.28$ 
		\\ 
		\hline

		
		Personal PCFG 
		& $50\%$ of 
		& $50\%$ of 
		& N, B, E, 
		& \multirow{2}{*}{$5$} 
		& \multirow{2}{*}{$0.048$} 
		\\ 
		Li \ea \ \cite{li2016study}
		& 12306 
		& 12306
		& A, H, I 
		& 
		& 
		\\
		\hline
		
		
		\multirow{2}{*}{Das \ea \ \cite{das2014tangled}}
		& $3,463$ passwords 	
		& \multirow{2}{*}{$-$} 		
		& \multirow{2}{*}{P}	
		& \multirow{2}{*}{$10$}
		& \multirow{2}{*}{$0.1$}
		\\ 
		& of 10 sites 
		& 
		& 	
		& 
		& 
		\\ 
		\hline
		
		TarGuess-I
		& 50\% of 
		& 50\% of 
		& B, N, E,
		& \multirow{2}{*}{$5$} 
		& \multirow{2}{*}{$0.085$} 
		\\ 
		Wang \ea \ \cite{wang2016fuzzypsm}
		& 12306	
		& 12306 
		& A, H, I
		& 
		& 
		\\
		\hline
		
		TarGuess-II
		& $308,045$ by 
		& $66,573$ by 
		& \multirow{2}{*}{P} 
		& \multirow{2}{*}{$5$}
		& \multirow{2}{*}{$0.065$}
		\\ 
		Wang \ea \ \cite{wang2016fuzzypsm}
		& $\text{Dodonew} \cap \text{CSDN}$ 
		& $126 \cap\text{CSDN}$
		& 
		& 
		& 
		\\ 
		\hline
		
		TarGuess-III
		& $308,045$ by  
		& $66,573$ by  
		& P, N, B, 
		& \multirow{2}{*}{$5$}
		& \multirow{2}{*}{$0.066$}
		\\
		Wang \ea \ \cite{wang2016fuzzypsm}
		& $\text{Dodonew}\cap\text{CSDN}$
		&$126 \cap \text{CSDN}$ 
		& H, I 
		& 
		& 
		\\
		\hline
		
		TarGuess-IV
		& $308,045$ by 
		& $66,573$ 
		& P, N, B, 
		& \multirow{2}{*}{$5$}
		& \multirow{2}{*}{$0.076$}
		\\
		
		%
		Wang \ea \ \cite{wang2016fuzzypsm}
		& $\text{Dodonew} \cap \text{CSDN}$  
		& $126 \cap \text{CSDN}$ 
		& H, I, G %
		& 
		& 
		\\
		\hline

		Pass2Path 
		& $29.2$ mils 
		& $146.4$ mil
		& \multirow{2}{*}{P} 
		& $10$
		& $0.099$
		\\\cline{5-6}
		
		Pal \ea \ \cite{pal2019beyond}
		& users
		& passwords
		& 
		& $10^2$
		& $0.0131$
		\\		
		
		\hline
		Pass2Path 
		& $15,776$ users of 
		& $146.4$ mil
		& \multirow{2}{*}{P} 
		& \multirow{2}{*}{$10$}
		& \multirow{2}{*}{$0.033$}
		\\
		
		Pal \ea \ \cite{pal2019beyond}
		& Cornell Uni
		& passwords
		& 
		& 
		& 
		\\		
		
		\hline

		PaMLGuess %
		& $50\%$ of 
		& $50\%$ of 
		& \multirow{2}{*}{P}
		& $10^2$
		& $0.135$
		\\ \cline{5-6}
		
		Zhang \ea \ \cite{zhang2020preliminary}
		& 12306 
		& 12306 
		& 
		& $10^3$
		& $0.18$
		\\		
		\hline
		\multicolumn{6}{c}{\small `$\cap$' \ti{means merging $2$ datasets to get the users having passwords in both datasets}}\\
		\multicolumn{6}{c}{\small $^{(\S)}$\ti{Target's Infor:} {`P' - \ti{leaked password;} `N' - \ti{name;} `B' - \ti{birthdate;} `E' - \ti{email;}}}\\
		
		\multicolumn{6}{c}{\small {`A' - \ti{account name,} `H' - \ti{phone number,} `I' - \ti{government issued number.}}}\\
		%
		%
		%
		
	\end{tabular}
	
\end{table}

In this section, we summarized the performance reported in existing password guessing models which are categorized into two groups: online and offline.
Note that, to be unified, we used $\beta$-\ti{success rate} (see \ref{sssec_eva_criteria}) to describe the guessing performance.
For the studies that used other metrics, we convert the reported performance to $\beta$-success rate (if possible) or provide additional notice.
In addition, we used $\Lambda_{\beta}$ as the notation of $\beta$-\ti{success rate} (instead of $\Lambda_{\beta}(\mathcal{T})$) also for simplicity.

\tb{Online Model:} Evaluating the performance of an online password guessing model is such a challenging task due to the inability to collect benchmark datasets.
%
%
This comes from the fact that most online guessing models require the leaked passwords and personal information of the targeted user to guess the unknown passwords of that user.
However, in practice, such information is extremely sensitive, and almost impossible to be obtained from the volunteers.
Most leaked password datasets have one password per user, and none or very little user information, thus, are not suitable for evaluating the online guessing models.
As a result, each study tried to obtain the evaluation dataset in a different way (\eg, using private database, combining different datasets) which is hard to reproduce or compare.

Zhang \ea \ \cite{zhang2010security} used a dataset containing $51,141$ hashed passwords from $10,374$ inactive users of the ONYEN service\footnote{\href{https://its.unc.edu/onyen-services/}{ONYEN service, The University of California}}, where each user had $4$ to $15$ passwords.
They first used various password cracking methods (\eg, Hashcat \cite{jens2009hashcat}, John the Ripper \cite{peslyak2014john}) to crack and obtain $31,075$ plaintext passwords of $7,936$ users.
Then, they used $7,752$ users who had at least two cracked passwords for their experiment.
With each user, given a password, they applied some mangling functions to get the unknown passwords of that user.
On average, the reported performance is $\Lambda_{5}=0.13$ and $\Lambda_{13}=0.18$ (\ie, $13\%$ of passwords can be guessed within $5$ attempts, and $18\%$ with $10$ attempts).
%
%
%
Li \ea \ \cite{li2016study} evaluated their model on the 12306 dataset.
This dataset contained more than $130,000$ passwords (along with some information such as user name, the government-issued unique ID number) leaked from the official site of an online railway ticket reservation in China\footnote{\href{https://www.12306.cn/index/}{https://www.12306.cn/index/}}. 
They used half of the dataset for training the PCFG (\ie, global data), and evaluated the trained model with the remaining half.
To guess the password of a user, they fed that user's personal information (\ie, target's data) to the PCFG's grammar to generate the candidate passwords.
They reported that their model achieved the performance as $\Lambda_{5}=0.048$. 
%
Instead of using passwords leaked from a service/organization, some studies combined datasets leaked from different sources to obtain new datasets having multiple leaked passwords per each user (\eg, \cite{das2014tangled,wang2016targeted}).
For instance, Das \ea \ \cite{das2014tangled} created a password dataset having user information and at least two passwords by combining the leaked datasets of $10$ sites (about $7,962,678$ passwords).
Then, they filtered and obtained a dataset of $6,077$ users who had at least two passwords used for different sites.
Furthermore, they removed all identical password pairs.
After all, the remaining dataset has only users that used different passwords for different sites ($\approx 3,463$ users).
Given a leaked password of a user, they performed some transformations to generate candidate passwords to guess the unknown passwords of that user. 
They reported that their method could achieve the performance as $\Lambda_{10}=0.1$ when guessing the nonidentical passwords.
%
Wang \ea \ \cite{wang2016targeted} also combined different leaked password databases to create sets of users having multiple passwords on different sites, then used these sets to evaluate their online password guessing models (\ie, TarGuess-I, -II, -III, -IV) under different scenarios (see Table \ref{table_onpg_per} for representative results).
%
%
%
%
Overall, their guessing performance on normal users was $\Lambda_{100}=0.73$, and $\Lambda_{100}=0.32$ for security-savvy users.
%
Pal \ea \ \cite{pal2019beyond} used a database consisting of $463$ million unique passwords of $1.1$ billion unique emails leaked from multiple sites (\eg, LinkedIn, Myspace, Badoo, Yahoo, Twitter, Zoosk, Neopet, etc).
The authors performed some data cleansing steps (\eg, remove dummy accounts, discard passwords in hex characters, etc), then, merged the accounts to find a set of passwords belonging to an individual user.
Finally, they formed a dataset that includes users having at least two passwords.
From the obtained dataset, they used $146.4$ million passwords ($116.8$ million users) for training and $42$ million passwords ($29.2$ million users) for testing.
On average, without counting the cracked passwords which appeared in the training set, their method achieved the $\Lambda_{10}$, $\Lambda_{10^2}$, and $\Lambda_{10^3}$ as $0.099$, $0.131$, and $0.158$, respectively.
The $\Lambda_{\beta}$ increased to $0.448$, $0.467$, and $0.483$ when including the passwords appeared in both training and testing sets.
When using multiple leaked passwords (at least $2$), the achieved performance was $\Lambda_{10^3}=0.23$, without counting the repeated passwords.
They additionally evaluated their model on the real email accounts of Cornell University, with the support of the IT Security Office of Cornell University (ITSO).
Cornell University adopted the password policy that required forming a password from at least $8$ characters of more than three classes chosen from upper-case, and lower-case letters, digits, and symbols.
The experiment on $15,776$ valid emails of Cornell accounts showed that the pass2path could crack $3.3\%$ of accounts within $10$ guesses.
%
Zhang \ea \ \cite{zhang2020preliminary} also used the 12306 dataset to evaluate the PaMLGuess model.
They also trained their model with $50\%$ dataset and tested on the remaining $50\%$.
The reported performances were $\Lambda_{100}=0.135$ and $\Lambda_{10^3}=0.18$.
%
We summarize in Table \ref{table_onpg_per} the reported performance of existing online password guessing models.

\tb{Offline Model:}
The offline password guessing models were evaluated on a large number of leaked password datasets available on the Internet (see \ref{sssec_real_password}).
However, these datasets were obtained by illegal attacks and leaked to the public under different versions.
There is no official organization responsible for maintaining and validating the content of such datasets.
Then, each study may use a different data cleaning process.
Thus, the datasets that were actually used for evaluation in each study are not identical.
Moreover, depending on each approach, the process of password generating may require additional information sources (\eg, dictionary, keyboard pattern, user information...).
So, we try to include as much as possible the important information involved in the experiment of each study to provide a detailed picture of the evaluation.
The experimental guessing performances of representative offline password guessing studies were summarized in Table \ref{table_offpg_per}.

\begin{table}[htp]	
	\centering
	\caption{Experimental settings and performances of the offline password guessing models.}
	\label{table_offpg_per}
	\def\arraystretch{1.2}
	
	\begin{tabular}{|C{1.8cm}C{1.8cm}cccccc|} \hline
		\multirow{2}{*}{\ti{\tb{Study}}}
		& \multirow{2}{*}{\ti{\tb{Method}}}
		& \multicolumn{2}{c}{\ti{\tb{Referenced Data}}}
		& $^{(\S)}$\ti{\tb{Attacked}} 
		& \ti{\tb{No. Guesses}}
		& \ti{\tb{$\beta$-success Rate} }
		&\ti{\tb{Note}}  
		\\ \cline{3-4}
		& 
		& \ti{Password}
		& $^{(\dagger)}$\ti{Others}
		& \ti{\tb{Set}}
		& $(\beta)$ 
		& $(\Lambda_\beta)$
		& 
		\\
		\hline
		
		%
		Narayanan \ea, 2005 \cite{narayanan2005fast}&
		Mangling Rules&
		$-$& 				
		L& 
		$142$ & 			
		$2B$ & 				
		$0.676$ & 			
		\\
		\hline
		
		\multirow{4}{1.8cm}{\centering Weir \ea, 2009 \cite{weir2009password}} &
		\multirow{4}{1.8cm}{\centering PCFG} &					
		$\text{MS}_{(33.5K)}$& 	
		\multirow{4}{*}{D}&						
		$\text{MS}_{(33.5K)}$& 	
		$37.8M$& 				
		$0.329$& 			
		\\ 
		&
		& 						
		$\text{MS}_{(33.5K)}$& 	
		&
		$\text{SW}_{(33.5K)}$ & 
		$37.8M$& 				
		$0.069$& 				
		\\		
		&
		&						
		$\text{FN}_{(15.7K)}$ & 
		&
		$\text{MS}_{(33.5K)}$& 	
		$37.8M$& 				
		$0.069$& 				
		\\		
		&
		&						
		$\text{FN}_{(15.7K)}$& 	
		&
		$\text{FN}_{(22.7K)}$& 	
		$37.8M$& 				
		$0.069$& 				
		unav					
		\\	
		\hline
		
		Zhang \ea, 2010 \cite{zhang2010security}&
		Mangling rules&					
		$\text{OY}_{(7.7K)}$ & 			
		D& 
		$\text{OY}_{(\approx20K)}$ & 	
		3 seconds&  					
		$0.41$&						 	
		pri 							
		\\\hline
		
		
		Rao \ea, 2013 \cite{rao2013effect}&
		POS&
		$-$& 					
		L& 
		$1434$& 				
		$2,500B$& 				
		$0.18$& 				
		una -spe				
		\\\hline
		\multirow{2}{1.8cm}{\centering Veras \ea, 2014 \cite{veras2014semantic}}&
		\multirow{2}{1.8cm}{\centering PCFG}& 			
		$\text{RY}_{(32M)}$ & 		
		\multirow{2}{*}{D, L}& 
		$\text{LI}_{(5.8M)}$ & 		
		$3B$& 						
		$0.28$ & 					
		\\
		
		&
		&
		$\text{RY}_{(32M)}$ & 	
		&
		$\text{MS}_{(49.6K)}$ & 
		$3B$& 					
		$0.74$ & 				
		\\\hline

		\multirow{2}{1.8cm}{\centering Li \ea, 2014 \cite{li2014large}}&
		\multirow{2}{1.8cm}{\centering PCFG}& 
		$\text{RY}_{(2M)}$& 
		\multirow{2}{*}{D, K, L}& 
		$\text{CS}$& 
		$10M$& 
		$0.127$& 
		\\

		&
		&
		$\text{DU}_{(2M}$& 
		&
		CS& 
		$10M$& 
		$0.173$ & 
		
		\\\hline
		\multirow{3}{1.8cm}{\centering Durmuth\ea, 2015 \cite{durmuth2015omen}}&
		\multirow{3}{1.8cm}{\centering Markov}&
		$\text{RY}_{(30M)}$& 				
		\multirow{3}{*}{\centering $-$}& 
		$\text{RY}_{(2.6M)}$& 				
		$1B$& 								
		$0.69$& 							
		\\
		&
		&										
		$\text{RY}_{(30M)}$&				 
		&
		$\text{MS}_{(50K)}$ & 				
		$1B$& 								
		$0.65$& 							
		\\
		&
		&										
		$\text{RY}_{(30M)}$& 				
		&
		$\text{FB}$& 						
		$1B$& 								
		$0.6$& 							
		\\
		\hline
		
		Li \ea, 2016 \cite{li2016study}& 	
		PCFG& 								
		$\text{OT}_{(50\%)}$& 				
		D, U& 	
		$\text{OT}_{(50\%)}$& 				
		$2M$& 								
		$0.49$& 							
		\\
		\hline
		\multirow{8}{1.8cm}{\centering Zhang \ea, 2018 \cite{zhang2018password}}&
		\multirow{8}{1.8cm}{\centering PCFG\& BiLSTM} & 
		$\text{DO}_{(14M)}$& 			
		\multirow{8}{*}{$-$}&
		$\text{DO}_{(2M)}$& 			
		$10M$& 							
		$0.35$& 						
		\\
		
		&
		&
		$\text{RY}_{(25M)}$& 			
		&
		$\text{RY}_{(7M)}$& 			
		$10M$& 							
		$0.53$& 						
		\\
		&
		&						
		$\text{RY}$& 			
		&
		$\text{DO}$& 			
		$10M$ & 				
		$0.355$& 				
		\\
		&
		&						
		$\text{DO}$& 			
		&
		$\text{RY}$& 			
		$10M$& 					
		$0.49$& 				
		\\

		&
		&
		$\text{DO}$& 			
		&
		$\text{CS} \cup \text{JD}$& 	
		$10M$& 					
		$0.38$& 				
		\\
		&
		&						
		$\text{RY}$& 			
		&
		$\text{PB}\cup \text{YH}$& 	
		$10M$& 					
		$0.56$& 				
		 
		\\
		&
		&						
		$\text{DO}$& 			
		&
		$\text{PB}\cup \text{YH}$ & 	
		$10M$	& 				
		$0.315$& 				
		\\
		&
		&						
		$\text{RY}$& 			
		&
		$\text{CS}\cup \text{JD}$& 	
		$10M$& 					
		$0.47$& 				
		\\
		\hline
		
		\multirow{4}{1.8cm}{\centering Xia \ea, 2019 \cite{xia2019genpass}}&
		\multirow{4}{1.8cm}{\centering GAN}&
		$\text{MS}$& 				
		\multirow{4}{*}{$-$}& 
		$\text{MS}$& 				
		$10M$& 						
		$0.556$& 					
		\\
		

		&
		&
		$\text{PB}$& 			
		&
		$\text{PB}$& 			
		$10M$& 					
		$0.579$& 				
		\\
		&
		&
		$\text{MS} \cup \text{PB}$& 	
		&
		$\text{RY}_{(10\%)}$& 	
		$1000B$& 				
		$0.34$& 				
		\\
		&
		&
		$\text{MS}\cup \text{PB}$& 		
		&
		$\text{LI}_{(10\%)}$& 		
		$1000B$& 					
		$0.22$& 					
		\\\hline
		Hita \ea, 2019 \cite{hitaj2019passgan}&
		GAN& 
		$\text{RY}_{(80\%)}$ & 			
		$-$& 
		$\text{RY}_{(1.98M)}$& 			
		$10B$& 							
		$0.26$& 					
		\\ 
		\hline

%
%
		
%
		
		Zhang \ea, 2020 \cite{zhang2020preliminary}&
		PaLSTM& 				
		$\text{OT}_{(50\%)}$ & 	
		U& 
		$\text{OT}_{(50\%)}$& 	
		$2M$& 					
		$0.255$& 				
		\\
		\hline

		\multirow{3}{1.8cm}{\centering Zhang \ea, 2020 \cite{zhang2020csnn}}&
		\multirow{3}{1.8cm}{\centering CSNN}&
		$\text{NE}_{(20.63M)}$& 		
		\multirow{3}{*}{$-$}&
		$\text{OT}$& 			
		$1M$& 					
		$0.135$& 				
		\\
		&
		&						
		$\text{NE}_{(20.63M)}$& 
		&
		$\text{DO}$& 			
		$1M$& 					
		$0.22$& 				
		\\
		&
		&
		$\text{NE}_{(20.63M)}$& 
		&
		$\text{CS}$& 			
		$1M$& 					
		$0.31$& 				
		\\
		\hline
		
		\multicolumn{8}{c}{\ti{to be continued}}
	\end{tabular}
\end{table}

\begin{table}[htp]	
	\ContinuedFloat
	\centering
	\caption{(continued) Experimental settings and performances of the offline password guessing models.}
	
	\def\arraystretch{1.2}
	
	\begin{tabular}{|C{1.8cm}C{1.8cm}cccccc|} \hline
		\multirow{2}{*}{\ti{\tb{Study}}}
		& \multirow{2}{*}{\ti{\tb{Method}}}
		& \multicolumn{2}{c}{\ti{\tb{Referenced Data}}}
		& $^{(\S)}$\ti{\tb{Attacked}} 
		& \ti{\tb{No. Guesses}}
		& \ti{\tb{$\beta$-success Rate} }
		&\ti{\tb{Note}}  
		\\ \cline{3-4}
		& 
		& \ti{Password}
		& $^{(\dagger)}$\ti{Others}
		& \ti{\tb{Set}}
		& $(\beta)$ 
		& $(\Lambda_\beta)$
		& 
		\\
		\hline
		\multirow{3}{1.8cm}{\centering Pasquini \ea, 2021 \cite{pasquini2021improving}}&
		\multirow{3}{1.8cm}{\centering GAN}& 
		$\text{RY}$& 				
		\multirow{3}{*}{D} &
		$\text{LI}$& 				
		$10B$& 						
		$0.62$&	 					
		\\
		&
		&
		$\text{RY}$& 				
		&
		$\text{YK}$& 				
		$10B$& 						
		$0.61$& 					
		\\
		&
		& 
		$\text{RY}$& 				
		&
		$\text{ZO}$& 				
		$10B$& 						
		$0.68$& 					
		\\
		\hline

		\multirow{14}{1.8cm}{\centering Pasquini \ea, 2021 \cite{pasquini2021reducing}}& 
		\multirow{14}{1.8cm}{\centering AdaMs}& 			
		MH&				
		\multirow{14}{*}{D}&				
		AN& 				
		$10^{11}$& 						
		$0.83$&	 					
		\\  

		& 
		& 			
		MH&				
		&				
		PB& 				
		$10^{11}$& 						
		$0.85$&	 					
		\\
		
		& 
		& 			
		MH&				
		&				
		RY& 				
		$10^{11}$& 						
		$0.83$&	 					
		\\

		& 
		& 			
		ZK&				
		&				
		AN& 			
		$8\cdot 10^{10}$& 						
		$0.4$&	 					
		\\
		
		& 
		& 			
		ZK&				
		&				
		PB& 				
		$4\cdot 10^{10}$& 						
		$0.37$&	 					
		\\
		
		& 
		& 			
		ZK&				
		&				
		RY& 				
		$6\cdot 10^{10}$& 						
		$0.395$&	 					
		\\
		
		& 
		& 			
		ZK&				
		&				
		MH& 				
		$10^{11}$& 						
		$0.44$&	 					
		\\

		& 
		& 			
		YK&				
		&				
		AN&				
		$8\cdot 10^{10}$& 						
		$0.52$&	 					
		\\
		
		& 
		& 			
		YK&				
		&				
		PB& 				
		$10^{11}$& 						
		$0.53$&	 					
		\\
		
		& 
		& 			
		YK&				
		&				
		RY& 				
		$5\cdot 10^{10}$& 						
		$0.48$&	 					
		\\

		& 
		& 			
		YK&				
		&				
		MH& 				
		$10^{11}$& 						
		$0.52$&	 					
		\\

		& 
		& 			
		RY&				
		&				
		AN& 				
		$9\cdot 10^{10}$& 						
		$0.62$&	 					
		\\
		
		& 
		& 			
		RY&				
		&				
		PB& 				
		$6\cdot 10^{10}$& 						
		$0.62$&	 					
		\\
		
		& 
		& 			
		RY&				
		&				
		MH& 				
		$3 \cdot 10^{11}$& 						
		$62.5$&	 					
		\\
		\hline
		
		\multirow{2}{1.8cm}{\centering Guo \ea, 2021 \cite{guo2021pggan}}&
		\multirow{2}{1.8cm}{\centering GAN}& 
		$\text{RY}_{(17M)}$& 		
		\multirow{2}{*}{$ -$}& 
		$\text{RY}_{(4.3M)}$& 		
		$100M$& 					
		$0.932$& 					
		\\
		&
		&
		$\text{CS}_{(3.6M)}$& 		
		&
		$\text{CS}_{(900K)}$& 		
		$100M$& 					
		$0.256$& 					
		\\
		\hline
		
		\multirow{6}{1.8cm}{\centering Cheng \ea, 2021 \cite{cheng2021improved}}&
		\multirow{6}{1.8cm}{\centering WordPCFG} & 					
		$\text{RY}_{(50\%)}$& 		
		\multirow{6}{*}{D, K, L}& 
		$\text{RY}_{(50\%)}$& 		
		$1B$& 
		$0.726$& 
		\\
		&
		&
		$\text{WH}_{(50\%)}$& 
		&
		$\text{WH}_{(50\%)}$& 
		$1B$& 
		$0.351$& 
		\\
		&
		& 
		$\text{CL}_{(50\%)}$& 
		&
		$\text{CL}_{(50\%)}$& 
		$1B$& 
		$0.518$& 
		\\
		&
		& 
		$\text{CS}_{(50\%)}$& 
		&
		$\text{CS}_{(50\%)}$& 
		$1B$& 
		$0.523$& 
		\\
		&
		& 
		$\text{DO}_{(50\%)}$& 
		&
		$\text{DO}_{(50\%)}$& 
		$1B$& 
		$0.586$& 
		\\
		&
		& 
		$\text{DW}_{(50\%)}$& 
		&
		$\text{DW}_{(50\%)}$& 
		$1B$& 
		$0.632$& 
		\\\hline

		\multirow{2}{1.8cm}{\centering Zhou \ea, 2022 \cite{zhou2022password}}&
		\multirow{2}{1.8cm}{\centering GAN}&
		$\text{RY}_{(4.8M)}$& 
		\multirow{2}{*}{$-$}& 
		$\text{RY}_{(1.2M)}$& 
		$1B$ & 
		$0.248$& 
		\\
		&
		&
		$\text{RY}_{(4.8M)}$& 
		&
		$\text{LI}_{(2.56M)}$ & 
		$1B$ & 
		$0.162$& 
		\\
		\hline
		
		\multicolumn{8}{l}{{\small $^{(\dagger)}$\ti{Other data:} `D'- \ti{dictionary;} `K'- \ti{keyboard;} `U'- \ti{user information;} `L'- \ti{language property;} `R'- \ti{password policy.}}}\\
		\multicolumn{8}{l}{\small $^{(\S)}$\ti{Dataset:} 
				`AN'- \ti{Animoto;}
				`CL'- \ti{Clixsense;}
				`CS'- \ti{CSDN;}
				`DO'- \ti{Dodonew;}
				`DU'- \ti{Duduniu;}
				`DW'- \ti{Duowan;}
				`FB'- \ti{Facebook;}
			}\\
		\multicolumn{8}{l}{{\small
				\tab
				`FN'- \ti{Finnish;}
				`JD'- \ti{JingDong;}
				`LI'- \ti{LinkedIn;}
				`MH'- \ti{MyHeritage;}
				`MS'- \ti{MySpace;}
				`NE'- \ti{Netease;}
				`OT'- \ti{12306;}
				`OY'- \ti{ONYEN;}
				}}\\
		\multicolumn{8}{l}{{\small
				\tab
				`PB'- \ti{PhbBB;}	
				`RY'- \ti{Rockyou;}
				`SW'- \ti{SilentWhisper;}
				`WH'- \ti{000webhost;}
				`YK'- \ti{Youku;}
				`ZK'- \ti{Zooks;}
				`ZO'- \ti{Zomato.}
					}}\\
		\multicolumn{8}{l}{{\small\tab$\text{AB}$ \ti{means entire dataset abbreviated by} \quotes{AB}, \eg, $\text{RY}$ \ti{means entire Rockyou dataset.}
				}}\\
		\multicolumn{8}{l}{{\small \tab$\text{AB}_{(p\%)}$ \ti{means} $p\%$ \ti{of dataset abbreviated by} \quotes{AB}, \eg, $\text{RY}_{(50\%)}$ \ti{means} $50\%$ \ti{of Rockyou dataset.}
		}}\\
		\multicolumn{8}{l}{{\small \tab$\text{AB}_{(N)}$ \ti{means} $N$ \ti{passwords in dataset abbreviated by} \quotes{AB}, \eg, $\text{RY}_{(5M)}$ \ti{means} $5$ \ti{million passwords of Rockyou dataset.}
		}}\\	
	\end{tabular}
	
\end{table}

%

\section{Future Research}\label{sssec_future_work}


With the emergence of machine learning and artificial intelligence techniques, the arts of password guessing have significantly improved and achieved a higher level compared to the beginning state, thanks to prior research efforts.
%
There are several limitations and unsolved works for future research:

\begin{itemize}
	
	\item \ti{(i) Improving Guessing Performance:}
	Overall, the guessing performance of current solutions highly relied on the training dataset, information source, and user configuration.
	Such solutions require expertise with a solid background to operate.
	A fully automatic setting is overwhelmed by a real attacker.
	Moreover, the guessing techniques used by the attackers continuously evolve and get smarter over time.
	Thus, there is always a need to improve the guessing performance, especially for automatic launching. 
	
	\item \ti{(ii) Targeted Guessing and Metering:}
	Most existing studies focused on offline password guessing, and only a few online guessing models were proposed in the literature. 
	However, in practice, offline password guessing only poses security concerns in limited scenarios (\eg, password database is leaked) which mainly comes from the misconfiguration of the security administrator \cite{florencio2014administrator}.
	The online/targeted attack can threaten any public services or personal devices which are protected by weak passwords. 
	Similarly, most current password strength meters employ trawling password guessing to evaluate the strength of an input password.
	It means each password is decided with a fixed strength regardless of who is going to use it.
	%
	%
	However, it is easy to see that, a weak password of a user (\eg, having personal information of this user) could be a strong password for some others.
	Thus, it is inappropriate to apply the same strength values for all users, and we need personalized password strength meter.

%
%
%
%
	
	
	
	\item \ti{(iii) Benchmark Datasets:} 
	Most existing password guessing models were evaluated using the leaked password datasets available on the Internet or private datasets that could not be shared for public research.
	There is no official organization that maintains standard datasets for research purposes.
	This raises several obstacles which restrain the research activities.
	First, the time used for collecting and preprocessing datasets is increased. 
	Second, as these datasets are not maintained by any official organization, the reliability of new research can not be guaranteed.
	However, unlike other fields, it is challenging to collect and maintain such datasets, mainly caused by the extreme sensitivity of passwords.
	Some mechanisms need to be designed to ensure that the passwords and other information contained in the dataset only can be used for ethical research.

	
	
	\item \ti{(iv) Measuring the Security Impacts of Password Strength Meter and Password Advisor:} It has been shown that users are influenced by the password strength meter/password advisor \cite{proctor2002improving,egelman2013does,ur2012does}.
	This raises a question: \quotes{\ti{is it possible for the attacker to infer or narrow down the guessing space by studying the employed PSM/PA model?}}.
	Specifically, under a PSM, the user will not be able to use some passwords that are considered weak.
	Thus, it is straightforward that attackers will not try with such passwords.
	For the password advisor, it would be a security problem if the model generates advice that always suggests user some identifiable or repeated password patterns.
	The attacker could generate high-hitting-chance guesses when trying with the passwords suggested by the advisors.
	Existing studies have exploited password policy to narrow down the guessing space in targeted password attacks \cite{wang2016targeted}.
	However, it is still an open question regarding the use of PSM/PA for password guessing.
	%
		
%
%
%
%
	
	
	\item \ti{(v) Password Strength Meter/Password Advisor Based on the Strategy of Attacker:}
	It is not a new approach that a defensive team try to learn the attacker's strategies and establish appropriate counters. 
	However, in password security, most PA/PSM enhances the password strength by exploiting the weaknesses in the password composing behavior of the user.
	We suggest a new direction to construct PA/PSM that learns the strategies used by the real attackers, and recommends the user use passwords that are beyond the expectation of the attacker.
	We believe that this approach is promising to improve password strength against guessing attacks.
	In addition, it should have a mechanism to protect the behavior of the PA/PSM, so that the attacker could not learn and adapt to his strategy.

%
%
%
%
%
%
%
%
%
%
	
	\item \ti{(vi) Metric to Evaluate the Performance of PSM/PA:} 
	%
	%
	Different PSM/PA approaches have been proposed in the literature and deployed in practice (\eg, heuristics \cite{wheeler2016zxcvbn, burr2018electronic}, PCFG \cite{houshmand2012building}, neural network \cite{melicher2017better,ur2017design}).  
	It has been shown that such PSMs/PAs have a significant impact on the password selection of users \cite{ur2012does,egelman2013does,tan2020practical}.
	%
	%
	However, most existing PSM/PA studies focused on proposing metrics and methods to measure the strength of passwords, it is still an open problem regarding \ti{how to measure the accuracy of a PSM/PA} \cite{golla2018accuracy}.
	The current state-of-the-art PSMs score passwords inconsistently \cite{golla2018accuracy,pereira2020evaluating}.
	Most existing studies evaluated their methods by heuristic rules designed by common sense and experts' knowledge with some representative examples.
	We are lacking a sound metric to determine the accuracy of a PSM.
	%
	%
	Thus, it is also unable to compare different PSMs/PAs and find a suitable model for a specific application.
	It is crucial to establish standard methods and metrics to verify the performance of a PSM/PA.
	This allows overall comparison to determine a suitable PSM/PS for a specific purpose.
	Moreover, it also enables validating the performance of a new proposed PSM.

	
	

%
%
%
	

	
%
%
%
%

\end{itemize}

%
%
%
%
%
%
%
%
%
%
%
%
%
%
%
%
%

\section{Conclusion}\label{sec_conclusion}
With the explosion of modern technology, smart devices (\eg, laptop, smart phone, smart watch) become so popular and common with everyone nowadays.
These devices usually contain a wide range of user information and mostly are connected with sensitive personal online accounts (\eg, Internet banking, online shopping, email).
So, designing mechanisms to ensure that unauthorized users could not access such information/services is more critical than ever.
Despite the emergence of other non-password authentication approaches, password is still a widely used authentication method. 
In order to encourage and support users to select strong passwords, many password security solutions have been proposed in the literature.
Among them, simulating password guessing attacks plays a crucial role in designing such solutions.
To summarize the achievement of existing password guessing methods, we surveyed all password guessing studies published in the literature from 1979 to 2022.
From the survey, we draw several conclusions:
\begin{itemize}
	\item 
	A large number of techniques have been adopted for password guessing ranging from heuristic to machine learning and deep learning.
	They were designed to output candidate passwords ordered by the likelihood of being chosen by the user by modeling the user behavior in password selection. 
	
	\item 
	Constructing a robust password guessing model involves not only designing a password generating algorithm that can closely simulate the attacker's strategies, but also collecting and aggregating a large number of unstructured data (\eg, leaked passwords, user information, native language property, keyboard patterns).
	
	\item 
	As passwords are extremely sensitive, it is unable to construct a password dataset from volunteers. 
	Thus, the real password datasets that have been leaked by some attacks were widely used to evaluate the performance of password guessing models.
	
	\item 	
	Passwords are still at a high risk of guessing attacks.
	To further improve the password strength against guessing attacks, we suggested several open research directions (see Section \ref{sssec_future_work}).

\end{itemize}

%


%


\section{Acknowledgments}

This work was supported in part by the Institute of Information \& Communications Technology Planning \& Evaluation (IITP) grant by the Korean Government, Ministry of Science and ICT (MSIT) under Grant 2020-0-00126.

%
%
%
%
%

\bibliographystyle{ACM-Reference-Format}
\bibliography{reference}


\begin{thebibliography}{138}


\ifx \showCODEN    \undefined \def \showCODEN     #1{\unskip}     \fi
\ifx \showDOI      \undefined \def \showDOI       #1{#1}\fi
\ifx \showISBNx    \undefined \def \showISBNx     #1{\unskip}     \fi
\ifx \showISBNxiii \undefined \def \showISBNxiii  #1{\unskip}     \fi
\ifx \showISSN     \undefined \def \showISSN      #1{\unskip}     \fi
\ifx \showLCCN     \undefined \def \showLCCN      #1{\unskip}     \fi
\ifx \shownote     \undefined \def \shownote      #1{#1}          \fi
\ifx \showarticletitle \undefined \def \showarticletitle #1{#1}   \fi
\ifx \showURL      \undefined \def \showURL       {\relax}        \fi
\providecommand\bibfield[2]{#2}
\providecommand\bibinfo[2]{#2}
\providecommand\natexlab[1]{#1}
\providecommand\showeprint[2][]{arXiv:#2}

\bibitem[\protect\citeauthoryear{Alsaleh, Mannan, and Van~Oorschot}{Alsaleh
  et~al\mbox{.}}{2011}]%
        {alsaleh2011revisiting}
\bibfield{author}{\bibinfo{person}{Mansour Alsaleh}, \bibinfo{person}{Mohammad
  Mannan}, {and} \bibinfo{person}{Paul~C Van~Oorschot}.}
  \bibinfo{year}{2011}\natexlab{}.
\newblock \showarticletitle{Revisiting defenses against large-scale online
  password guessing attacks}.
\newblock \bibinfo{journal}{\emph{IEEE Transactions on dependable and secure
  computing}} \bibinfo{volume}{9}, \bibinfo{number}{1} (\bibinfo{year}{2011}),
  \bibinfo{pages}{128--141}.
\newblock


\bibitem[\protect\citeauthoryear{Aumasson}{Aumasson}{2017}]%
        {aumasson2017serious}
\bibfield{author}{\bibinfo{person}{Jean-Philippe Aumasson}.}
  \bibinfo{year}{2017}\natexlab{}.
\newblock \bibinfo{booktitle}{\emph{Serious cryptography: a practical
  introduction to modern encryption}}.
\newblock \bibinfo{publisher}{No Starch Press}.
\newblock


\bibitem[\protect\citeauthoryear{Bergadano, Crispo, and Ruffo}{Bergadano
  et~al\mbox{.}}{1998}]%
        {bergadano1998high}
\bibfield{author}{\bibinfo{person}{Francesco Bergadano}, \bibinfo{person}{Bruno
  Crispo}, {and} \bibinfo{person}{Giancarlo Ruffo}.}
  \bibinfo{year}{1998}\natexlab{}.
\newblock \showarticletitle{High dictionary compression for proactive password
  checking}.
\newblock \bibinfo{journal}{\emph{ACM Transactions on Information and System
  Security (TISSEC)}} \bibinfo{volume}{1}, \bibinfo{number}{1}
  (\bibinfo{year}{1998}), \bibinfo{pages}{3--25}.
\newblock


\bibitem[\protect\citeauthoryear{Biesner, Cvejoski, Georgiev, Sifa, and
  Krupicka}{Biesner et~al\mbox{.}}{2021}]%
        {biesner2021advances}
\bibfield{author}{\bibinfo{person}{David Biesner}, \bibinfo{person}{Kostadin
  Cvejoski}, \bibinfo{person}{Bogdan Georgiev}, \bibinfo{person}{Rafet Sifa},
  {and} \bibinfo{person}{Erik Krupicka}.} \bibinfo{year}{2021}\natexlab{}.
\newblock \showarticletitle{Advances in Password Recovery Using Generative Deep
  Learning Techniques}. In \bibinfo{booktitle}{\emph{International Conference
  on Artificial Neural Networks}}. Springer, \bibinfo{pages}{15--27}.
\newblock


\bibitem[\protect\citeauthoryear{Bishop}{Bishop}{1990}]%
        {bishop1990proactive}
\bibfield{author}{\bibinfo{person}{Matt Bishop}.}
  \bibinfo{year}{1990}\natexlab{}.
\newblock \bibinfo{booktitle}{\emph{A proactive password checker}}.
\newblock \bibinfo{publisher}{National Aeronautics and Space Administration}.
\newblock


\bibitem[\protect\citeauthoryear{Bonneau, Herley, Van~Oorschot, and
  Stajano}{Bonneau et~al\mbox{.}}{2015}]%
        {bonneau2015passwords}
\bibfield{author}{\bibinfo{person}{Joseph Bonneau}, \bibinfo{person}{Cormac
  Herley}, \bibinfo{person}{Paul~C Van~Oorschot}, {and} \bibinfo{person}{Frank
  Stajano}.} \bibinfo{year}{2015}\natexlab{}.
\newblock \showarticletitle{Passwords and the evolution of imperfect
  authentication}.
\newblock \bibinfo{journal}{\emph{Commun. ACM}} \bibinfo{volume}{58},
  \bibinfo{number}{7} (\bibinfo{year}{2015}), \bibinfo{pages}{78--87}.
\newblock


\bibitem[\protect\citeauthoryear{Bonneau, Preibusch, and Anderson}{Bonneau
  et~al\mbox{.}}{2012}]%
        {bonneau2012birthday}
\bibfield{author}{\bibinfo{person}{Joseph Bonneau}, \bibinfo{person}{S{\"o}ren
  Preibusch}, {and} \bibinfo{person}{Ross Anderson}.}
  \bibinfo{year}{2012}\natexlab{}.
\newblock \showarticletitle{A birthday present every eleven wallets? the
  security of customer-chosen banking pins}. In
  \bibinfo{booktitle}{\emph{International Conference on Financial Cryptography
  and Data Security}}. Springer, \bibinfo{pages}{25--40}.
\newblock


\bibitem[\protect\citeauthoryear{Bonneau and Shutova}{Bonneau and
  Shutova}{2012}]%
        {bonneau2012linguistic}
\bibfield{author}{\bibinfo{person}{Joseph Bonneau} {and}
  \bibinfo{person}{Ekaterina Shutova}.} \bibinfo{year}{2012}\natexlab{}.
\newblock \showarticletitle{Linguistic properties of multi-word passphrases}.
  In \bibinfo{booktitle}{\emph{International Conference on Financial
  Cryptography and Data Security}}. Springer, \bibinfo{pages}{1--12}.
\newblock


\bibitem[\protect\citeauthoryear{Boztas}{Boztas}{1999}]%
        {boztas1999entropies}
\bibfield{author}{\bibinfo{person}{S Boztas}.} \bibinfo{year}{1999}\natexlab{}.
\newblock \showarticletitle{Entropies, guessing, and cryptography}.
\newblock \bibinfo{journal}{\emph{Department of Mathematics, Royal Melbourne
  Institute of Technology, Tech. Rep}}  \bibinfo{volume}{6}
  (\bibinfo{year}{1999}), \bibinfo{pages}{2--3}.
\newblock


\bibitem[\protect\citeauthoryear{Bryant, Campbell, et~al\mbox{.}}{Bryant
  et~al\mbox{.}}{2006}]%
        {bryant2006user}
\bibfield{author}{\bibinfo{person}{Kay Bryant}, \bibinfo{person}{John
  Campbell}, {et~al\mbox{.}}} \bibinfo{year}{2006}\natexlab{}.
\newblock \showarticletitle{User behaviours associated with password security
  and management}.
\newblock \bibinfo{journal}{\emph{Australasian Journal of Information Systems}}
  \bibinfo{volume}{14}, \bibinfo{number}{1} (\bibinfo{year}{2006}).
\newblock


\bibitem[\protect\citeauthoryear{Burr, Dodson, and Polk}{Burr
  et~al\mbox{.}}{2018}]%
        {burr2018electronic}
\bibfield{author}{\bibinfo{person}{William~E Burr}, \bibinfo{person}{Donna~F
  Dodson}, {and} \bibinfo{person}{W~Timothy Polk}.}
  \bibinfo{year}{2018}\natexlab{}.
\newblock \bibinfo{title}{Electronic Authentication Guideline: NIST SP 800-63
  Ver. 1.0 (2004) to 800-63-2 (2013)}.
\newblock
\newblock


\bibitem[\protect\citeauthoryear{Campbell, Ma, and Kleeman}{Campbell
  et~al\mbox{.}}{2011}]%
        {campbell2011impact}
\bibfield{author}{\bibinfo{person}{John Campbell}, \bibinfo{person}{Wanli Ma},
  {and} \bibinfo{person}{Dale Kleeman}.} \bibinfo{year}{2011}\natexlab{}.
\newblock \showarticletitle{Impact of restrictive composition policy on user
  password choices}.
\newblock \bibinfo{journal}{\emph{Behaviour \& Information Technology}}
  \bibinfo{volume}{30}, \bibinfo{number}{3} (\bibinfo{year}{2011}),
  \bibinfo{pages}{379--388}.
\newblock


\bibitem[\protect\citeauthoryear{Carnavalet and Mannan}{Carnavalet and
  Mannan}{2015}]%
        {carnavalet2015large}
\bibfield{author}{\bibinfo{person}{Xavier De Carn{\'e}~De Carnavalet} {and}
  \bibinfo{person}{Mohammad Mannan}.} \bibinfo{year}{2015}\natexlab{}.
\newblock \showarticletitle{A large-scale evaluation of high-impact password
  strength meters}.
\newblock \bibinfo{journal}{\emph{ACM Transactions on Information and System
  Security (TISSEC)}} \bibinfo{volume}{18}, \bibinfo{number}{1}
  (\bibinfo{year}{2015}), \bibinfo{pages}{1--32}.
\newblock


\bibitem[\protect\citeauthoryear{Castelluccia, D{\"u}rmuth, and
  Perito}{Castelluccia et~al\mbox{.}}{2012}]%
        {castelluccia2012adaptive}
\bibfield{author}{\bibinfo{person}{Claude Castelluccia},
  \bibinfo{person}{Markus D{\"u}rmuth}, {and} \bibinfo{person}{Daniele
  Perito}.} \bibinfo{year}{2012}\natexlab{}.
\newblock \showarticletitle{Adaptive password-strength meters from markov
  models.}. In \bibinfo{booktitle}{\emph{NDSS}}.
\newblock


\bibitem[\protect\citeauthoryear{Cheng, Li, Wang, and Liang}{Cheng
  et~al\mbox{.}}{2021}]%
        {cheng2021improved}
\bibfield{author}{\bibinfo{person}{Haibo Cheng}, \bibinfo{person}{Wenting Li},
  \bibinfo{person}{Ping Wang}, {and} \bibinfo{person}{Kaitai Liang}.}
  \bibinfo{year}{2021}\natexlab{}.
\newblock \showarticletitle{Improved Probabilistic Context-Free Grammars for
  Passwords Using Word Extraction}. In \bibinfo{booktitle}{\emph{ICASSP
  2021-2021 IEEE International Conference on Acoustics, Speech and Signal
  Processing (ICASSP)}}. IEEE, \bibinfo{pages}{2690--2694}.
\newblock


\bibitem[\protect\citeauthoryear{Chou, Lee, Hsueh, and Lai}{Chou
  et~al\mbox{.}}{2012}]%
        {chou2012password}
\bibfield{author}{\bibinfo{person}{Hsien-Cheng Chou},
  \bibinfo{person}{Hung-Chang Lee}, \bibinfo{person}{Chih-Wen Hsueh}, {and}
  \bibinfo{person}{Fei-Pei Lai}.} \bibinfo{year}{2012}\natexlab{}.
\newblock \showarticletitle{Password cracking based on special keyboard
  patterns}.
\newblock \bibinfo{journal}{\emph{International Journal of Innovative
  Computing, Information and Control}} \bibinfo{volume}{8}, \bibinfo{number}{1}
  (\bibinfo{year}{2012}), \bibinfo{pages}{387--402}.
\newblock


\bibitem[\protect\citeauthoryear{Chou, Lee, Yu, Lai, Huang, Hsueh,
  et~al\mbox{.}}{Chou et~al\mbox{.}}{2013}]%
        {chou2013password}
\bibfield{author}{\bibinfo{person}{Hsien-Cheng Chou},
  \bibinfo{person}{Hung-Chang Lee}, \bibinfo{person}{Hwan-Jeu Yu},
  \bibinfo{person}{Fei-Pei Lai}, \bibinfo{person}{Kuo-Hsuan Huang},
  \bibinfo{person}{Chih-Wen Hsueh}, {et~al\mbox{.}}}
  \bibinfo{year}{2013}\natexlab{}.
\newblock \showarticletitle{Password cracking based on learned patterns from
  disclosed passwords}.
\newblock \bibinfo{journal}{\emph{IJICIC}} \bibinfo{volume}{9},
  \bibinfo{number}{2} (\bibinfo{year}{2013}), \bibinfo{pages}{821--839}.
\newblock


\bibitem[\protect\citeauthoryear{Clair, Johansen, Enck, Pirretti, Traynor,
  McDaniel, and Jaeger}{Clair et~al\mbox{.}}{2006}]%
        {clair2006password}
\bibfield{author}{\bibinfo{person}{Luke~St Clair}, \bibinfo{person}{Lisa
  Johansen}, \bibinfo{person}{William Enck}, \bibinfo{person}{Matthew
  Pirretti}, \bibinfo{person}{Patrick Traynor}, \bibinfo{person}{Patrick
  McDaniel}, {and} \bibinfo{person}{Trent Jaeger}.}
  \bibinfo{year}{2006}\natexlab{}.
\newblock \showarticletitle{Password exhaustion: Predicting the end of password
  usefulness}. In \bibinfo{booktitle}{\emph{International Conference on
  Information Systems Security}}. Springer, \bibinfo{pages}{37--55}.
\newblock


\bibitem[\protect\citeauthoryear{Das, Bonneau, Caesar, Borisov, and Wang}{Das
  et~al\mbox{.}}{2014}]%
        {das2014tangled}
\bibfield{author}{\bibinfo{person}{Anupam Das}, \bibinfo{person}{Joseph
  Bonneau}, \bibinfo{person}{Matthew Caesar}, \bibinfo{person}{Nikita Borisov},
  {and} \bibinfo{person}{XiaoFeng Wang}.} \bibinfo{year}{2014}\natexlab{}.
\newblock \showarticletitle{The tangled web of password reuse.}. In
  \bibinfo{booktitle}{\emph{NDSS}}, Vol.~\bibinfo{volume}{14}.
  \bibinfo{pages}{23--26}.
\newblock


\bibitem[\protect\citeauthoryear{de~Carn{\'e}~de Carnavalet and
  Mannan}{de~Carn{\'e}~de Carnavalet and Mannan}{2014}]%
        {de2014very}
\bibfield{author}{\bibinfo{person}{Xavier de~Carn{\'e}~de Carnavalet} {and}
  \bibinfo{person}{Mohammad Mannan}.} \bibinfo{year}{2014}\natexlab{}.
\newblock \showarticletitle{From very weak to very strong: Analyzing
  password-strength meters}. In \bibinfo{booktitle}{\emph{Network and
  Distributed System Security Symposium (NDSS 2014)}}. Internet Society.
\newblock


\bibitem[\protect\citeauthoryear{Dell'Amico and Filippone}{Dell'Amico and
  Filippone}{2015}]%
        {dell2015monte}
\bibfield{author}{\bibinfo{person}{Matteo Dell'Amico} {and}
  \bibinfo{person}{Maurizio Filippone}.} \bibinfo{year}{2015}\natexlab{}.
\newblock \showarticletitle{Monte Carlo strength evaluation: Fast and reliable
  password checking}. In \bibinfo{booktitle}{\emph{Proceedings of the 22nd ACM
  SIGSAC conference on computer and communications security}}.
  \bibinfo{pages}{158--169}.
\newblock


\bibitem[\protect\citeauthoryear{Devlin, Chang, Lee, and Toutanova}{Devlin
  et~al\mbox{.}}{2018}]%
        {devlin2018bert}
\bibfield{author}{\bibinfo{person}{Jacob Devlin}, \bibinfo{person}{Ming-Wei
  Chang}, \bibinfo{person}{Kenton Lee}, {and} \bibinfo{person}{Kristina
  Toutanova}.} \bibinfo{year}{2018}\natexlab{}.
\newblock \showarticletitle{Bert: Pre-training of deep bidirectional
  transformers for language understanding}.
\newblock \bibinfo{journal}{\emph{arXiv preprint arXiv:1810.04805}}
  (\bibinfo{year}{2018}).
\newblock


\bibitem[\protect\citeauthoryear{D{\"u}rmuth, Angelstorf, Castelluccia, Perito,
  and Chaabane}{D{\"u}rmuth et~al\mbox{.}}{2015}]%
        {durmuth2015omen}
\bibfield{author}{\bibinfo{person}{Markus D{\"u}rmuth}, \bibinfo{person}{Fabian
  Angelstorf}, \bibinfo{person}{Claude Castelluccia}, \bibinfo{person}{Daniele
  Perito}, {and} \bibinfo{person}{Abdelberi Chaabane}.}
  \bibinfo{year}{2015}\natexlab{}.
\newblock \showarticletitle{OMEN: Faster password guessing using an ordered
  markov enumerator}. In \bibinfo{booktitle}{\emph{International symposium on
  engineering secure software and systems}}. Springer,
  \bibinfo{pages}{119--132}.
\newblock


\bibitem[\protect\citeauthoryear{Egelman, Sotirakopoulos, Muslukhov, Beznosov,
  and Herley}{Egelman et~al\mbox{.}}{2013}]%
        {egelman2013does}
\bibfield{author}{\bibinfo{person}{Serge Egelman}, \bibinfo{person}{Andreas
  Sotirakopoulos}, \bibinfo{person}{Ildar Muslukhov},
  \bibinfo{person}{Konstantin Beznosov}, {and} \bibinfo{person}{Cormac
  Herley}.} \bibinfo{year}{2013}\natexlab{}.
\newblock \showarticletitle{Does my password go up to eleven? The impact of
  password meters on password selection}. In
  \bibinfo{booktitle}{\emph{Proceedings of the SIGCHI Conference on Human
  Factors in Computing Systems}}. \bibinfo{pages}{2379--2388}.
\newblock


\bibitem[\protect\citeauthoryear{Fang, Liu, Jing, and Zuo}{Fang
  et~al\mbox{.}}{2018}]%
        {fang2018password}
\bibfield{author}{\bibinfo{person}{Yong Fang}, \bibinfo{person}{Kai Liu},
  \bibinfo{person}{Fan Jing}, {and} \bibinfo{person}{Zheng Zuo}.}
  \bibinfo{year}{2018}\natexlab{}.
\newblock \showarticletitle{Password guessing based on semantic analysis and
  neural networks}. In \bibinfo{booktitle}{\emph{Chinese Conference on Trusted
  Computing and Information Security}}. Springer, \bibinfo{pages}{84--98}.
\newblock


\bibitem[\protect\citeauthoryear{Fine, Singer, and Tishby}{Fine
  et~al\mbox{.}}{1998}]%
        {fine1998hierarchical}
\bibfield{author}{\bibinfo{person}{Shai Fine}, \bibinfo{person}{Yoram Singer},
  {and} \bibinfo{person}{Naftali Tishby}.} \bibinfo{year}{1998}\natexlab{}.
\newblock \showarticletitle{The hierarchical hidden Markov model: Analysis and
  applications}.
\newblock \bibinfo{journal}{\emph{Machine learning}} \bibinfo{volume}{32},
  \bibinfo{number}{1} (\bibinfo{year}{1998}), \bibinfo{pages}{41--62}.
\newblock


\bibitem[\protect\citeauthoryear{Florencio and Herley}{Florencio and
  Herley}{2007}]%
        {florencio2007large}
\bibfield{author}{\bibinfo{person}{Dinei Florencio} {and}
  \bibinfo{person}{Cormac Herley}.} \bibinfo{year}{2007}\natexlab{}.
\newblock \showarticletitle{A large-scale study of web password habits}. In
  \bibinfo{booktitle}{\emph{Proceedings of the 16th international conference on
  World Wide Web}}. \bibinfo{pages}{657--666}.
\newblock


\bibitem[\protect\citeauthoryear{Flor{\^e}ncio, Herley, and
  Van~Oorschot}{Flor{\^e}ncio et~al\mbox{.}}{2014}]%
        {florencio2014administrator}
\bibfield{author}{\bibinfo{person}{Dinei Flor{\^e}ncio},
  \bibinfo{person}{Cormac Herley}, {and} \bibinfo{person}{Paul~C
  Van~Oorschot}.} \bibinfo{year}{2014}\natexlab{}.
\newblock \showarticletitle{An $\{$Administrator’s$\}$ Guide to Internet
  Password Research}. In \bibinfo{booktitle}{\emph{28th large installation
  system administration conference (LISA14)}}. \bibinfo{pages}{44--61}.
\newblock


\bibitem[\protect\citeauthoryear{Gaw and Felten}{Gaw and Felten}{2006}]%
        {gaw2006password}
\bibfield{author}{\bibinfo{person}{Shirley Gaw} {and} \bibinfo{person}{Edward~W
  Felten}.} \bibinfo{year}{2006}\natexlab{}.
\newblock \showarticletitle{Password management strategies for online
  accounts}. In \bibinfo{booktitle}{\emph{Proceedings of the second symposium
  on Usable privacy and security}}. \bibinfo{pages}{44--55}.
\newblock


\bibitem[\protect\citeauthoryear{Golla, Beuscher, and D{\"u}rmuth}{Golla
  et~al\mbox{.}}{2016}]%
        {golla2016security}
\bibfield{author}{\bibinfo{person}{Maximilian Golla}, \bibinfo{person}{Benedict
  Beuscher}, {and} \bibinfo{person}{Markus D{\"u}rmuth}.}
  \bibinfo{year}{2016}\natexlab{}.
\newblock \showarticletitle{On the security of cracking-resistant password
  vaults}. In \bibinfo{booktitle}{\emph{Proceedings of the 2016 ACM SIGSAC
  conference on computer and communications security}}.
  \bibinfo{pages}{1230--1241}.
\newblock


\bibitem[\protect\citeauthoryear{Golla and D{\"u}rmuth}{Golla and
  D{\"u}rmuth}{2018}]%
        {golla2018accuracy}
\bibfield{author}{\bibinfo{person}{Maximilian Golla} {and}
  \bibinfo{person}{Markus D{\"u}rmuth}.} \bibinfo{year}{2018}\natexlab{}.
\newblock \showarticletitle{On the accuracy of password strength meters}. In
  \bibinfo{booktitle}{\emph{Proceedings of the 2018 ACM SIGSAC Conference on
  Computer and Communications Security}}. \bibinfo{pages}{1567--1582}.
\newblock


\bibitem[\protect\citeauthoryear{Goodfellow, Pouget-Abadie, Mirza, Xu,
  Warde-Farley, Ozair, Courville, and Bengio}{Goodfellow et~al\mbox{.}}{2014}]%
        {goodfellow2014generative}
\bibfield{author}{\bibinfo{person}{Ian Goodfellow}, \bibinfo{person}{Jean
  Pouget-Abadie}, \bibinfo{person}{Mehdi Mirza}, \bibinfo{person}{Bing Xu},
  \bibinfo{person}{David Warde-Farley}, \bibinfo{person}{Sherjil Ozair},
  \bibinfo{person}{Aaron Courville}, {and} \bibinfo{person}{Yoshua Bengio}.}
  \bibinfo{year}{2014}\natexlab{}.
\newblock \showarticletitle{Generative adversarial nets}.
\newblock \bibinfo{journal}{\emph{Advances in neural information processing
  systems}}  \bibinfo{volume}{27} (\bibinfo{year}{2014}).
\newblock


\bibitem[\protect\citeauthoryear{Gulrajani, Ahmed, Arjovsky, Dumoulin, and
  Courville}{Gulrajani et~al\mbox{.}}{2017}]%
        {gulrajani2017improved}
\bibfield{author}{\bibinfo{person}{Ishaan Gulrajani}, \bibinfo{person}{Faruk
  Ahmed}, \bibinfo{person}{Martin Arjovsky}, \bibinfo{person}{Vincent
  Dumoulin}, {and} \bibinfo{person}{Aaron~C Courville}.}
  \bibinfo{year}{2017}\natexlab{}.
\newblock \showarticletitle{Improved training of wasserstein gans}.
\newblock \bibinfo{journal}{\emph{Advances in neural information processing
  systems}}  \bibinfo{volume}{30} (\bibinfo{year}{2017}).
\newblock


\bibitem[\protect\citeauthoryear{Guo, Liu, Tan, Jin, and Lu}{Guo
  et~al\mbox{.}}{2021a}]%
        {guo2021pggan}
\bibfield{author}{\bibinfo{person}{Xiaozhou Guo}, \bibinfo{person}{Yi Liu},
  \bibinfo{person}{Kaijun Tan}, \bibinfo{person}{Min Jin}, {and}
  \bibinfo{person}{Huaxiang Lu}.} \bibinfo{year}{2021}\natexlab{a}.
\newblock \showarticletitle{PGGAN: Improve Password Cover Rate Using the
  Controller}. In \bibinfo{booktitle}{\emph{Journal of Physics: Conference
  Series}}, Vol.~\bibinfo{volume}{1856}. IOP Publishing,
  \bibinfo{pages}{012012}.
\newblock


\bibitem[\protect\citeauthoryear{Guo, Liu, Tan, Mao, Jin, and Lu}{Guo
  et~al\mbox{.}}{2021b}]%
        {guo2021dynamic}
\bibfield{author}{\bibinfo{person}{Xiaozhou Guo}, \bibinfo{person}{Yi Liu},
  \bibinfo{person}{Kaijun Tan}, \bibinfo{person}{Wenyu Mao},
  \bibinfo{person}{Min Jin}, {and} \bibinfo{person}{Huaxiang Lu}.}
  \bibinfo{year}{2021}\natexlab{b}.
\newblock \showarticletitle{Dynamic Markov Model: Password Guessing Using
  Probability Adjustment Method}.
\newblock \bibinfo{journal}{\emph{Applied Sciences}} \bibinfo{volume}{11},
  \bibinfo{number}{10} (\bibinfo{year}{2021}), \bibinfo{pages}{4607}.
\newblock


\bibitem[\protect\citeauthoryear{Habib, Naeini, Devlin, Oates, Swoopes, Bauer,
  Christin, and Cranor}{Habib et~al\mbox{.}}{2018}]%
        {habib2018user}
\bibfield{author}{\bibinfo{person}{Hana Habib}, \bibinfo{person}{Pardis~Emami
  Naeini}, \bibinfo{person}{Summer Devlin}, \bibinfo{person}{Maggie Oates},
  \bibinfo{person}{Chelse Swoopes}, \bibinfo{person}{Lujo Bauer},
  \bibinfo{person}{Nicolas Christin}, {and} \bibinfo{person}{Lorrie~Faith
  Cranor}.} \bibinfo{year}{2018}\natexlab{}.
\newblock \showarticletitle{User behaviors and attitudes under password
  expiration policies}. In \bibinfo{booktitle}{\emph{Fourteenth Symposium on
  Usable Privacy and Security (SOUPS 2018)}}. \bibinfo{pages}{13--30}.
\newblock


\bibitem[\protect\citeauthoryear{Han, Wong, and Chao}{Han
  et~al\mbox{.}}{2014}]%
        {han2014password}
\bibfield{author}{\bibinfo{person}{Aaron L-F Han}, \bibinfo{person}{Derek~F
  Wong}, {and} \bibinfo{person}{Lidia~S Chao}.}
  \bibinfo{year}{2014}\natexlab{}.
\newblock \showarticletitle{Password cracking and countermeasures in computer
  security: A survey}.
\newblock \bibinfo{journal}{\emph{arXiv preprint arXiv:1411.7803}}
  (\bibinfo{year}{2014}).
\newblock


\bibitem[\protect\citeauthoryear{Han, Li, Yuan, and Xu}{Han
  et~al\mbox{.}}{2015}]%
        {han2015regional}
\bibfield{author}{\bibinfo{person}{Weili Han}, \bibinfo{person}{Zhigong Li},
  \bibinfo{person}{Lang Yuan}, {and} \bibinfo{person}{Wenyuan Xu}.}
  \bibinfo{year}{2015}\natexlab{}.
\newblock \showarticletitle{Regional patterns and vulnerability analysis of
  chinese web passwords}.
\newblock \bibinfo{journal}{\emph{IEEE Transactions on Information Forensics
  and Security}} \bibinfo{volume}{11}, \bibinfo{number}{2}
  (\bibinfo{year}{2015}), \bibinfo{pages}{258--272}.
\newblock


\bibitem[\protect\citeauthoryear{Haque, Wright, and Scielzo}{Haque
  et~al\mbox{.}}{2013}]%
        {haque2013study}
\bibfield{author}{\bibinfo{person}{SM~Taiabul Haque}, \bibinfo{person}{Matthew
  Wright}, {and} \bibinfo{person}{Shannon Scielzo}.}
  \bibinfo{year}{2013}\natexlab{}.
\newblock \showarticletitle{A study of user password strategy for multiple
  accounts}. In \bibinfo{booktitle}{\emph{Proceedings of the third ACM
  conference on Data and application security and privacy}}.
  \bibinfo{pages}{173--176}.
\newblock


\bibitem[\protect\citeauthoryear{He, Zhang, Ren, and Sun}{He
  et~al\mbox{.}}{2016}]%
        {he2016deep}
\bibfield{author}{\bibinfo{person}{Kaiming He}, \bibinfo{person}{Xiangyu
  Zhang}, \bibinfo{person}{Shaoqing Ren}, {and} \bibinfo{person}{Jian Sun}.}
  \bibinfo{year}{2016}\natexlab{}.
\newblock \showarticletitle{Deep residual learning for image recognition}. In
  \bibinfo{booktitle}{\emph{Proceedings of the IEEE conference on computer
  vision and pattern recognition}}. \bibinfo{pages}{770--778}.
\newblock


\bibitem[\protect\citeauthoryear{He and Zhu}{He and Zhu}{2001}]%
        {he2001bootstrap}
\bibfield{author}{\bibinfo{person}{Shan He} {and} \bibinfo{person}{Jie Zhu}.}
  \bibinfo{year}{2001}\natexlab{}.
\newblock \showarticletitle{Bootstrap method for Chinese new words extraction}.
  In \bibinfo{booktitle}{\emph{2001 IEEE International Conference on Acoustics,
  Speech, and Signal Processing. Proceedings (Cat. No. 01CH37221)}},
  Vol.~\bibinfo{volume}{1}. IEEE, \bibinfo{pages}{581--584}.
\newblock


\bibitem[\protect\citeauthoryear{Helkala}{Helkala}{2011}]%
        {helkala2011password}
\bibfield{author}{\bibinfo{person}{Kirsi Helkala}.}
  \bibinfo{year}{2011}\natexlab{}.
\newblock \showarticletitle{Password education based on guidelines tailored to
  different password categories}.
\newblock  (\bibinfo{year}{2011}).
\newblock


\bibitem[\protect\citeauthoryear{Herley and Van~Oorschot}{Herley and
  Van~Oorschot}{2011}]%
        {herley2011research}
\bibfield{author}{\bibinfo{person}{Cormac Herley} {and} \bibinfo{person}{Paul
  Van~Oorschot}.} \bibinfo{year}{2011}\natexlab{}.
\newblock \showarticletitle{A research agenda acknowledging the persistence of
  passwords}.
\newblock \bibinfo{journal}{\emph{IEEE Security \& privacy}}
  \bibinfo{volume}{10}, \bibinfo{number}{1} (\bibinfo{year}{2011}),
  \bibinfo{pages}{28--36}.
\newblock


\bibitem[\protect\citeauthoryear{Hinton, Vinyals, Dean, et~al\mbox{.}}{Hinton
  et~al\mbox{.}}{2015}]%
        {hinton2015distilling}
\bibfield{author}{\bibinfo{person}{Geoffrey Hinton}, \bibinfo{person}{Oriol
  Vinyals}, \bibinfo{person}{Jeff Dean}, {et~al\mbox{.}}}
  \bibinfo{year}{2015}\natexlab{}.
\newblock \showarticletitle{Distilling the knowledge in a neural network}.
\newblock \bibinfo{journal}{\emph{arXiv preprint arXiv:1503.02531}}
  \bibinfo{volume}{2}, \bibinfo{number}{7} (\bibinfo{year}{2015}).
\newblock


\bibitem[\protect\citeauthoryear{Hitaj, Gasti, Ateniese, and Perez-Cruz}{Hitaj
  et~al\mbox{.}}{2019}]%
        {hitaj2019passgan}
\bibfield{author}{\bibinfo{person}{Briland Hitaj}, \bibinfo{person}{Paolo
  Gasti}, \bibinfo{person}{Giuseppe Ateniese}, {and} \bibinfo{person}{Fernando
  Perez-Cruz}.} \bibinfo{year}{2019}\natexlab{}.
\newblock \showarticletitle{Passgan: A deep learning approach for password
  guessing}. In \bibinfo{booktitle}{\emph{International Conference on Applied
  Cryptography and Network Security}}. Springer, \bibinfo{pages}{217--237}.
\newblock


\bibitem[\protect\citeauthoryear{Hochreiter}{Hochreiter}{1998}]%
        {hochreiter1998vanishing}
\bibfield{author}{\bibinfo{person}{Sepp Hochreiter}.}
  \bibinfo{year}{1998}\natexlab{}.
\newblock \showarticletitle{The vanishing gradient problem during learning
  recurrent neural nets and problem solutions}.
\newblock \bibinfo{journal}{\emph{International Journal of Uncertainty,
  Fuzziness and Knowledge-Based Systems}} \bibinfo{volume}{6},
  \bibinfo{number}{02} (\bibinfo{year}{1998}), \bibinfo{pages}{107--116}.
\newblock


\bibitem[\protect\citeauthoryear{Hochreiter and Schmidhuber}{Hochreiter and
  Schmidhuber}{1997}]%
        {hochreiter1997long}
\bibfield{author}{\bibinfo{person}{Sepp Hochreiter} {and}
  \bibinfo{person}{J{\"u}rgen Schmidhuber}.} \bibinfo{year}{1997}\natexlab{}.
\newblock \showarticletitle{Long short-term memory}.
\newblock \bibinfo{journal}{\emph{Neural computation}} \bibinfo{volume}{9},
  \bibinfo{number}{8} (\bibinfo{year}{1997}), \bibinfo{pages}{1735--1780}.
\newblock


\bibitem[\protect\citeauthoryear{Houshmand and Aggarwal}{Houshmand and
  Aggarwal}{2012}]%
        {houshmand2012building}
\bibfield{author}{\bibinfo{person}{Shiva Houshmand} {and}
  \bibinfo{person}{Sudhir Aggarwal}.} \bibinfo{year}{2012}\natexlab{}.
\newblock \showarticletitle{Building better passwords using probabilistic
  techniques}. In \bibinfo{booktitle}{\emph{Proceedings of the 28th Annual
  Computer Security Applications Conference}}. \bibinfo{pages}{109--118}.
\newblock


\bibitem[\protect\citeauthoryear{Houshmand, Aggarwal, and Flood}{Houshmand
  et~al\mbox{.}}{2015}]%
        {houshmand2015next}
\bibfield{author}{\bibinfo{person}{Shiva Houshmand}, \bibinfo{person}{Sudhir
  Aggarwal}, {and} \bibinfo{person}{Randy Flood}.}
  \bibinfo{year}{2015}\natexlab{}.
\newblock \showarticletitle{Next gen PCFG password cracking}.
\newblock \bibinfo{journal}{\emph{IEEE Transactions on Information Forensics
  and Security}} \bibinfo{volume}{10}, \bibinfo{number}{8}
  (\bibinfo{year}{2015}), \bibinfo{pages}{1776--1791}.
\newblock


\bibitem[\protect\citeauthoryear{Hu}{Hu}{2017}]%
        {hu2017password}
\bibfield{author}{\bibinfo{person}{Gongzhu Hu}.}
  \bibinfo{year}{2017}\natexlab{}.
\newblock \showarticletitle{On password strength: a survey and analysis}. In
  \bibinfo{booktitle}{\emph{International Conference on Software Engineering,
  Artificial Intelligence, Networking and Parallel/Distributed Computing}}.
  Springer, \bibinfo{pages}{165--186}.
\newblock


\bibitem[\protect\citeauthoryear{Huh, Oh, Kim, Beznosov, Mohan, and
  Rajagopalan}{Huh et~al\mbox{.}}{2015}]%
        {huh2015surpass}
\bibfield{author}{\bibinfo{person}{Jun~Ho Huh}, \bibinfo{person}{Seongyeol Oh},
  \bibinfo{person}{Hyoungshick Kim}, \bibinfo{person}{Konstantin Beznosov},
  \bibinfo{person}{Apurva Mohan}, {and} \bibinfo{person}{S~Raj Rajagopalan}.}
  \bibinfo{year}{2015}\natexlab{}.
\newblock \showarticletitle{Surpass: System-initiated user-replaceable
  passwords}. In \bibinfo{booktitle}{\emph{Proceedings of the 22nd ACM SIGSAC
  Conference on Computer and Communications Security}}.
  \bibinfo{pages}{170--181}.
\newblock


\bibitem[\protect\citeauthoryear{Inglesant and Sasse}{Inglesant and
  Sasse}{2010}]%
        {inglesant2010true}
\bibfield{author}{\bibinfo{person}{Philip~G Inglesant} {and}
  \bibinfo{person}{M~Angela Sasse}.} \bibinfo{year}{2010}\natexlab{}.
\newblock \showarticletitle{The true cost of unusable password policies:
  password use in the wild}. In \bibinfo{booktitle}{\emph{Proceedings of the
  sigchi conference on human factors in computing systems}}.
  \bibinfo{pages}{383--392}.
\newblock


\bibitem[\protect\citeauthoryear{Jang, Gu, and Poole}{Jang
  et~al\mbox{.}}{2016}]%
        {jang2016categorical}
\bibfield{author}{\bibinfo{person}{Eric Jang}, \bibinfo{person}{Shixiang Gu},
  {and} \bibinfo{person}{Ben Poole}.} \bibinfo{year}{2016}\natexlab{}.
\newblock \showarticletitle{Categorical reparameterization with
  gumbel-softmax}.
\newblock \bibinfo{journal}{\emph{arXiv preprint arXiv:1611.01144}}
  (\bibinfo{year}{2016}).
\newblock


\bibitem[\protect\citeauthoryear{Ji, Yang, Hu, Han, Li, and Beyah}{Ji
  et~al\mbox{.}}{2015}]%
        {ji2015zero}
\bibfield{author}{\bibinfo{person}{Shouling Ji}, \bibinfo{person}{Shukun Yang},
  \bibinfo{person}{Xin Hu}, \bibinfo{person}{Weili Han},
  \bibinfo{person}{Zhigong Li}, {and} \bibinfo{person}{Raheem Beyah}.}
  \bibinfo{year}{2015}\natexlab{}.
\newblock \showarticletitle{Zero-sum password cracking game: A large-scale
  empirical study on the crackability, correlation, and security of passwords}.
\newblock \bibinfo{journal}{\emph{IEEE transactions on dependable and secure
  computing}} \bibinfo{volume}{14}, \bibinfo{number}{5} (\bibinfo{year}{2015}),
  \bibinfo{pages}{550--564}.
\newblock


\bibitem[\protect\citeauthoryear{Kamilaris and Prenafeta-Bold{\'u}}{Kamilaris
  and Prenafeta-Bold{\'u}}{2018}]%
        {kamilaris2018deep}
\bibfield{author}{\bibinfo{person}{Andreas Kamilaris} {and}
  \bibinfo{person}{Francesc~X Prenafeta-Bold{\'u}}.}
  \bibinfo{year}{2018}\natexlab{}.
\newblock \showarticletitle{Deep learning in agriculture: A survey}.
\newblock \bibinfo{journal}{\emph{Computers and electronics in agriculture}}
  \bibinfo{volume}{147} (\bibinfo{year}{2018}), \bibinfo{pages}{70--90}.
\newblock


\bibitem[\protect\citeauthoryear{K{\'a}roly, Galambos, Kuti, and
  Rudas}{K{\'a}roly et~al\mbox{.}}{2020}]%
        {karoly2020deep}
\bibfield{author}{\bibinfo{person}{Art{\'u}r~Istv{\'a}n K{\'a}roly},
  \bibinfo{person}{P{\'e}ter Galambos}, \bibinfo{person}{J{\'o}zsef Kuti},
  {and} \bibinfo{person}{Imre~J Rudas}.} \bibinfo{year}{2020}\natexlab{}.
\newblock \showarticletitle{Deep learning in robotics: Survey on model
  structures and training strategies}.
\newblock \bibinfo{journal}{\emph{IEEE Transactions on Systems, Man, and
  Cybernetics: Systems}} \bibinfo{volume}{51}, \bibinfo{number}{1}
  (\bibinfo{year}{2020}), \bibinfo{pages}{266--279}.
\newblock


\bibitem[\protect\citeauthoryear{K{\"a}vrestad, Zaxmy, and
  Nohlberg}{K{\"a}vrestad et~al\mbox{.}}{2020}]%
        {kavrestad2020analyzing}
\bibfield{author}{\bibinfo{person}{Joakim K{\"a}vrestad},
  \bibinfo{person}{Johan Zaxmy}, {and} \bibinfo{person}{Marcus Nohlberg}.}
  \bibinfo{year}{2020}\natexlab{}.
\newblock \showarticletitle{Analyzing the usage of character groups and
  keyboard patterns in password creation}.
\newblock \bibinfo{journal}{\emph{Information \& Computer Security}}
  (\bibinfo{year}{2020}).
\newblock


\bibitem[\protect\citeauthoryear{Kelley, Komanduri, Mazurek, Shay, Vidas,
  Bauer, Christin, Cranor, and Lopez}{Kelley et~al\mbox{.}}{2012}]%
        {kelley2012guess}
\bibfield{author}{\bibinfo{person}{Patrick~Gage Kelley},
  \bibinfo{person}{Saranga Komanduri}, \bibinfo{person}{Michelle~L Mazurek},
  \bibinfo{person}{Richard Shay}, \bibinfo{person}{Timothy Vidas},
  \bibinfo{person}{Lujo Bauer}, \bibinfo{person}{Nicolas Christin},
  \bibinfo{person}{Lorrie~Faith Cranor}, {and} \bibinfo{person}{Julio Lopez}.}
  \bibinfo{year}{2012}\natexlab{}.
\newblock \showarticletitle{Guess again (and again and again): Measuring
  password strength by simulating password-cracking algorithms}. In
  \bibinfo{booktitle}{\emph{2012 IEEE symposium on security and privacy}}.
  IEEE, \bibinfo{pages}{523--537}.
\newblock


\bibitem[\protect\citeauthoryear{Kim, Stuart, Hsiao, Lin, Zhang, Dabbish, and
  Kiesler}{Kim et~al\mbox{.}}{2014}]%
        {kim2014yourpassword}
\bibfield{author}{\bibinfo{person}{Tiffany Hyun-Jin Kim},
  \bibinfo{person}{H~Colleen Stuart}, \bibinfo{person}{Hsu-Chun Hsiao},
  \bibinfo{person}{Yue-Hsun Lin}, \bibinfo{person}{Leon Zhang},
  \bibinfo{person}{Laura Dabbish}, {and} \bibinfo{person}{Sara Kiesler}.}
  \bibinfo{year}{2014}\natexlab{}.
\newblock \showarticletitle{YourPassword: applying feedback loops to improve
  security behavior of managing multiple passwords}. In
  \bibinfo{booktitle}{\emph{Proceedings of the 9th ACM symposium on
  Information, computer and communications security}}.
  \bibinfo{pages}{513--518}.
\newblock


\bibitem[\protect\citeauthoryear{Kingma and Welling}{Kingma and
  Welling}{2013}]%
        {kingma2013auto}
\bibfield{author}{\bibinfo{person}{Diederik~P Kingma} {and}
  \bibinfo{person}{Max Welling}.} \bibinfo{year}{2013}\natexlab{}.
\newblock \showarticletitle{Auto-encoding variational bayes}.
\newblock \bibinfo{journal}{\emph{arXiv preprint arXiv:1312.6114}}
  (\bibinfo{year}{2013}).
\newblock


\bibitem[\protect\citeauthoryear{Knight and Graehl}{Knight and Graehl}{2005}]%
        {knight2005overview}
\bibfield{author}{\bibinfo{person}{Kevin Knight} {and}
  \bibinfo{person}{Jonathan Graehl}.} \bibinfo{year}{2005}\natexlab{}.
\newblock \showarticletitle{An overview of probabilistic tree transducers for
  natural language processing}. In \bibinfo{booktitle}{\emph{International
  Conference on Intelligent Text Processing and Computational Linguistics}}.
  Springer, \bibinfo{pages}{1--24}.
\newblock


\bibitem[\protect\citeauthoryear{Krishnamurthy and Wills}{Krishnamurthy and
  Wills}{2009}]%
        {krishnamurthy2009leakage}
\bibfield{author}{\bibinfo{person}{Balachander Krishnamurthy} {and}
  \bibinfo{person}{Craig~E Wills}.} \bibinfo{year}{2009}\natexlab{}.
\newblock \showarticletitle{On the leakage of personally identifiable
  information via online social networks}. In
  \bibinfo{booktitle}{\emph{Proceedings of the 2nd ACM workshop on Online
  social networks}}. \bibinfo{pages}{7--12}.
\newblock


\bibitem[\protect\citeauthoryear{Kuo, Romanosky, and Cranor}{Kuo
  et~al\mbox{.}}{2006}]%
        {kuo2006human}
\bibfield{author}{\bibinfo{person}{Cynthia Kuo}, \bibinfo{person}{Sasha
  Romanosky}, {and} \bibinfo{person}{Lorrie~Faith Cranor}.}
  \bibinfo{year}{2006}\natexlab{}.
\newblock \showarticletitle{Human selection of mnemonic phrase-based
  passwords}. In \bibinfo{booktitle}{\emph{Proceedings of the second symposium
  on Usable privacy and security}}. \bibinfo{pages}{67--78}.
\newblock


\bibitem[\protect\citeauthoryear{Lamport}{Lamport}{1981}]%
        {lamport1981password}
\bibfield{author}{\bibinfo{person}{Leslie Lamport}.}
  \bibinfo{year}{1981}\natexlab{}.
\newblock \showarticletitle{Password authentication with insecure
  communication}.
\newblock \bibinfo{journal}{\emph{Commun. ACM}} \bibinfo{volume}{24},
  \bibinfo{number}{11} (\bibinfo{year}{1981}), \bibinfo{pages}{770--772}.
\newblock


\bibitem[\protect\citeauthoryear{LeCun, Bengio, and Hinton}{LeCun
  et~al\mbox{.}}{2015}]%
        {lecun2015deep}
\bibfield{author}{\bibinfo{person}{Yann LeCun}, \bibinfo{person}{Yoshua
  Bengio}, {and} \bibinfo{person}{Geoffrey Hinton}.}
  \bibinfo{year}{2015}\natexlab{}.
\newblock \showarticletitle{Deep learning}.
\newblock \bibinfo{journal}{\emph{nature}} \bibinfo{volume}{521},
  \bibinfo{number}{7553} (\bibinfo{year}{2015}), \bibinfo{pages}{436--444}.
\newblock


\bibitem[\protect\citeauthoryear{Lee, Sj{\"o}berg, and Narayanan}{Lee
  et~al\mbox{.}}{2022}]%
        {lee2022password}
\bibfield{author}{\bibinfo{person}{Kevin Lee}, \bibinfo{person}{Sten
  Sj{\"o}berg}, {and} \bibinfo{person}{Arvind Narayanan}.}
  \bibinfo{year}{2022}\natexlab{}.
\newblock \showarticletitle{Password policies of most top websites fail to
  follow best practices}. In \bibinfo{booktitle}{\emph{Eighteenth Symposium on
  Usable Privacy and Security (SOUPS 2022)}}. \bibinfo{pages}{561--580}.
\newblock


\bibitem[\protect\citeauthoryear{Li, Chen, Yan, Jia, and Li}{Li
  et~al\mbox{.}}{2019}]%
        {li2019password}
\bibfield{author}{\bibinfo{person}{Hang Li}, \bibinfo{person}{Mengqi Chen},
  \bibinfo{person}{Shengbo Yan}, \bibinfo{person}{Chunfu Jia}, {and}
  \bibinfo{person}{Zhaohui Li}.} \bibinfo{year}{2019}\natexlab{}.
\newblock \showarticletitle{Password guessing via neural language modeling}. In
  \bibinfo{booktitle}{\emph{International Conference on Machine Learning for
  Cyber Security}}. Springer, \bibinfo{pages}{78--93}.
\newblock


\bibitem[\protect\citeauthoryear{Li, Wang, and Sun}{Li et~al\mbox{.}}{2016}]%
        {li2016study}
\bibfield{author}{\bibinfo{person}{Yue Li}, \bibinfo{person}{Haining Wang},
  {and} \bibinfo{person}{Kun Sun}.} \bibinfo{year}{2016}\natexlab{}.
\newblock \showarticletitle{A study of personal information in human-chosen
  passwords and its security implications}. In \bibinfo{booktitle}{\emph{IEEE
  INFOCOM 2016-The 35th Annual IEEE International Conference on Computer
  Communications}}. IEEE, \bibinfo{pages}{1--9}.
\newblock


\bibitem[\protect\citeauthoryear{Li, Han, and Xu}{Li et~al\mbox{.}}{2014}]%
        {li2014large}
\bibfield{author}{\bibinfo{person}{Zhigong Li}, \bibinfo{person}{Weili Han},
  {and} \bibinfo{person}{Wenyuan Xu}.} \bibinfo{year}{2014}\natexlab{}.
\newblock \showarticletitle{A $\{$Large-Scale$\}$ Empirical Analysis of Chinese
  Web Passwords}. In \bibinfo{booktitle}{\emph{23rd USENIX Security Symposium
  (USENIX Security 14)}}. \bibinfo{pages}{559--574}.
\newblock


\bibitem[\protect\citeauthoryear{Litjens, Kooi, Bejnordi, Setio, Ciompi,
  Ghafoorian, Van Der~Laak, Van~Ginneken, and S{\'a}nchez}{Litjens
  et~al\mbox{.}}{2017}]%
        {litjens2017survey}
\bibfield{author}{\bibinfo{person}{Geert Litjens}, \bibinfo{person}{Thijs
  Kooi}, \bibinfo{person}{Babak~Ehteshami Bejnordi}, \bibinfo{person}{Arnaud
  Arindra~Adiyoso Setio}, \bibinfo{person}{Francesco Ciompi},
  \bibinfo{person}{Mohsen Ghafoorian}, \bibinfo{person}{Jeroen~Awm Van
  Der~Laak}, \bibinfo{person}{Bram Van~Ginneken}, {and}
  \bibinfo{person}{Clara~I S{\'a}nchez}.} \bibinfo{year}{2017}\natexlab{}.
\newblock \showarticletitle{A survey on deep learning in medical image
  analysis}.
\newblock \bibinfo{journal}{\emph{Medical image analysis}}
  \bibinfo{volume}{42} (\bibinfo{year}{2017}), \bibinfo{pages}{60--88}.
\newblock


\bibitem[\protect\citeauthoryear{Liu, Xia, Yi, Yao, Xie, Wang, and Zhu}{Liu
  et~al\mbox{.}}{2018}]%
        {liu2018genpass}
\bibfield{author}{\bibinfo{person}{Yunyu Liu}, \bibinfo{person}{Zhiyang Xia},
  \bibinfo{person}{Ping Yi}, \bibinfo{person}{Yao Yao},
  \bibinfo{person}{Tiantian Xie}, \bibinfo{person}{Wei Wang}, {and}
  \bibinfo{person}{Ting Zhu}.} \bibinfo{year}{2018}\natexlab{}.
\newblock \showarticletitle{GENPass: A general deep learning model for password
  guessing with PCFG rules and adversarial generation}. In
  \bibinfo{booktitle}{\emph{2018 IEEE International Conference on
  Communications (ICC)}}. IEEE, \bibinfo{pages}{1--6}.
\newblock


\bibitem[\protect\citeauthoryear{Luo, Deng, Lu, and Liu}{Luo
  et~al\mbox{.}}{2019}]%
        {luo2019recurrent}
\bibfield{author}{\bibinfo{person}{Jun Luo}, \bibinfo{person}{Jin Deng},
  \bibinfo{person}{Chu Lu}, {and} \bibinfo{person}{Hong Liu}.}
  \bibinfo{year}{2019}\natexlab{}.
\newblock \showarticletitle{Recurrent Neural Network Based Password Generation
  for Group Attribute Context-Ware Applications}. In
  \bibinfo{booktitle}{\emph{2019 IEEE 21st International Conference on High
  Performance Computing and Communications; IEEE 17th International Conference
  on Smart City; IEEE 5th International Conference on Data Science and Systems
  (HPCC/SmartCity/DSS)}}. IEEE, \bibinfo{pages}{2688--2693}.
\newblock


\bibitem[\protect\citeauthoryear{Ma, Yang, Luo, and Li}{Ma
  et~al\mbox{.}}{2014}]%
        {ma2014study}
\bibfield{author}{\bibinfo{person}{Jerry Ma}, \bibinfo{person}{Weining Yang},
  \bibinfo{person}{Min Luo}, {and} \bibinfo{person}{Ninghui Li}.}
  \bibinfo{year}{2014}\natexlab{}.
\newblock \showarticletitle{A study of probabilistic password models}. In
  \bibinfo{booktitle}{\emph{2014 IEEE Symposium on Security and Privacy}}.
  IEEE, \bibinfo{pages}{689--704}.
\newblock


\bibitem[\protect\citeauthoryear{Makhzani, Shlens, Jaitly, Goodfellow, and
  Frey}{Makhzani et~al\mbox{.}}{2015}]%
        {makhzani2015adversarial}
\bibfield{author}{\bibinfo{person}{Alireza Makhzani}, \bibinfo{person}{Jonathon
  Shlens}, \bibinfo{person}{Navdeep Jaitly}, \bibinfo{person}{Ian Goodfellow},
  {and} \bibinfo{person}{Brendan Frey}.} \bibinfo{year}{2015}\natexlab{}.
\newblock \showarticletitle{Adversarial autoencoders}.
\newblock \bibinfo{journal}{\emph{arXiv preprint arXiv:1511.05644}}
  (\bibinfo{year}{2015}).
\newblock


\bibitem[\protect\citeauthoryear{Mazurek, Komanduri, Vidas, Bauer, Christin,
  Cranor, Kelley, Shay, and Ur}{Mazurek et~al\mbox{.}}{2013}]%
        {mazurek2013measuring}
\bibfield{author}{\bibinfo{person}{Michelle~L Mazurek},
  \bibinfo{person}{Saranga Komanduri}, \bibinfo{person}{Timothy Vidas},
  \bibinfo{person}{Lujo Bauer}, \bibinfo{person}{Nicolas Christin},
  \bibinfo{person}{Lorrie~Faith Cranor}, \bibinfo{person}{Patrick~Gage Kelley},
  \bibinfo{person}{Richard Shay}, {and} \bibinfo{person}{Blase Ur}.}
  \bibinfo{year}{2013}\natexlab{}.
\newblock \showarticletitle{Measuring password guessability for an entire
  university}. In \bibinfo{booktitle}{\emph{Proceedings of the 2013 ACM SIGSAC
  conference on Computer \& communications security}}.
  \bibinfo{pages}{173--186}.
\newblock


\bibitem[\protect\citeauthoryear{Melicher, Ur, Segreti, Bauer, Christin, and
  Cranor}{Melicher et~al\mbox{.}}{2017}]%
        {melicher2017better}
\bibfield{author}{\bibinfo{person}{William Melicher}, \bibinfo{person}{Blase
  Ur}, \bibinfo{person}{Sean~M Segreti}, \bibinfo{person}{Lujo Bauer},
  \bibinfo{person}{Nicolas Christin}, {and} \bibinfo{person}{Lorrie~Faith
  Cranor}.} \bibinfo{year}{2017}\natexlab{}.
\newblock \showarticletitle{Better passwords through science (and neural
  networks)}. In \bibinfo{booktitle}{\emph{USENIX}}, Vol.~\bibinfo{volume}{42}.
  \bibinfo{pages}{1--7}.
\newblock


\bibitem[\protect\citeauthoryear{Melicher, Ur, Segreti, Komanduri, Bauer,
  Christin, and Cranor}{Melicher et~al\mbox{.}}{2016}]%
        {melicher2016fast}
\bibfield{author}{\bibinfo{person}{William Melicher}, \bibinfo{person}{Blase
  Ur}, \bibinfo{person}{Sean~M Segreti}, \bibinfo{person}{Saranga Komanduri},
  \bibinfo{person}{Lujo Bauer}, \bibinfo{person}{Nicolas Christin}, {and}
  \bibinfo{person}{Lorrie~Faith Cranor}.} \bibinfo{year}{2016}\natexlab{}.
\newblock \showarticletitle{Fast, lean, and accurate: Modeling password
  guessability using neural networks}. In \bibinfo{booktitle}{\emph{25th USENIX
  Security Symposium (USENIX Security 16)}}. \bibinfo{pages}{175--191}.
\newblock


\bibitem[\protect\citeauthoryear{Miljanovic}{Miljanovic}{[n.\,d.]}]%
        {branka2022ken}
\bibfield{author}{\bibinfo{person}{Branka Miljanovic}.}
  \bibinfo{year}{[n.\,d.]}\natexlab{}.
\newblock \bibinfo{title}{‘Ken Sent Me’ – A Brief History of Password
  Hacking}.
\newblock
\newblock
\urldef\tempurl%
\url{https://www.nevis.net/en/blog/a-brief-history-of-password-hacking}
\showURL{%
Retrieved September 6, 2022 from \tempurl}


\bibitem[\protect\citeauthoryear{Morris and Thompson}{Morris and
  Thompson}{1979}]%
        {morris1979password}
\bibfield{author}{\bibinfo{person}{Robert Morris} {and} \bibinfo{person}{Ken
  Thompson}.} \bibinfo{year}{1979}\natexlab{}.
\newblock \showarticletitle{Password security: A case history}.
\newblock \bibinfo{journal}{\emph{Commun. ACM}} \bibinfo{volume}{22},
  \bibinfo{number}{11} (\bibinfo{year}{1979}), \bibinfo{pages}{594--597}.
\newblock


\bibitem[\protect\citeauthoryear{Narayanan and Shmatikov}{Narayanan and
  Shmatikov}{2005}]%
        {narayanan2005fast}
\bibfield{author}{\bibinfo{person}{Arvind Narayanan} {and}
  \bibinfo{person}{Vitaly Shmatikov}.} \bibinfo{year}{2005}\natexlab{}.
\newblock \showarticletitle{Fast dictionary attacks on passwords using
  time-space tradeoff}. In \bibinfo{booktitle}{\emph{Proceedings of the 12th
  ACM conference on Computer and communications security}}.
  \bibinfo{pages}{364--372}.
\newblock


\bibitem[\protect\citeauthoryear{Oechslin}{Oechslin}{2003}]%
        {oechslin2003making}
\bibfield{author}{\bibinfo{person}{Philippe Oechslin}.}
  \bibinfo{year}{2003}\natexlab{}.
\newblock \showarticletitle{Making a faster cryptanalytic time-memory
  trade-off}. In \bibinfo{booktitle}{\emph{Annual International Cryptology
  Conference}}. Springer, \bibinfo{pages}{617--630}.
\newblock


\bibitem[\protect\citeauthoryear{Pal, Daniel, Chatterjee, and Ristenpart}{Pal
  et~al\mbox{.}}{2019}]%
        {pal2019beyond}
\bibfield{author}{\bibinfo{person}{Bijeeta Pal}, \bibinfo{person}{Tal Daniel},
  \bibinfo{person}{Rahul Chatterjee}, {and} \bibinfo{person}{Thomas
  Ristenpart}.} \bibinfo{year}{2019}\natexlab{}.
\newblock \showarticletitle{Beyond credential stuffing: Password similarity
  models using neural networks}. In \bibinfo{booktitle}{\emph{2019 IEEE
  Symposium on Security and Privacy (SP)}}. IEEE, \bibinfo{pages}{417--434}.
\newblock


\bibitem[\protect\citeauthoryear{Pasquini, Cianfriglia, Ateniese, and
  Bernaschi}{Pasquini et~al\mbox{.}}{2021a}]%
        {pasquini2021reducing}
\bibfield{author}{\bibinfo{person}{Dario Pasquini}, \bibinfo{person}{Marco
  Cianfriglia}, \bibinfo{person}{Giuseppe Ateniese}, {and}
  \bibinfo{person}{Massimo Bernaschi}.} \bibinfo{year}{2021}\natexlab{a}.
\newblock \showarticletitle{Reducing bias in modeling real-world password
  strength via deep learning and dynamic dictionaries}. In
  \bibinfo{booktitle}{\emph{30th USENIX Security Symposium (USENIX Security
  21)}}. \bibinfo{pages}{821--838}.
\newblock


\bibitem[\protect\citeauthoryear{Pasquini, Gangwal, Ateniese, Bernaschi, and
  Conti}{Pasquini et~al\mbox{.}}{2021b}]%
        {pasquini2021improving}
\bibfield{author}{\bibinfo{person}{Dario Pasquini}, \bibinfo{person}{Ankit
  Gangwal}, \bibinfo{person}{Giuseppe Ateniese}, \bibinfo{person}{Massimo
  Bernaschi}, {and} \bibinfo{person}{Mauro Conti}.}
  \bibinfo{year}{2021}\natexlab{b}.
\newblock \showarticletitle{Improving password guessing via representation
  learning}. In \bibinfo{booktitle}{\emph{2021 IEEE Symposium on Security and
  Privacy (SP)}}. IEEE, \bibinfo{pages}{1382--1399}.
\newblock


\bibitem[\protect\citeauthoryear{Pearman, Thomas, Naeini, Habib, Bauer,
  Christin, Cranor, Egelman, and Forget}{Pearman et~al\mbox{.}}{2017}]%
        {pearman2017let}
\bibfield{author}{\bibinfo{person}{Sarah Pearman}, \bibinfo{person}{Jeremy
  Thomas}, \bibinfo{person}{Pardis~Emami Naeini}, \bibinfo{person}{Hana Habib},
  \bibinfo{person}{Lujo Bauer}, \bibinfo{person}{Nicolas Christin},
  \bibinfo{person}{Lorrie~Faith Cranor}, \bibinfo{person}{Serge Egelman}, {and}
  \bibinfo{person}{Alain Forget}.} \bibinfo{year}{2017}\natexlab{}.
\newblock \showarticletitle{Let's go in for a closer look: Observing passwords
  in their natural habitat}. In \bibinfo{booktitle}{\emph{Proceedings of the
  2017 ACM SIGSAC Conference on Computer and Communications Security}}.
  \bibinfo{pages}{295--310}.
\newblock


\bibitem[\protect\citeauthoryear{Pereira, Ferreira, and Mendes}{Pereira
  et~al\mbox{.}}{2020}]%
        {pereira2020evaluating}
\bibfield{author}{\bibinfo{person}{David Pereira}, \bibinfo{person}{Joao~F
  Ferreira}, {and} \bibinfo{person}{Alexandra Mendes}.}
  \bibinfo{year}{2020}\natexlab{}.
\newblock \showarticletitle{Evaluating the accuracy of password strength meters
  using off-the-shelf guessing attacks}. In \bibinfo{booktitle}{\emph{2020 IEEE
  International Symposium on Software Reliability Engineering Workshops
  (ISSREW)}}. IEEE, \bibinfo{pages}{237--242}.
\newblock


\bibitem[\protect\citeauthoryear{Peslyak}{Peslyak}{[n.\,d.]}]%
        {peslyak2014john}
\bibfield{author}{\bibinfo{person}{Alexander Peslyak}.}
  \bibinfo{year}{[n.\,d.]}\natexlab{}.
\newblock \bibinfo{title}{John the Ripper}.
\newblock
\newblock
\urldef\tempurl%
\url{https://www.openwall.com/john/}
\showURL{%
Retrieved March 3, 2022 from \tempurl}


\bibitem[\protect\citeauthoryear{Pliam}{Pliam}{2000}]%
        {pliam2000incomparability}
\bibfield{author}{\bibinfo{person}{John~O Pliam}.}
  \bibinfo{year}{2000}\natexlab{}.
\newblock \showarticletitle{On the incomparability of entropy and marginal
  guesswork in brute-force attacks}. In \bibinfo{booktitle}{\emph{International
  conference on cryptology in India}}. Springer, \bibinfo{pages}{67--79}.
\newblock


\bibitem[\protect\citeauthoryear{Proctor, Lien, Vu, Schultz, and
  Salvendy}{Proctor et~al\mbox{.}}{2002}]%
        {proctor2002improving}
\bibfield{author}{\bibinfo{person}{Robert~W Proctor},
  \bibinfo{person}{Mei-Ching Lien}, \bibinfo{person}{Kim-Phuong~L Vu},
  \bibinfo{person}{E~Eugene Schultz}, {and} \bibinfo{person}{Gavriel
  Salvendy}.} \bibinfo{year}{2002}\natexlab{}.
\newblock \showarticletitle{Improving computer security for authentication of
  users: Influence of proactive password restrictions}.
\newblock \bibinfo{journal}{\emph{Behavior Research Methods, Instruments, \&
  Computers}} \bibinfo{volume}{34}, \bibinfo{number}{2} (\bibinfo{year}{2002}),
  \bibinfo{pages}{163--169}.
\newblock


\bibitem[\protect\citeauthoryear{Rao, Jha, and Kini}{Rao et~al\mbox{.}}{2013}]%
        {rao2013effect}
\bibfield{author}{\bibinfo{person}{Ashwini Rao}, \bibinfo{person}{Birendra
  Jha}, {and} \bibinfo{person}{Gananand Kini}.}
  \bibinfo{year}{2013}\natexlab{}.
\newblock \showarticletitle{Effect of grammar on security of long passwords}.
  In \bibinfo{booktitle}{\emph{Proceedings of the third ACM conference on Data
  and application security and privacy}}. \bibinfo{pages}{317--324}.
\newblock


\bibitem[\protect\citeauthoryear{Scarfone and Souppaya}{Scarfone and
  Souppaya}{2009}]%
        {scarfone2009guide}
\bibfield{author}{\bibinfo{person}{Karen Scarfone} {and}
  \bibinfo{person}{Murugiah Souppaya}.} \bibinfo{year}{2009}\natexlab{}.
\newblock \showarticletitle{Guide to enterprise password management (draft)}.
\newblock \bibinfo{journal}{\emph{NIST special publication}}
  \bibinfo{volume}{800}, \bibinfo{number}{118} (\bibinfo{year}{2009}),
  \bibinfo{pages}{800--118}.
\newblock


\bibitem[\protect\citeauthoryear{Schuster and Paliwal}{Schuster and
  Paliwal}{1997}]%
        {schuster1997bidirectional}
\bibfield{author}{\bibinfo{person}{Mike Schuster} {and}
  \bibinfo{person}{Kuldip~K Paliwal}.} \bibinfo{year}{1997}\natexlab{}.
\newblock \showarticletitle{Bidirectional recurrent neural networks}.
\newblock \bibinfo{journal}{\emph{IEEE transactions on Signal Processing}}
  \bibinfo{volume}{45}, \bibinfo{number}{11} (\bibinfo{year}{1997}),
  \bibinfo{pages}{2673--2681}.
\newblock


\bibitem[\protect\citeauthoryear{Schweitzer, Boleng, Hughes, and
  Murphy}{Schweitzer et~al\mbox{.}}{2011}]%
        {schweitzer2011visualizing}
\bibfield{author}{\bibinfo{person}{Dino Schweitzer}, \bibinfo{person}{Jeff
  Boleng}, \bibinfo{person}{Colin Hughes}, {and} \bibinfo{person}{Louis
  Murphy}.} \bibinfo{year}{2011}\natexlab{}.
\newblock \showarticletitle{Visualizing keyboard pattern passwords}.
\newblock \bibinfo{journal}{\emph{Information Visualization}}
  \bibinfo{volume}{10}, \bibinfo{number}{2} (\bibinfo{year}{2011}),
  \bibinfo{pages}{127--133}.
\newblock


\bibitem[\protect\citeauthoryear{Shay, Bauer, Christin, Cranor, Forget,
  Komanduri, Mazurek, Melicher, Segreti, and Ur}{Shay et~al\mbox{.}}{2015}]%
        {shay2015spoonful}
\bibfield{author}{\bibinfo{person}{Richard Shay}, \bibinfo{person}{Lujo Bauer},
  \bibinfo{person}{Nicolas Christin}, \bibinfo{person}{Lorrie~Faith Cranor},
  \bibinfo{person}{Alain Forget}, \bibinfo{person}{Saranga Komanduri},
  \bibinfo{person}{Michelle~L Mazurek}, \bibinfo{person}{William Melicher},
  \bibinfo{person}{Sean~M Segreti}, {and} \bibinfo{person}{Blase Ur}.}
  \bibinfo{year}{2015}\natexlab{}.
\newblock \showarticletitle{A spoonful of sugar? The impact of guidance and
  feedback on password-creation behavior}. In
  \bibinfo{booktitle}{\emph{Proceedings of the 33rd annual ACM conference on
  human factors in computing systems}}. \bibinfo{pages}{2903--2912}.
\newblock


\bibitem[\protect\citeauthoryear{Shay, Komanduri, Kelley, Leon, Mazurek, Bauer,
  Christin, and Cranor}{Shay et~al\mbox{.}}{2010}]%
        {shay2010encountering}
\bibfield{author}{\bibinfo{person}{Richard Shay}, \bibinfo{person}{Saranga
  Komanduri}, \bibinfo{person}{Patrick~Gage Kelley},
  \bibinfo{person}{Pedro~Giovanni Leon}, \bibinfo{person}{Michelle~L Mazurek},
  \bibinfo{person}{Lujo Bauer}, \bibinfo{person}{Nicolas Christin}, {and}
  \bibinfo{person}{Lorrie~Faith Cranor}.} \bibinfo{year}{2010}\natexlab{}.
\newblock \showarticletitle{Encountering stronger password requirements: user
  attitudes and behaviors}. In \bibinfo{booktitle}{\emph{Proceedings of the
  sixth symposium on usable privacy and security}}. \bibinfo{pages}{1--20}.
\newblock


\bibitem[\protect\citeauthoryear{Shi, Zhou, Li, and Han}{Shi
  et~al\mbox{.}}{2021}]%
        {shi2021understanding}
\bibfield{author}{\bibinfo{person}{Ruixin Shi}, \bibinfo{person}{Yongbin Zhou},
  \bibinfo{person}{Yong Li}, {and} \bibinfo{person}{Weili Han}.}
  \bibinfo{year}{2021}\natexlab{}.
\newblock \showarticletitle{Understanding Offline Password-Cracking Methods: A
  Large-Scale Empirical Study}.
\newblock \bibinfo{journal}{\emph{Security and Communication Networks}}
  \bibinfo{volume}{2021} (\bibinfo{year}{2021}).
\newblock


\bibitem[\protect\citeauthoryear{Spafford}{Spafford}{1992}]%
        {spafford1992opus}
\bibfield{author}{\bibinfo{person}{Eugene~H Spafford}.}
  \bibinfo{year}{1992}\natexlab{}.
\newblock \showarticletitle{Opus: Preventing weak password choices}.
\newblock \bibinfo{journal}{\emph{Computers \& Security}} \bibinfo{volume}{11},
  \bibinfo{number}{3} (\bibinfo{year}{1992}), \bibinfo{pages}{273--278}.
\newblock


\bibitem[\protect\citeauthoryear{Steube}{Steube}{[n.\,d.]}]%
        {jens2009hashcat}
\bibfield{author}{\bibinfo{person}{Jens Steube}.}
  \bibinfo{year}{[n.\,d.]}\natexlab{}.
\newblock \bibinfo{title}{Hashcat}.
\newblock
\newblock
\urldef\tempurl%
\url{https://hashcat.net/hashcat/}
\showURL{%
Retrieved March 3, 2022 from \tempurl}


\bibitem[\protect\citeauthoryear{Stobert and Biddle}{Stobert and
  Biddle}{2014}]%
        {stobert2014password}
\bibfield{author}{\bibinfo{person}{Elizabeth Stobert} {and}
  \bibinfo{person}{Robert Biddle}.} \bibinfo{year}{2014}\natexlab{}.
\newblock \showarticletitle{The password life cycle: user behaviour in managing
  passwords}. In \bibinfo{booktitle}{\emph{10th Symposium On Usable Privacy and
  Security (SOUPS 2014)}}. \bibinfo{pages}{243--255}.
\newblock


\bibitem[\protect\citeauthoryear{Stobert and Biddle}{Stobert and
  Biddle}{2018}]%
        {stobert2018password}
\bibfield{author}{\bibinfo{person}{Elizabeth Stobert} {and}
  \bibinfo{person}{Robert Biddle}.} \bibinfo{year}{2018}\natexlab{}.
\newblock \showarticletitle{The password life cycle}.
\newblock \bibinfo{journal}{\emph{ACM Transactions on Privacy and Security
  (TOPS)}} \bibinfo{volume}{21}, \bibinfo{number}{3} (\bibinfo{year}{2018}),
  \bibinfo{pages}{1--32}.
\newblock


\bibitem[\protect\citeauthoryear{Summers and Bosworth}{Summers and
  Bosworth}{2004}]%
        {summers2004password}
\bibfield{author}{\bibinfo{person}{Wayne~C Summers} {and}
  \bibinfo{person}{Edward Bosworth}.} \bibinfo{year}{2004}\natexlab{}.
\newblock \showarticletitle{Password policy: the good, the bad, and the ugly}.
  In \bibinfo{booktitle}{\emph{Proceedings of the winter international
  synposium on Information and communication technologies}}.
  \bibinfo{pages}{1--6}.
\newblock


\bibitem[\protect\citeauthoryear{Sutskever, Vinyals, and Le}{Sutskever
  et~al\mbox{.}}{2014}]%
        {sutskever2014sequence}
\bibfield{author}{\bibinfo{person}{Ilya Sutskever}, \bibinfo{person}{Oriol
  Vinyals}, {and} \bibinfo{person}{Quoc~V Le}.}
  \bibinfo{year}{2014}\natexlab{}.
\newblock \showarticletitle{Sequence to sequence learning with neural
  networks}.
\newblock \bibinfo{journal}{\emph{Advances in neural information processing
  systems}}  \bibinfo{volume}{27} (\bibinfo{year}{2014}).
\newblock


\bibitem[\protect\citeauthoryear{Tan, Bauer, Christin, and Cranor}{Tan
  et~al\mbox{.}}{2020}]%
        {tan2020practical}
\bibfield{author}{\bibinfo{person}{Joshua Tan}, \bibinfo{person}{Lujo Bauer},
  \bibinfo{person}{Nicolas Christin}, {and} \bibinfo{person}{Lorrie~Faith
  Cranor}.} \bibinfo{year}{2020}\natexlab{}.
\newblock \showarticletitle{Practical recommendations for stronger, more usable
  passwords combining minimum-strength, minimum-length, and blocklist
  requirements}. In \bibinfo{booktitle}{\emph{Proceedings of the 2020 ACM
  SIGSAC Conference on Computer and Communications Security}}.
  \bibinfo{pages}{1407--1426}.
\newblock


\bibitem[\protect\citeauthoryear{Taneski, Heri{\v{c}}ko, and Brumen}{Taneski
  et~al\mbox{.}}{2019}]%
        {taneski2019systematic}
\bibfield{author}{\bibinfo{person}{Viktor Taneski}, \bibinfo{person}{Marjan
  Heri{\v{c}}ko}, {and} \bibinfo{person}{Bo{\v{s}}tjan Brumen}.}
  \bibinfo{year}{2019}\natexlab{}.
\newblock \showarticletitle{Systematic overview of password security problems}.
\newblock \bibinfo{journal}{\emph{Acta Polytechnica Hungarica}}
  \bibinfo{volume}{16}, \bibinfo{number}{3} (\bibinfo{year}{2019}).
\newblock


\bibitem[\protect\citeauthoryear{Tatl{\i}}{Tatl{\i}}{2015}]%
        {tatli2015cracking}
\bibfield{author}{\bibinfo{person}{Emin~Islam Tatl{\i}}.}
  \bibinfo{year}{2015}\natexlab{}.
\newblock \showarticletitle{Cracking more password hashes with patterns}.
\newblock \bibinfo{journal}{\emph{IEEE Transactions on Information Forensics
  and Security}} \bibinfo{volume}{10}, \bibinfo{number}{8}
  (\bibinfo{year}{2015}), \bibinfo{pages}{1656--1665}.
\newblock


\bibitem[\protect\citeauthoryear{Tolstikhin, Bousquet, Gelly, and
  Schoelkopf}{Tolstikhin et~al\mbox{.}}{2017}]%
        {tolstikhin2017wasserstein}
\bibfield{author}{\bibinfo{person}{Ilya Tolstikhin}, \bibinfo{person}{Olivier
  Bousquet}, \bibinfo{person}{Sylvain Gelly}, {and} \bibinfo{person}{Bernhard
  Schoelkopf}.} \bibinfo{year}{2017}\natexlab{}.
\newblock \showarticletitle{Wasserstein auto-encoders}.
\newblock \bibinfo{journal}{\emph{arXiv preprint arXiv:1711.01558}}
  (\bibinfo{year}{2017}).
\newblock


\bibitem[\protect\citeauthoryear{Ur, Alfieri, Aung, Bauer, Christin, Colnago,
  Cranor, Dixon, Emami~Naeini, Habib, et~al\mbox{.}}{Ur et~al\mbox{.}}{2017}]%
        {ur2017design}
\bibfield{author}{\bibinfo{person}{Blase Ur}, \bibinfo{person}{Felicia
  Alfieri}, \bibinfo{person}{Maung Aung}, \bibinfo{person}{Lujo Bauer},
  \bibinfo{person}{Nicolas Christin}, \bibinfo{person}{Jessica Colnago},
  \bibinfo{person}{Lorrie~Faith Cranor}, \bibinfo{person}{Henry Dixon},
  \bibinfo{person}{Pardis Emami~Naeini}, \bibinfo{person}{Hana Habib},
  {et~al\mbox{.}}} \bibinfo{year}{2017}\natexlab{}.
\newblock \showarticletitle{Design and evaluation of a data-driven password
  meter}. In \bibinfo{booktitle}{\emph{Proceedings of the 2017 chi conference
  on human factors in computing systems}}. \bibinfo{pages}{3775--3786}.
\newblock


\bibitem[\protect\citeauthoryear{Ur, Kelley, Komanduri, Lee, Maass, Mazurek,
  Passaro, Shay, Vidas, Bauer, et~al\mbox{.}}{Ur et~al\mbox{.}}{2012}]%
        {ur2012does}
\bibfield{author}{\bibinfo{person}{Blase Ur}, \bibinfo{person}{Patrick~Gage
  Kelley}, \bibinfo{person}{Saranga Komanduri}, \bibinfo{person}{Joel Lee},
  \bibinfo{person}{Michael Maass}, \bibinfo{person}{Michelle~L Mazurek},
  \bibinfo{person}{Timothy Passaro}, \bibinfo{person}{Richard Shay},
  \bibinfo{person}{Timothy Vidas}, \bibinfo{person}{Lujo Bauer},
  {et~al\mbox{.}}} \bibinfo{year}{2012}\natexlab{}.
\newblock \showarticletitle{How does your password measure up? The effect of
  strength meters on password creation}. In \bibinfo{booktitle}{\emph{21st
  $\{$USENIX$\}$ Security Symposium ($\{$USENIX$\}$ Security 12)}}.
  \bibinfo{pages}{65--80}.
\newblock


\bibitem[\protect\citeauthoryear{Ur, Noma, Bees, Segreti, Shay, Bauer,
  Christin, and Cranor}{Ur et~al\mbox{.}}{2015a}]%
        {ur2015added}
\bibfield{author}{\bibinfo{person}{Blase Ur}, \bibinfo{person}{Fumiko Noma},
  \bibinfo{person}{Jonathan Bees}, \bibinfo{person}{Sean~M Segreti},
  \bibinfo{person}{Richard Shay}, \bibinfo{person}{Lujo Bauer},
  \bibinfo{person}{Nicolas Christin}, {and} \bibinfo{person}{Lorrie~Faith
  Cranor}.} \bibinfo{year}{2015}\natexlab{a}.
\newblock \showarticletitle{" I Added'!'at the End to Make It Secure":
  Observing Password Creation in the Lab}. In
  \bibinfo{booktitle}{\emph{Eleventh Symposium On Usable Privacy and Security
  (SOUPS 2015)}}. \bibinfo{pages}{123--140}.
\newblock


\bibitem[\protect\citeauthoryear{Ur, Segreti, Bauer, Christin, Cranor,
  Komanduri, Kurilova, Mazurek, Melicher, and Shay}{Ur et~al\mbox{.}}{2015b}]%
        {ur2015measuring}
\bibfield{author}{\bibinfo{person}{Blase Ur}, \bibinfo{person}{Sean~M Segreti},
  \bibinfo{person}{Lujo Bauer}, \bibinfo{person}{Nicolas Christin},
  \bibinfo{person}{Lorrie~Faith Cranor}, \bibinfo{person}{Saranga Komanduri},
  \bibinfo{person}{Darya Kurilova}, \bibinfo{person}{Michelle~L Mazurek},
  \bibinfo{person}{William Melicher}, {and} \bibinfo{person}{Richard Shay}.}
  \bibinfo{year}{2015}\natexlab{b}.
\newblock \showarticletitle{Measuring $\{$Real-World$\}$ Accuracies and Biases
  in Modeling Password Guessability}. In \bibinfo{booktitle}{\emph{24th USENIX
  Security Symposium (USENIX Security 15)}}. \bibinfo{pages}{463--481}.
\newblock


\bibitem[\protect\citeauthoryear{Veras, Collins, and Thorpe}{Veras
  et~al\mbox{.}}{2014}]%
        {veras2014semantic}
\bibfield{author}{\bibinfo{person}{Rafael Veras}, \bibinfo{person}{Christopher
  Collins}, {and} \bibinfo{person}{Julie Thorpe}.}
  \bibinfo{year}{2014}\natexlab{}.
\newblock \showarticletitle{On Semantic Patterns of Passwords and their
  Security Impact.}. In \bibinfo{booktitle}{\emph{NDSS}}. Citeseer.
\newblock


\bibitem[\protect\citeauthoryear{Veras, Thorpe, and Collins}{Veras
  et~al\mbox{.}}{2012}]%
        {veras2012visualizing}
\bibfield{author}{\bibinfo{person}{Rafael Veras}, \bibinfo{person}{Julie
  Thorpe}, {and} \bibinfo{person}{Christopher Collins}.}
  \bibinfo{year}{2012}\natexlab{}.
\newblock \showarticletitle{Visualizing semantics in passwords: The role of
  dates}. In \bibinfo{booktitle}{\emph{Proceedings of the ninth international
  symposium on visualization for cyber security}}. \bibinfo{pages}{88--95}.
\newblock


\bibitem[\protect\citeauthoryear{Voutilainen}{Voutilainen}{2003}]%
        {voutilainen2003part}
\bibfield{author}{\bibinfo{person}{Atro Voutilainen}.}
  \bibinfo{year}{2003}\natexlab{}.
\newblock \bibinfo{booktitle}{\emph{Part-of-speech tagging}}.
  Vol.~\bibinfo{volume}{219}.
\newblock \bibinfo{publisher}{The Oxford handbook of computational
  linguistics}.
\newblock


\bibitem[\protect\citeauthoryear{Wang, He, Cheng, and Wang}{Wang
  et~al\mbox{.}}{2016a}]%
        {wang2016fuzzypsm}
\bibfield{author}{\bibinfo{person}{Ding Wang}, \bibinfo{person}{Debiao He},
  \bibinfo{person}{Haibo Cheng}, {and} \bibinfo{person}{Ping Wang}.}
  \bibinfo{year}{2016}\natexlab{a}.
\newblock \showarticletitle{fuzzyPSM: A new password strength meter using fuzzy
  probabilistic context-free grammars}. In \bibinfo{booktitle}{\emph{2016 46th
  Annual IEEE/IFIP International Conference on Dependable Systems and Networks
  (DSN)}}. IEEE, \bibinfo{pages}{595--606}.
\newblock


\bibitem[\protect\citeauthoryear{Wang and Wang}{Wang and Wang}{2015}]%
        {wang2015emperor}
\bibfield{author}{\bibinfo{person}{Ding Wang} {and} \bibinfo{person}{Ping
  Wang}.} \bibinfo{year}{2015}\natexlab{}.
\newblock \showarticletitle{The emperor’s new password creation policies}. In
  \bibinfo{booktitle}{\emph{European Symposium on Research in Computer
  Security}}. Springer, \bibinfo{pages}{456--477}.
\newblock


\bibitem[\protect\citeauthoryear{Wang, Wang, He, and Tian}{Wang
  et~al\mbox{.}}{2019}]%
        {wang2019birthday}
\bibfield{author}{\bibinfo{person}{Ding Wang}, \bibinfo{person}{Ping Wang},
  \bibinfo{person}{Debiao He}, {and} \bibinfo{person}{Yuan Tian}.}
  \bibinfo{year}{2019}\natexlab{}.
\newblock \showarticletitle{Birthday, name and bifacial-security: understanding
  passwords of Chinese web users}. In \bibinfo{booktitle}{\emph{28th USENIX
  Security Symposium (USENIX Security 19)}}. \bibinfo{pages}{1537--1555}.
\newblock


\bibitem[\protect\citeauthoryear{Wang, Zhang, Wang, Yan, and Huang}{Wang
  et~al\mbox{.}}{2016b}]%
        {wang2016targeted}
\bibfield{author}{\bibinfo{person}{Ding Wang}, \bibinfo{person}{Zijian Zhang},
  \bibinfo{person}{Ping Wang}, \bibinfo{person}{Jeff Yan}, {and}
  \bibinfo{person}{Xinyi Huang}.} \bibinfo{year}{2016}\natexlab{b}.
\newblock \showarticletitle{Targeted online password guessing: An
  underestimated threat}. In \bibinfo{booktitle}{\emph{Proceedings of the 2016
  ACM SIGSAC conference on computer and communications security}}.
  \bibinfo{pages}{1242--1254}.
\newblock


\bibitem[\protect\citeauthoryear{Wash, Rader, Berman, and Wellmer}{Wash
  et~al\mbox{.}}{2016}]%
        {wash2016understanding}
\bibfield{author}{\bibinfo{person}{Rick Wash}, \bibinfo{person}{Emilee Rader},
  \bibinfo{person}{Ruthie Berman}, {and} \bibinfo{person}{Zac Wellmer}.}
  \bibinfo{year}{2016}\natexlab{}.
\newblock \showarticletitle{Understanding password choices: How frequently
  entered passwords are re-used across websites}. In
  \bibinfo{booktitle}{\emph{Twelfth Symposium on Usable Privacy and Security
  (SOUPS 2016)}}. \bibinfo{pages}{175--188}.
\newblock


\bibitem[\protect\citeauthoryear{Wei, Golla, and Ur}{Wei et~al\mbox{.}}{2018}]%
        {wei2018password}
\bibfield{author}{\bibinfo{person}{Miranda Wei}, \bibinfo{person}{Maximilian
  Golla}, {and} \bibinfo{person}{Blase Ur}.} \bibinfo{year}{2018}\natexlab{}.
\newblock \showarticletitle{The password doesn’t fall far: How service
  influences password choice}.
\newblock \bibinfo{journal}{\emph{Who Are You}}  \bibinfo{volume}{87}
  (\bibinfo{year}{2018}), \bibinfo{pages}{108--112}.
\newblock


\bibitem[\protect\citeauthoryear{Weir, Aggarwal, Collins, and Stern}{Weir
  et~al\mbox{.}}{2010}]%
        {weir2010testing}
\bibfield{author}{\bibinfo{person}{Matt Weir}, \bibinfo{person}{Sudhir
  Aggarwal}, \bibinfo{person}{Michael Collins}, {and} \bibinfo{person}{Henry
  Stern}.} \bibinfo{year}{2010}\natexlab{}.
\newblock \showarticletitle{Testing metrics for password creation policies by
  attacking large sets of revealed passwords}. In
  \bibinfo{booktitle}{\emph{Proceedings of the 17th ACM conference on Computer
  and communications security}}. \bibinfo{pages}{162--175}.
\newblock


\bibitem[\protect\citeauthoryear{Weir, Aggarwal, De~Medeiros, and Glodek}{Weir
  et~al\mbox{.}}{2009}]%
        {weir2009password}
\bibfield{author}{\bibinfo{person}{Matt Weir}, \bibinfo{person}{Sudhir
  Aggarwal}, \bibinfo{person}{Breno De~Medeiros}, {and} \bibinfo{person}{Bill
  Glodek}.} \bibinfo{year}{2009}\natexlab{}.
\newblock \showarticletitle{Password cracking using probabilistic context-free
  grammars}. In \bibinfo{booktitle}{\emph{2009 30th IEEE Symposium on Security
  and Privacy}}. IEEE, \bibinfo{pages}{391--405}.
\newblock


\bibitem[\protect\citeauthoryear{Wheeler}{Wheeler}{2016}]%
        {wheeler2016zxcvbn}
\bibfield{author}{\bibinfo{person}{Daniel~Lowe Wheeler}.}
  \bibinfo{year}{2016}\natexlab{}.
\newblock \showarticletitle{zxcvbn:$\{$Low-Budget$\}$ Password Strength
  Estimation}. In \bibinfo{booktitle}{\emph{25th USENIX Security Symposium
  (USENIX Security 16)}}. \bibinfo{pages}{157--173}.
\newblock


\bibitem[\protect\citeauthoryear{Wilt, Thayer, and Ruml}{Wilt
  et~al\mbox{.}}{2010}]%
        {wilt2010comparison}
\bibfield{author}{\bibinfo{person}{Christopher~Makoto Wilt},
  \bibinfo{person}{Jordan~Tyler Thayer}, {and} \bibinfo{person}{Wheeler Ruml}.}
  \bibinfo{year}{2010}\natexlab{}.
\newblock \showarticletitle{A comparison of greedy search algorithms}. In
  \bibinfo{booktitle}{\emph{third annual symposium on combinatorial search}}.
\newblock


\bibitem[\protect\citeauthoryear{Woo and Mirkovic}{Woo and Mirkovic}{2018}]%
        {woo2018guidedpass}
\bibfield{author}{\bibinfo{person}{Simon~S Woo} {and} \bibinfo{person}{Jelena
  Mirkovic}.} \bibinfo{year}{2018}\natexlab{}.
\newblock \showarticletitle{GuidedPass: helping users to create strong and
  memorable passwords}. In \bibinfo{booktitle}{\emph{International Symposium on
  Research in Attacks, Intrusions, and Defenses}}. Springer,
  \bibinfo{pages}{250--270}.
\newblock


\bibitem[\protect\citeauthoryear{Xia, Yi, Liu, Jiang, Wang, and Zhu}{Xia
  et~al\mbox{.}}{2019}]%
        {xia2019genpass}
\bibfield{author}{\bibinfo{person}{Zhiyang Xia}, \bibinfo{person}{Ping Yi},
  \bibinfo{person}{Yunyu Liu}, \bibinfo{person}{Bo Jiang}, \bibinfo{person}{Wei
  Wang}, {and} \bibinfo{person}{Ting Zhu}.} \bibinfo{year}{2019}\natexlab{}.
\newblock \showarticletitle{GENPass: A multi-source deep learning model for
  password guessing}.
\newblock \bibinfo{journal}{\emph{IEEE Transactions on Multimedia}}
  \bibinfo{volume}{22}, \bibinfo{number}{5} (\bibinfo{year}{2019}),
  \bibinfo{pages}{1323--1332}.
\newblock


\bibitem[\protect\citeauthoryear{Xie, Zhang, Yin, and Li}{Xie
  et~al\mbox{.}}{2020}]%
        {xie2020new}
\bibfield{author}{\bibinfo{person}{Zhijie Xie}, \bibinfo{person}{Min Zhang},
  \bibinfo{person}{Anqi Yin}, {and} \bibinfo{person}{Zhenhan Li}.}
  \bibinfo{year}{2020}\natexlab{}.
\newblock \showarticletitle{A new targeted password guessing model}. In
  \bibinfo{booktitle}{\emph{Australasian Conference on Information Security and
  Privacy}}. Springer, \bibinfo{pages}{350--368}.
\newblock


\bibitem[\protect\citeauthoryear{Xu, Ge, Qiu, Huang, Gong, Guo, and Lian}{Xu
  et~al\mbox{.}}{2017}]%
        {xu2017password}
\bibfield{author}{\bibinfo{person}{Lingzhi Xu}, \bibinfo{person}{Can Ge},
  \bibinfo{person}{Weidong Qiu}, \bibinfo{person}{Zheng Huang},
  \bibinfo{person}{Zheng Gong}, \bibinfo{person}{Jie Guo}, {and}
  \bibinfo{person}{Huijuan Lian}.} \bibinfo{year}{2017}\natexlab{}.
\newblock \showarticletitle{Password guessing based on LSTM recurrent neural
  networks}. In \bibinfo{booktitle}{\emph{2017 IEEE International Conference on
  Computational Science and Engineering (CSE) and IEEE International Conference
  on Embedded and Ubiquitous Computing (EUC)}}, Vol.~\bibinfo{volume}{1}. IEEE,
  \bibinfo{pages}{785--788}.
\newblock


\bibitem[\protect\citeauthoryear{Yang, Hu, Zhang, Wei, and Liu}{Yang
  et~al\mbox{.}}{2021}]%
        {yang2021studies}
\bibfield{author}{\bibinfo{person}{Kunyu Yang}, \bibinfo{person}{Xuexian Hu},
  \bibinfo{person}{Qihui Zhang}, \bibinfo{person}{Jianghong Wei}, {and}
  \bibinfo{person}{Wenfen Liu}.} \bibinfo{year}{2021}\natexlab{}.
\newblock \showarticletitle{Studies of Keyboard Patterns in Passwords:
  Recognition, Characteristics and Strength Evolution}. In
  \bibinfo{booktitle}{\emph{International Conference on Information and
  Communications Security}}. Springer, \bibinfo{pages}{153--168}.
\newblock


\bibitem[\protect\citeauthoryear{Yang, Li, Molloy, Park, and Chari}{Yang
  et~al\mbox{.}}{2016}]%
        {yang2016comparing}
\bibfield{author}{\bibinfo{person}{Weining Yang}, \bibinfo{person}{Ninghui Li},
  \bibinfo{person}{Ian~M Molloy}, \bibinfo{person}{Youngja Park}, {and}
  \bibinfo{person}{Suresh~N Chari}.} \bibinfo{year}{2016}\natexlab{}.
\newblock \showarticletitle{Comparing password ranking algorithms on real-world
  password datasets}. In \bibinfo{booktitle}{\emph{European Symposium on
  Research in Computer Security}}. Springer, \bibinfo{pages}{69--90}.
\newblock


\bibitem[\protect\citeauthoryear{Yu}{Yu}{2022}]%
        {yu2022deep}
\bibfield{author}{\bibinfo{person}{Fangyi Yu}.}
  \bibinfo{year}{2022}\natexlab{}.
\newblock \showarticletitle{On Deep Learning in Password Guessing, a Survey}.
\newblock \bibinfo{journal}{\emph{arXiv preprint arXiv:2208.10413}}
  (\bibinfo{year}{2022}).
\newblock


\bibitem[\protect\citeauthoryear{Zhang, Yang, Zheng, You, Su, and Ma}{Zhang
  et~al\mbox{.}}{2020c}]%
        {zhang2020preliminary}
\bibfield{author}{\bibinfo{person}{Jing Zhang}, \bibinfo{person}{Chao Yang},
  \bibinfo{person}{Yu Zheng}, \bibinfo{person}{Wei You},
  \bibinfo{person}{Ruidan Su}, {and} \bibinfo{person}{Jianfeng Ma}.}
  \bibinfo{year}{2020}\natexlab{c}.
\newblock \showarticletitle{A preliminary analysis of password guessing
  algorithm}. In \bibinfo{booktitle}{\emph{2020 29th International Conference
  on Computer Communications and Networks (ICCCN)}}. IEEE,
  \bibinfo{pages}{1--9}.
\newblock


\bibitem[\protect\citeauthoryear{Zhang, Zhang, Hu, and Liu}{Zhang
  et~al\mbox{.}}{2018}]%
        {zhang2018password}
\bibfield{author}{\bibinfo{person}{Mengli Zhang}, \bibinfo{person}{Qihui
  Zhang}, \bibinfo{person}{Xuexian Hu}, {and} \bibinfo{person}{Wenfen Liu}.}
  \bibinfo{year}{2018}\natexlab{}.
\newblock \showarticletitle{A Password Cracking Method Based On Structure
  Partition and BiLSTM Recurrent Neural Network}. In
  \bibinfo{booktitle}{\emph{Proceedings of the 8th International Conference on
  Communication and Network Security}}. \bibinfo{pages}{79--83}.
\newblock


\bibitem[\protect\citeauthoryear{Zhang, Cheng, Qin, Li, and Shi}{Zhang
  et~al\mbox{.}}{2020a}]%
        {zhang2020deep}
\bibfield{author}{\bibinfo{person}{Tao Zhang}, \bibinfo{person}{Zelei Cheng},
  \bibinfo{person}{Yi Qin}, \bibinfo{person}{Qiang Li}, {and}
  \bibinfo{person}{Lin Shi}.} \bibinfo{year}{2020}\natexlab{a}.
\newblock \showarticletitle{Deep Learning for Password Guessing and Password
  Strength Evaluation, A Survey}. In \bibinfo{booktitle}{\emph{2020 IEEE 19th
  International Conference on Trust, Security and Privacy in Computing and
  Communications (TrustCom)}}. IEEE, \bibinfo{pages}{1162--1166}.
\newblock


\bibitem[\protect\citeauthoryear{Zhang, Monrose, and Reiter}{Zhang
  et~al\mbox{.}}{2010}]%
        {zhang2010security}
\bibfield{author}{\bibinfo{person}{Yinqian Zhang}, \bibinfo{person}{Fabian
  Monrose}, {and} \bibinfo{person}{Michael~K Reiter}.}
  \bibinfo{year}{2010}\natexlab{}.
\newblock \showarticletitle{The security of modern password expiration: An
  algorithmic framework and empirical analysis}. In
  \bibinfo{booktitle}{\emph{Proceedings of the 17th ACM conference on Computer
  and communications security}}. \bibinfo{pages}{176--186}.
\newblock


\bibitem[\protect\citeauthoryear{Zhang, Xian, and Yu}{Zhang
  et~al\mbox{.}}{2020b}]%
        {zhang2020csnn}
\bibfield{author}{\bibinfo{person}{Yi Zhang}, \bibinfo{person}{Hequn Xian},
  {and} \bibinfo{person}{Aimin Yu}.} \bibinfo{year}{2020}\natexlab{b}.
\newblock \showarticletitle{CSNN: Password guessing method based on Chinese
  syllables and neural network}.
\newblock \bibinfo{journal}{\emph{Peer-to-Peer Networking and Applications}}
  \bibinfo{volume}{13}, \bibinfo{number}{6} (\bibinfo{year}{2020}),
  \bibinfo{pages}{2237--2250}.
\newblock


\bibitem[\protect\citeauthoryear{Zhang-Kennedy, Chiasson, and
  Biddle}{Zhang-Kennedy et~al\mbox{.}}{2013}]%
        {zhang2013password}
\bibfield{author}{\bibinfo{person}{Leah Zhang-Kennedy}, \bibinfo{person}{Sonia
  Chiasson}, {and} \bibinfo{person}{Robert Biddle}.}
  \bibinfo{year}{2013}\natexlab{}.
\newblock \showarticletitle{Password advice shouldn't be boring: Visualizing
  password guessing attacks}. In \bibinfo{booktitle}{\emph{2013 APWG eCrime
  Researchers Summit}}. IEEE, \bibinfo{pages}{1--11}.
\newblock


\bibitem[\protect\citeauthoryear{Zhao and Yue}{Zhao and Yue}{2013}]%
        {zhao2013all}
\bibfield{author}{\bibinfo{person}{Rui Zhao} {and} \bibinfo{person}{Chuan
  Yue}.} \bibinfo{year}{2013}\natexlab{}.
\newblock \showarticletitle{All your browser-saved passwords could belong to
  us: A security analysis and a cloud-based new design}. In
  \bibinfo{booktitle}{\emph{Proceedings of the third ACM conference on Data and
  application security and privacy}}. \bibinfo{pages}{333--340}.
\newblock


\bibitem[\protect\citeauthoryear{Zhou, Wu, Lu, Xu, and Cheung}{Zhou
  et~al\mbox{.}}{2022}]%
        {zhou2022password}
\bibfield{author}{\bibinfo{person}{Tao Zhou}, \bibinfo{person}{Hao-Tian Wu},
  \bibinfo{person}{Hui Lu}, \bibinfo{person}{Peiming Xu}, {and}
  \bibinfo{person}{Yiu-Ming Cheung}.} \bibinfo{year}{2022}\natexlab{}.
\newblock \showarticletitle{Password Guessing Based on GAN with
  Gumbel-Softmax}.
\newblock \bibinfo{journal}{\emph{Security and Communication Networks}}
  \bibinfo{volume}{2022} (\bibinfo{year}{2022}).
\newblock


\end{thebibliography}

%
%
%
%
%
%
%
%

\end{document}